\documentclass[aps,pre,twocolumn,longbibliography,amsmath,amssymb,amsfonts,superscriptaddress,nofootinbib]{revtex4-2}
\usepackage{graphicx}
\usepackage{dcolumn}
\usepackage{bm}
\usepackage[usenames,dvipsnames]{color}
\usepackage{multirow}
\usepackage{hyperref}
\usepackage{soul}
\usepackage{natbib}
\usepackage{amsmath, amssymb}
\usepackage{bbold}
\usepackage{physics}
\usepackage{cancel}

\usepackage{times}

\usepackage{tikz}
\usetikzlibrary{"arrows","shapes.geometric"}
\usetikzlibrary{"calc"}

\definecolor{magenta}{rgb}{0.8,0.2,0.8}

\begin{document}

\title{Nonconserved critical dynamics of the two-dimensional Ising model as a surface kinetic roughening process}

\author{H\'ector Vaquero del Pino}
\email{hector001@e.ntu.edu.sg}
\affiliation{Division of Physics and Applied Physics, School of Physical and Mathematical Sciences,
Nanyang Technological University, 21 Nanyang Link, Singapore 637371}
\author{Rodolfo Cuerno}
\email{cuerno@math.uc3m.es}
\affiliation{Universidad Carlos III de Madrid, Departamento de Matemáticas and Grupo Interdisciplinar de Sistemas Complejos (GISC), Avenida de la Universidad 30, 28911 Legan{\'e}s (Madrid), Spain}

\begin{abstract}
We have revisited the non-conserved (or model A) critical dynamics of the two-dimensional Ising model through numerical simulations of its lattice and continuum formulations ---Glauber dynamics and the time-dependent Ginzburg-Landau (TDGL) equation, respectively---, to analyze them with current tools from surface kinetic roughening. Our study examines two critical quenches, one from an ordered and a different one from a disordered initial state, for both of which we assess the dynamic scaling ansatz, the critical exponent values, and the fluctuation field statistics that occur. 
Notably, the dynamic scaling ansatz followed by the system strongly depends on the initial condition: a critical quench from the ordered phase follows Family-Vicsek (FV) scaling, while a critical quench from the disordered phase shows an initial overgrowth regime with intrinsic anomalous roughening, followed by relaxation to equilibrium. This behavior is explained in terms of the dynamical instability of the stochastic Ginzburg-Landau equation at the critical temperature, whereby the linearly unstable term is eventually stabilized by nonlinear interactions. For both quenches we have determined the occurrence of probability distribution functions for the field fluctuations, which are time-independent along the non-equilibrium dynamics when suitably normalized by the time-dependent fluctuation amplitude. Additionally, we have developed a related interface model for a field which scales as the space integral of the TDGL field (integral GL model). Numerical simulations of this model reveal either FV or faceted anomalous roughening, depending on the critical quenched performed, as well as an emergent symmetry in the fluctuation statistics for a critical quench from the ordered phase.
\end{abstract}

\maketitle

\section{Introduction}
\label{sec:intro}

The Ising model stands out as the quintessential prototype of continuous phase transitions and critical phenomena \cite{Binney,Goldenfeld,kardar}, showcasing the emergence of universal collective behavior from a seemingly simple system \cite{Ising_2017}. Of particular interest is its two-dimensional (2D) instance, notable for its analytical solution in equilibrium, achieved by Lars Onsager in 1944 \cite{Onsager}. Away from equilibrium  \cite{hohenberg,ojalvo,Ma2018,Mazenko,kardar,Tauber}, the Ising model also illustrates paradigms of critical dynamics, like the classic models A and B for relaxation of a scalar order-parameter field in the absence or presence of conservation laws, respectively.

Historically, knowledge on equilibrium critical systems has enabled the study of a large variety of other systems with strong space-time correlations, whose behavior can be similarly classified in terms of universality classes \cite{Odor2004}. One such context is that of surface kinetic roughening \cite{Barabasi,krug}, where the physical object of interest is a rough surface or interface whose time evolution results from the interplay between deterministic and stochastic processes. Often such surfaces display fractal properties characterized by critical exponents which are related, e.g., with their fractal dimension \cite{Barabasi,Mozo22}. The critical behavior of these systems often satisfies the same dynamic scaling ansatz found for the classic models A and B, which become thus generalized far from equilibrium \cite{hohenberg,Tauber}.

A remarkable example of the way in which surface roughening is enriching the field of critical dynamics is provided by the celebrated Kardar-Parisi-Zhang (KPZ) equation \cite{Kardar1986}. Indeed, this system was originally proposed as an insightful generalization of model A to describe the dynamics of a kinetically rough surface. However, the ensuing universality class \cite{Kriecherbauer10,Halpin-Healy2015,Takeuchi18} is lately being found to encompass a wide range of far-from-equilibrium systems, which goes well beyond  interfacial phenomena. Recent examples include, e.g., driven-dissipative Bose-Einstein condensates \cite{Fontaine2022}, quantum spin chains \cite{Wei22}, Anderson localization in disordered conductors \cite{Mu2024}, or synchronization of nonlinear oscillator lattices with time-dependent noise \cite{Gutierrez24}. Notably also, the critical behavior displayed by the KPZ universality class is quite rich, implying universal behavior not only from the point of view of the critical exponent values, but also with respect to the probability distribution function (PDF) of one-point field fluctuations \cite{Kriecherbauer10,Halpin-Healy2015,Takeuchi18}. E.g.\ for 1D interfaces, the PDF is a member of the celebrated Tracy-Widom (TW) distribution family \cite{Bertin2006,Makey20}, which generalizes naturally into novel PDF families in higher dimensions \cite{Halpin-Healy2015,Takeuchi18}, as conveniently confirmed, e.g., in growth experiments of rough 2D films \cite{Almeida14,Halpin-Healy2014,Orrillo17}.

Work on surface kinetic roughening has further unveiled novel modes of critical behavior. Specifically, generalizations have been introduced of the standard dynamic scaling ansatz \cite{Tauber} ---usually referred to as the Family-Vicsek (FV) ansatz \cite{Barabasi,krug}--- satisfied e.g.\ by the KPZ equation. Such generalizations are collectively termed anomalous scaling or anomalous kinetic roughening \cite{krug,Schroeder2,Dassarma94,Lopez97,ifisc,Cuerno04} and are frequently found in the presence of particularly strong interface fluctuations, as recently seen, e.g., in models of coffee-ring formation \cite{Barreales22}, in the characterization of the tensionless KPZ equation \cite{Rodriguez22}, or in the synchronization of disordered oscillator rings \cite{Gutierrez23,Gutierrez24b}.

In this paper we revisit the original nonconserved critical dynamics of the Ising universality class in view of recent progress in surface kinetic roughening. To do this, we characterize the space-time correlations of the 2D Ising system (both in its discrete and in continuous formulations) as if they correspond to a two-dimensional rough surface evolving in 3D space. Interestingly, we find that both dynamic scaling ansatzs, FV and anomalous, hold, depending on the initial condition. Note at this that, although well established both, theoretically and experimentally (see e.g.\ Refs.\ \cite{krug,Cuerno04} and, more recently, \cite{Rodriguez22} and therein), anomalous surface roughening remains a peculiar form of dynamic scaling and we find it remarkable that it already shows up no less than for the Ising model A, the paradigm of dynamical critical systems. Also, in view of the importance of the PDF of field fluctuations to unambiguously characterize the kinetic roughening universality class (see recent discussions, e.g., in Refs.\ \cite{Rodriguez21,Marcos22} and other therein), we perform a numerical study of field statistics from a kinetic roughening perspective. Notably again, we find that the fluctuation statistics of the local order parameter in the TDGL model is time-independent within the nonlinear growth regime, once the fluctuations are suitably rescaled by time-dependent factors, in full analogy to the occurrence of Tracy-Widom statistics for the KPZ equation. The assessment of these probability distributions may thus find use in assessing critical properties in the dynamics of statistical mechanical systems, in analogy to their frequent application in equilibrium criticality \cite{Plascak2013,Landau_book}. Our present study of the critical dynamics of the Ising model is finally complemented by the definition and study of a related interface model (integral GL model) for a scalar variable which scales as the space integral of the TDGL field. Numerical simulations now reveal either FV or faceted anomalous roughening, depending on the critical quenched performed, as well as an emergent symmetry in the fluctuating field of this model for a critical quench from the ordered phase.

This paper is organized as follows. Section \ref{sec:mod} introduces the model under investigation. Section \ref{sec:kr} provides an overview of the surface kinetic roughening framework, which can be skipped by readers already familiar with the topic, except perhaps for Sec.\ \ref{sec:em_stat} which contains less standard material. Sections \ref{sec:T=0}, \ref{sec:T=infty}, and \ref{sec:im} present the results for critical quenches from $T=0$, $T=\infty$, and the integral GL model, respectively. Section \ref{sec:disc} offers a discussion of our findings, and Sec.\ \ref{sec:concl} summarizes the main conclusions and outlook of this study. Additional analytical and numerical results are organized into five appendices at the end.

\section{Model details}
\label{sec:mod}
Originally devised to describe the behavior of uniaxial magnets, the Ising model represents
a $d$-dimensional lattice of $N$ sites, each hosting a magnetic dipole moment of an
atomic spin \cite{Binney,kardar,Goldenfeld}. The spin of site $i$ is described by a discrete variable 
$\sigma_i=\pm 1$. 
The (classical) Hamiltonian of this system in the absence of an external magnetic field reads
\begin{equation}
    \mathcal{H}=-J\sum_{\langle ij\rangle}\sigma_i \sigma_j ,
\end{equation}
where $J>0$ is a ferromagnetic coupling constant and in this work $i$ takes as values the vertices of a 2D square lattice with periodic boundary conditions. The non-conserved dynamics of this model will be simulated numerically through the Markov chain Monte Carlo method with Glauber acceptance rule for the probability, $h(\sigma_i'|\sigma_i)$, of acceptance of a spin flip $\sigma_i \to \sigma'_i$ as \cite{newman,Toral}
\begin{equation}
    h(\sigma_i'|\sigma_i)=\left(1+e^{\beta \Delta \mathcal{H}}\right)^{-1},
\end{equation}
where $\beta=1/k_B T$ with $T$ being temperature and $k_B$ Boltzmann's constant, and $\Delta \mathcal{H}$ is the energy variation implied by that spin flip.

The Ising universality class can be alternatively addressed through a coarse-grained approach \cite{hohenberg,Goldenfeld,ojalvo,Ma2018,Mazenko,kardar,Tauber} leading to the real, stochastic time-dependent Ginzburg-Landau (TDGL, or Allen-Cahn) equation, namely,
\begin{equation}\label{eq:TDGL}
    \partial_t \phi=-r\phi-u\phi^{3}+\nu\nabla^{2}\phi+\zeta,
\end{equation}
where $\phi(x,y,t)\in \mathbb{R}$ is the local magnetization field and $\zeta(\textbf{r},t)$ represents thermal noise such that
\begin{align}
    &\langle\zeta(\textbf{r},t) \rangle = 0, \label{eq:noise}\\
    &\langle\zeta(\textbf{r},t)\zeta(\textbf{r}',t')\rangle=2\Gamma\delta^{(d)}(\textbf{r}-\textbf{r}')\delta(t-t'),
\end{align}
where $\textbf{r}\in \mathbb{R}^d$ and $\Gamma=k_B T$ is the fluctuation-dissipation condition \cite{hohenberg,Goldenfeld,ojalvo,Ma2018,Mazenko,kardar,Tauber}. We have simulated numerically Eq.\ \eqref{eq:TDGL} for $d=2$ on $L\times L$ substrates with periodic boundary conditions, using finite differences in space and the Euler-Maruyama method \cite{Toral} for the time evolution, in view of the stochastic term. Note, mathematically the continuum limit of the 2D stochastic TDGL or Allen-Cahn equation has been shown \cite{Hairer2012,Ryser12} not to retrieve a physically meaningful behavior. As frequently done in the literature, we will interpret our simulations as numerical approximations of an equation driven by a noise field which has a finite correlation length \cite{ojalvo,Monroy21}; as will be seen below, the ensuing results will underscore the universality of the behavior of our Glauber dynamics simulations of the discrete Ising model. Further parameters of our numerical simulations and the units employed are specified in Appendix \ref{AppC}.

\section{Surface kinetic roughening}
\label{sec:kr}
As mentioned in Sec.\ \ref{sec:intro}, surface kinetic roughening is an important instance of non-equilibrium critical behavior. The tools and methods developed in this context facilitate the exploration of critical properties extending beyond critical exponent values, such as probability distribution functions. Leveraging the general nature of this framework, we will employ these methods to study the out-of-equilibrium critical properties of the $2\text{D}$ Ising model.

\subsection{Observables}
\begin{figure*}[!t]
    \centering
    \includegraphics[width=\textwidth]{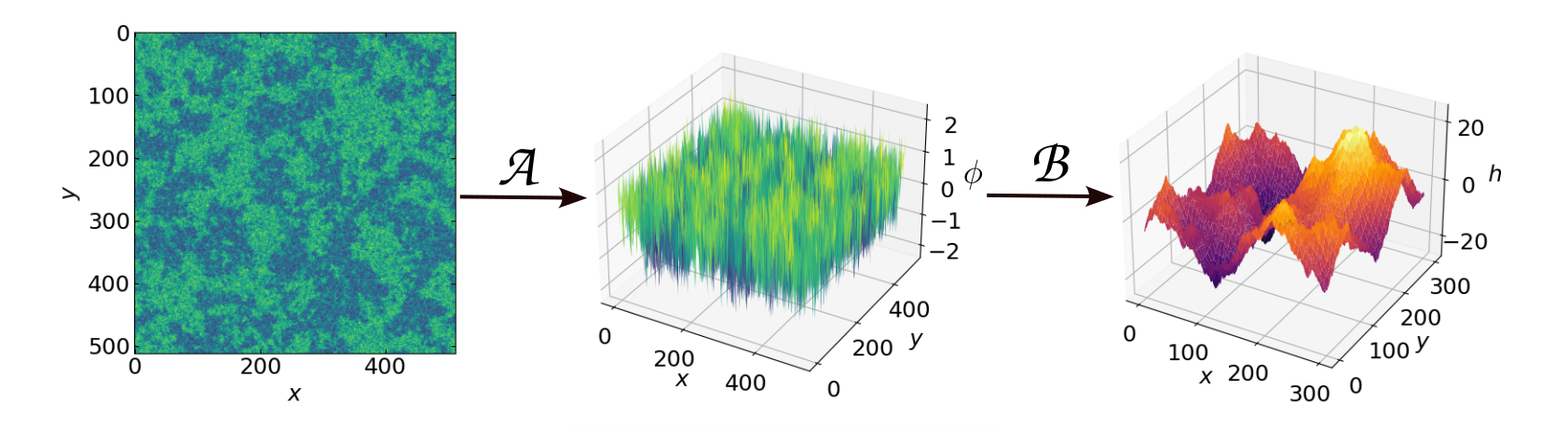}
    \caption{$\mathcal{A}$ denotes the mapping whereby the local magnetization $\phi(\mathbf{r},t)$ of the 2D Ising model (left panel) at the critical temperature is interpreted as the height field of a kinetically rough surface (middle panel). This mapping is studied in Secs.\ \ref{sec:T=0} and \ref{sec:T=infty}. Then, $\mathcal{B}$ denotes the mapping from such a surface to the integral GL model, a different but related rough surface $h(\mathbf{r},t)$, imaged on the right panel and addressed in Sec.\ \ref{sec:im}.}
    \label{fig:mapping}
\end{figure*}

From the point of view of surface kinetic roughening, the scalar field $\phi(\textbf{r},t)\in\mathbb{R}$ measures the height of a surface above point $\textbf{r}$ on a $d$-dimensional substrate at time $t$. See Fig.\ \ref{fig:mapping} for an example in which $d=2$.
Hence, fluctuations in $\phi$ correspond to height differences. Thus, it is natural to define the average amplitude of fluctuations as the surface roughness, $W(L,t)$. A perfectly smooth surface is flat and has a null roughness. Specifically,
\begin{equation}\label{eq:W}
    W^{2}(L,t)=\frac{1}{L^d}\int d^d \textbf{r} \left( \langle\phi^2(\textbf{r},t)\rangle-\langle\phi(\textbf{r},t)\rangle^2\right),
\end{equation}
where angular brackets denote average over realizations of the noise. The roughness can be related to the two-point correlation function in Fourier space or surface structure factor, denoted $S(\textbf{q},t)$, through Parseval's theorem as
\begin{equation}\label{eq:W_S_relation}
    W^{2}(L,t)=\frac{1}{L^d}\int \frac{d^d \textbf{q}}{(2\pi)^d}  S(\textbf{q},t),
\end{equation}
where
\begin{equation}\label{eq:SF}
\begin{split}
S(\textbf{q},t)&=\langle \hat{\phi}(\textbf{q},t)\hat{\phi}(-\textbf{q},t) \rangle=\langle |\hat{\phi}(\textbf{q},t)|^2 \rangle ,\\
\hat{\phi}(\textbf{q},t) &= \int d^{d}\textbf{r} \; \phi(\textbf{r},t)e^{i\textbf{q}\cdot\textbf{r}},
\end{split}
\end{equation}
and $\mathbf{q}$ is $d$-dimensional wave vector. Alternatively, two-point correlation functions can indeed be  evaluated in real space too, being mathematically equivalent to $S(\textbf{q},t)$ \cite{Barabasi,krug}.

\subsection{Dynamic scaling ansatzs}\label{sec:dsa}
Observables in kinetic roughening are scale-free both in space and time. However, for a finite lateral system size $L$ the correlation length between field values at different points, $\xi(t)$ ---which increases with time like in critical dynamical systems as $\xi(t)\sim t^z$, with $z$ the dynamic critical exponent \cite{Tauber}--- is bounded above by $L$. This leads the system to saturate to steady state at a time $t_{\rm sat}\sim L^z$ when $\xi(t_{\rm sat})\approx L$. While for earlier times the roughness is also seen to grow with time as $W(t) \sim t^{\beta}$, where $\beta$ is the growth exponent, saturation to steady state implies $W_{\rm sat}\sim L^{\alpha} \sim t_{\rm sat}^{\beta}\sim (L^z)^{\beta}$, so that $\beta=\alpha/z$.\footnote{Note, the letters $\alpha$ and $\beta$ are frequently employed for the critical exponents characterizing the critical behavior of the Ising model at its Curie phase transition, and/or to denote inverse temperature. From now on, they are to be understood as kinetic roughening exponents.}
All this behavior is encoded into the Family-Vicsek dynamic scaling ansatz for the roughness \cite{Barabasi,krug},
\begin{equation}
    W\sim L^{\alpha}f_{W}\left(\frac{t}{L^z}\right),
\end{equation}
where 
\begin{equation}
    f_{W}(u)\sim \begin{cases}
        u^{\beta}\quad &\text{if}\quad u\ll 1, \\
        \text{const.}\quad &\text{if}\quad u\gg 1.
    \end{cases}
\end{equation}

The dynamic scaling ansatz is also satisfied by correlation functions \cite{Barabasi,krug}. Thus, within FV scaling, the 
structure factor behaves as
\begin{equation}
    S(q,t)\sim q^{-(2\alpha+d)}f_{S}(t^{1/z}q),
    \label{ec:SFV}
\end{equation}
where \begin{equation}
    f_{S}(u)\sim \begin{cases}
        u^{2\alpha+d}\quad &\text{if}\quad u\ll 1, \\
        \text{const.}\quad &\text{if}\quad u\gg 1, \label{ec:SFVf}
    \end{cases}
\end{equation}
where isotropy in real space has been assumed so that $S(\mathbf{q},t) = S(q,t)$ with $q=|\mathbf{q}|$. Indeed, there is a time-dependent cut-off wavenumber $q_c(t) \sim 1/\xi(t) \sim t^{-1/z}$ above which $S(q,t)$ displays correlated $q$-dependent behavior and below which it is an uncorrelated, $q$-independent spectrum. 
Specifically, from Eqs.\ \eqref{ec:SFV} and \eqref{ec:SFVf},
\begin{equation}\label{ec:Sqexpl}
S(q,t)\sim
    \begin{cases}
        t^{(2\alpha+d)/z}\quad &\text{for}\quad q\ll t^{-1/z}, \\
        q^{-(2\alpha+d)}\quad &\text{for}\quad q\gg t^{-1/z}. 
    \end{cases}
\end{equation}
An example of FV dynamic scaling behavior for the structure factor is provided in Appendix \ref{App:EW} for the celebrated Edwards-Wilkinson (EW) equation \cite{Barabasi} which, while defining an important universality class of surface kinetic roughening on its own, is nothing else than the Gaussian approximation of the TDGL equation at its critical temperature \cite{kardar}.

In equilibrium critical phenomena, the convention for the behavior of the structure factor at criticality is
\begin{equation}
    S(q) \sim \frac{1}{q^{2-\eta}},
\end{equation}
which defines the anomalous dimension $\eta$ \cite{Binney,kardar,Goldenfeld,hohenberg} (``anomalous'' in this case is unrelated to the ``anomalous roughening'' to be described below). Comparing with Eq.\ \eqref{ec:Sqexpl}, this exponent is related with the roughness exponent as
\begin{equation}\label{alpha_eta_relation}
    \alpha = \frac{2-d-\eta}{2} .
\end{equation}

\subsubsection{Anomalous roughening} 
In some rough surface growth systems \cite{Schroeder2,Dassarma94,Lopez97,ifisc,Cuerno04}, the scaling behavior of local fluctuations differs from that of the global roughness. 
In such cases, the two-point correlation functions in real space may display a new independent exponent, $\alpha_{\rm loc}\neq\alpha$, termed local roughness exponent. This scenario is collectively termed anomalous roughening or anomalous scaling, and it has been observed in many models and experiments.
Anomalous roughening can be actually seen as a generalization of FV scaling and can arise in three different forms, depending on the number and relative values of new independent exponents. The most complete set of possibilities can be classified \cite{ifisc} in terms of the so-called spectral exponent, $\alpha_s$, that can be most conveniently identified in the behavior of the structure factor, thus 
    \begin{eqnarray}
        \alpha_s <1\Rightarrow \alpha_{\rm loc} = \alpha_s & & \hspace*{1mm} 
        \begin{cases}
            \alpha_s = \alpha \Rightarrow \text{Family-Vicsek}, \\
            \alpha_s \neq \alpha \Rightarrow \text{Intrinsic},
        \end{cases} \\
        \alpha_s >1\Rightarrow \alpha_{\rm loc} = 1 & & \hspace*{1mm}
        \begin{cases}
            \alpha_s = \alpha \Rightarrow \text{Super-rough}, \\
            \alpha_s \neq \alpha \Rightarrow \text{Faceted}.  
        \end{cases}
    \end{eqnarray} \label{eq:scaling_types}
Note that intrinsic and faceted anomalous scaling behaviors are then characterized by two independent roughness exponents, e.g.\ $\alpha$ and $\alpha_s$, while there is a single independent one under super-rough and FV scalings.

The most general anomalous dynamic scaling ansatz for the structure factor reads \cite{ifisc}
\begin{equation}\label{eq:S_anomalous_scaling}
    S(q,t)\sim q^{-(2\alpha+d)}f_{S'}(t^{1/z}q),
\end{equation}
where \begin{equation}\label{eq:anomalous_scaling}
    f_{S'}(u)\sim \begin{cases}
        u^{2\alpha+d}\quad &\text{if}\quad u\ll 1, \\
        u^{2(\alpha-\alpha_s )}\quad &\text{if}\quad u\gg 1,
    \end{cases}
\end{equation}
indeed a generalization of the FV ansatz, Eqs.\ \eqref{ec:SFV}-\eqref{ec:SFVf}, which becomes the particular case of Eq.\ \eqref{eq:anomalous_scaling} in which $\alpha_s=\alpha$. Note that, otherwise, in the presence of anomalous scaling the roughness exponent that characterizes the correlated behavior of the structure factor is not $\alpha$, but $\alpha_s$. Indeed, from Eqs.\ \eqref{eq:S_anomalous_scaling}-\eqref{eq:anomalous_scaling},
\begin{equation}
    S(q,t) \sim q^{-(2\alpha_s+d)} \;\;\; \mbox{for} \; q \gg 1/t^{1/z} .
\end{equation}

\subsection{Fluctuation statistics} \label{sec:PDF} 

While scaling exponents values were generally held to unambiguously determine kinetic roughening universality classes \cite{Barabasi,krug}, more recent developments, largely driven by results on the KPZ class \cite{Kriecherbauer10,Halpin-Healy2015,Takeuchi18}, are underscoring the need for additional universal traits, such as the PDF of height fluctuations, to remove possible ambiguities. In Appendix \ref{App_cKPZ} we describe an example on a non-linear equation ---the celebrated conserved KPZ (CKPZ) equation, representative of a paradigmatic universality class of rough surfaces evolving under conserved dynamics with non-conserved noise \cite{Barabasi,krug}--- and a linear equation \cite{Vivo2012}, both of which have the same set of exponent values and whose universality classes can only be told apart through the statistics of height fluctuations, Gaussian for the linear equation and non-Gaussian for the CKPZ equation \cite{Carrasco2016}. An analogous example is known \cite{Saito12} in the context of the KPZ equation; see further discussions e.g.\ in Refs.\ \cite{Rodriguez21,Marcos22}.

Fluctuation statistics are obtained from the fluctuation field \cite{Prahofer00,Kriecherbauer10,Halpin-Healy2015,Takeuchi18}
\begin{equation} \label{eq:Pra-Spohn}
    \mathcal{X} = \frac{\phi(\textbf{r},t)-\langle \phi(\textbf{r},t)\rangle}{\mathrm{std}(\phi(\textbf{r},t))},
\end{equation}
where $\mathrm{std}(\phi(\textbf{r},t))$ stands for standard deviation. The histogram for such a quantity is then computed and averaged, yielding the fluctuations PDF, denoted by $P(\mathcal{X})$. Note, under kinetic roughening conditions $\mathrm{std}(\phi) \sim t^{\beta}$, where the value of $\beta$ may change if, e.g., time crossover effects come into play.
Still, PDF curves computed from different $\mathcal{X}(\textbf{r},t)$ belonging to the same dynamical regime will collapse into the same time-independent distribution $P(\mathcal{X})$, as seminally shown by Pr\"ahofer and Spohn (PS) \cite{Prahofer00} in the context of the KPZ universality class. To facilitate comparison, we will also compute time-averaged PDFs by standardizing the data (normalizing to zero mean and unit variance) and smoothing through averaging over the temporal window corresponding to the sampled non equilibrium regime.

Finally, to further characterize the PDF, two of its cumulants, the skewness and kurtosis, are frequently analyzed \cite{Kriecherbauer10,Halpin-Healy2015,Takeuchi18}. For instance, given that the skewness measures the asymmetry of the PDF, a non-zero value for it indicates the relevance of nonlinear effects, as linear stochastic growth equations are associated with Gaussian statistics and hence zero skewness. Specifically, the skewness reads
\begin{equation}\label{eq:skew}
    \mathcal{S}(t)=\frac{1}{W^3 (L,t)}\left\langle \frac{1}{L^d}\int d^d \textbf{r} \left[\phi(\textbf{r},t)-\langle\phi(\textbf{r},t)\rangle \right]^3 \right\rangle .
\end{equation}
Likewise, the kurtosis is a measure of the size of the PDF tails, given by
\begin{equation}\label{eq:kur}
    \mathcal{K}(t)=\frac{1}{W^4 (L,t)}\left\langle \frac{1}{L^d}\int d^d \textbf{r} \left[\phi(\textbf{r},t)-\langle\phi(\textbf{r},t)\rangle \right]^4 \right\rangle,
\end{equation}
which equals exactly 3 for a Gaussian distribution. For $\mathcal{K}>3$, tails are heavier relative to those of a normal distribution, resulting into a higher frequency of outliers.

\subsubsection{Emergent statistics}
\label{sec:em_stat}

As already noted, one of the most relevant kinetic roughening universality classes is that of the KPZ equation. This continuum model describes surface dynamics when irreversible growth along the local normal direction competes with smoothening through surface tension, and time-dependent noise, and reads \cite{Kardar1986,Barabasi,krug}
\begin{equation}\label{eq:KPZ_eq}
    \partial_t h(\mathbf{r},t)=\nu \nabla^{2}h(\mathbf{r},t)+\frac{\lambda}{2}\left(\nabla h(\mathbf{r},t)\right)^2+\mathbf{\zeta}(\mathbf{r},t),
\end{equation}
where, as is customary, here we use $h(\mathbf{r},t)$ to denote the height of the surface above position $\mathbf{r}$ on a reference substrate at time $t$, $\zeta(\mathbf{r},t)$ is delta-correlated Gaussian noise, as in Eq.\ \eqref{eq:noise}, and $\nu>0$, $\lambda$, and $\Gamma$ are parameters quantifying surface tension, lateral growth, and randomness, respectively. The KPZ universality class has been observed in a wide range of systems \cite{Barabasi,Halpin-Healy2015,Takeuchi18}. 
It satisfies the FV dynamic scaling ansatz with the exact exponent values (for 1D interfaces) $\alpha = 1/2$, $z=3/2$, and $\beta=1/3$. The time evolution of e.g.\ a 1D interface described by this equation basically displays two growth regimes \cite{Prolhac11}: an early-time linear (hence, Gaussian) regime followed by a non-linear regime displaying TW statistics and the quoted scaling exponent values, previous to saturation to steady state. Interestingly, the sign of the skewness of the TW fluctuations is that of $\lambda$ in Eq.\ \eqref{eq:KPZ_eq} \cite{krug,Kriecherbauer10,Takeuchi18}.

Another interesting property of fluctuation statistics is the possibility for emergent behavior; for instance, the occurrence of symmetries in the large-scale behavior which are not present in the ``microscopic'' model. Using again the example of the 1D KPZ equation, 
related behavior is seen to occur in the related noisy Burgers equation. This continuum model can be obtained in 1D by differentiating both sides of Eq.\ \eqref{eq:KPZ_eq} and denoting $u(x,t)=\partial_x h(x,t)$, thus
\begin{equation} \label{eq:Burgers_cn}
    \partial_t u(x,t)=\nu\partial^{2}_x u+\lambda u\partial_x u+\partial_x \mathbf{\zeta}(x,t), 
\end{equation}
in which noise takes on a conserved form. A natural interpretation of the non-zero skewness of the KPZ equation is \cite{krug} to associate it with the lack of up-down ($h \leftrightarrow -h$) symmetry of the system, thus implementing irreversible interface motion along a preferred growth direction. Although one might naively then expect non-zero skewness for the related Eq.\ \eqref{eq:Burgers_cn}, such is not the case \cite{Rodriguez-Fernandez20}. Indeed, both Eq.\ \eqref{eq:Burgers_cn} and its variant with non-conserved noise feature nonlinear scaling exponent values, but Gaussian fluctuation PDF \cite{Rodriguez-Fernandez19,Rodriguez-Fernandez20}. The PDF behavior for these Burgers equations can actually be more properly understood in view of their symmetry under a combined ($x \leftrightarrow -x$, $h \leftrightarrow -h$) transformation, notwithstanding the remarkable fact they reveal that the sum [$h(x,t)=\int^x_{x_0} u(x',t) dx'$] of Gaussian variables is TW-distributed \cite{Rodriguez-Fernandez20}! At any rate, this indicates potential nontrivial relations between the PDF of a model and that of its space integral. In the context of the Ising model, these issues are addressed in Sec.\ \ref{sec:im}.

\subsection{Additional exponent measurements}
The dynamic exponent can be computed by finding the relation between the saturation time, inferred from the roughness, and system size, since, as discussed in Sec.\ \ref{sec:dsa}, $t_{\rm sat}\sim L^z$. There are additional, independent methods, such as the time power spectrum method or the domain size method, which will be employed in our study and are summarized next.

\subsubsection{Time power spectrum method}
The time power spectrum method \cite{Kent} takes advantage of the autocorrelation function of the space-averaged magnetization and of the Wiener-Khintchine relation, in order to obtain the power spectrum in time frequency $\omega$ as
\begin{equation}\label{eq:S_w}
    \Omega(\omega)=\lim_{t\rightarrow\infty}\frac{1}{t}\langle|\hat{m}(\omega)|^2\rangle,
\end{equation}
where
\begin{equation}
    \hat{m}(\omega)=\mathcal{F}[\omega]\left(\int d^d \textbf{r}\; \phi(\textbf{r},t)\right),
\end{equation}
and here $\mathcal{F}[\omega]$ denotes time Fourier transform. By applying the FV scaling relations of the two-point correlation function, the time power spectrum at the critical point scales as \cite{Kent}
\begin{equation}
    \Omega(\omega)\sim \frac{1}{V}\omega^{-\mu},
\end{equation}
where $V$ is system volume and
\begin{equation}
    \mu=1+\frac{2-\eta}{z}=1+\frac{2\alpha+d}{z}.
    \label{eq:mu_vs_eta}
\end{equation}
Hence, by fitting the exponent $\mu$ to the time power spectrum, and introducing the roughness exponent computed from the structure factor, the dynamic exponent can be obtained. Note, this relation assumes FV scaling and is not guaranteed for processes displaying anomalous scaling. Note also that this observable displays finite-size effects which can be seen as deviations for high-frequencies, analogous to those of the structure factor for discrete or discretized systems.

\subsubsection{Domain size method}
The domain size method has been employed for kinetic simulations of quenches starting in the completely disordered phase \cite{domain_size}. It consists on measuring the average linear size $R(t)$ of ordered domains, and relating it to the domain growth exponent $n_c$, obtained at the critical point from \cite{domain_size}
\begin{equation}\label{eq:R}
    R(t)\sim \sqrt{\langle M(t)^2\rangle} \sim t^{n_c},
\end{equation}
where 
\begin{equation}
    M(t)= \int d^d \textbf{r}\; \phi(\textbf{r},t).
\end{equation}
The dynamic exponent is then computed as
\begin{equation}
    z=\frac{2-\eta}{2n_c}=\frac{2\alpha+d}{2n_c}.
    \label{eq:z_vs_nc}
\end{equation}
The observable $R(t)$ displays finite size effects as well; hence, to obtain a good estimate, a time interval must be chosen to perform a power-law fit as in Eq.\ \eqref{eq:R}. This interval should be located after the initial time transient and prior to saturation.

\subsection{Analytical expectations on exponent values}\label{sec:DRG}
As is well known, the anomalous dimension $\eta$ can be computed analytically using the Renormalization Group  \cite{Binney,kardar,Goldenfeld}. The dynamic exponent $z$ can similarly be obtained through the Dynamic Renormalization Group (DRG) \cite{hohenberg,Mazenko,Tauber}, and also through alternative analytical approaches, such as the self-consistent screening approximation \cite{Campellone_1997} (although this particular reference addresses the $d=3$ case). Given the relation between the anomalous dimension $\eta$ and the roughness exponent $\alpha$, Eq.\ \eqref{alpha_eta_relation}, for the reader's convenience we rephrase the classical DRG analysis of the stochastic TDGL model into our present kinetic roughening setting, closer to similar derivations for e.g.\ the KPZ and related equations \cite{Medina89,Barabasi,ojalvo}. 

The DRG involves two main stages. First, a coarse-graining in Fourier space, in which the short-length-scale interactions (fast modes) are integrated as effective contributions to the long-length-scale physics (slow modes), yielding an effective macroscopic theory. Then, to be able to compare the transformation of the couplings with the original theory, a rescaling is performed that returns the original value of the assumed fixed wave-vector cut-off in Fourier space, namely,
\begin{align}
        \textbf{r}^{\prime}=\textbf{r}/b&,\quad 
        t^{\prime}=b^{-z} t, \label{eq:rescaling} \\
        \phi^{\prime}(\textbf{r}^{\prime},t^{\prime})=b^{-\alpha}\phi\left(\textbf{r}, t\right)&,\quad \zeta^{\prime}(\textbf{r}^{\prime},t^{\prime})=b^{z-\alpha} \zeta\left(\textbf{r}, t\right), \nonumber
\end{align}
where $b>0$ is arbitrary and the rescaling of time (the field) depends on the dynamic exponent $z$ (the roughness exponent $\alpha$).
With this, the dynamical equation for the primed field takes the exact same form as the original (bare) TDGL, but with redefined (renormalized) parameters. To obtain a differential flow in parameter space, we take $b=e^{l}$ for rescaling factor in the $l \rightarrow 0$ limit, to finally obtain the flow induced in the theory space by the DRG transformation. The detailed computation of these steps is provided in Appendix \ref{AppA}, finally giving the DRG flow equations,
\begin{equation}\label{eq:RG_flow}
\begin{split}
\frac{d r}{d l} &=r\left\{z-9\ln(4/3)\frac{u^2 \Gamma^2}{\nu^4}K_4^2 \right\}+3\frac{u\Gamma}{\nu} \; \frac{K_4}{1+r/ \nu} ,\\
\frac{d u}{d l} &=u\left\{(z+2\alpha)-9 \frac{u\Gamma}{\nu^2} K_4-9\ln(4/3)\frac{u^2 \Gamma^2}{\nu^4}K_4^2\right\} ,\\
\frac{d \nu}{d l} &=\nu\left\{(z-2)+\frac{3}{2} \frac{u^2\Gamma^2}{\nu^4} K_4^2 -9\ln(4/3)\frac{u^2 \Gamma^2}{\nu^4}K_4^2\right\} ,\\
\frac{d \Gamma}{d l} &=\Gamma\left\{(z-2\alpha-d)-9\ln(4/3)\frac{u^2 \Gamma^2}{\nu^4}K_4^2\right\},
\end{split}
\end{equation}
where $K_4=1/(8 \pi^2)$ is a number. Since the critical point is scale-invariant, it corresponds to a fixed point of the DRG flow. Before addressing this, the set of equations can still be reduced to an equivalent, smaller one which does not depend explicitly on the values of $\alpha$ and $z$. 
This is achieved by defining the coupling parameters
\begin{equation}
       \bar{u}=\frac{u\Gamma}{\nu^2},\quad\bar{r}=\frac{r}{\nu},
\end{equation}
whose flow can be found using the chain rule, thus obtaining
\begin{equation}
\begin{split}
    \frac{d\bar{u}}{dl}&=\bar{u}\left(\varepsilon-9\bar{u}K_4-3\bar{u}^2K_4^2\right),\\
    \frac{d\bar{r}}{dl}&=\bar{r}\left(2-\frac{3}{2}\bar{u}^2 K_4^2\right)+3\bar{u}K_4(1-\bar{r}+\mathcal{O}\left(\varepsilon^2)\right),
    \label{eq:red_rg_flow}
\end{split}
\end{equation}
where we have used $\varepsilon=4-d \ll 1$ as in Wilson's $\varepsilon$-expansion \cite{Ma2018,kardar,Mazenko,Tauber}. The celebrated Wilson-Fisher (WF) fixed point corresponds to the nontrivial fixed point of the reduced dynamical system, Eq.\ \eqref{eq:red_rg_flow}, and reads 
\begin{equation}
\left.\frac{d\bar{u}}{dl}\right|_{\bar{u}_{*},\bar{r}_{*}} =\left. \frac{d\bar{r}}{dl}\right|_{\bar{u}_{*},\bar{r}_{*}}=0\Rightarrow 
\left\{ \begin{array}{ll}
     \bar{u}_{*}=\varepsilon/(9K_4)+\mathcal{O}(\varepsilon^2),\\
     \bar{r}_{*}=-\varepsilon/6+\mathcal{O}(\varepsilon^2).
     \end{array} \right. 
\label{eq_sign_r}
\end{equation}
Using this into the flow equations for $\nu$ and $\Gamma$ from Eq.\ \eqref{eq:RG_flow}, one obtains a linear homogeneous system for $\alpha$ and $z$, namely,
\begin{equation}
    \begin{split}
(z-2)+\frac{3}{2} \bar{u}_*^{2} K_4^2 -9\ln(4/3)\bar{u}_*^{2}K_4^2=0,\\
(z-2\alpha-d)-9\ln(4/3)\bar{u}_*^{2}K_4^2=0 ,
    \end{split}
    \label{eq:u*r*}
\end{equation}
whose unique solution yields 
\begin{equation}
    \alpha=\frac{1}{2}\left(2-d-\frac{\varepsilon^2}{54}\right),
\end{equation}
or, equivalently, for the anomalous dimension,
\begin{equation}
    \eta = 2-2\alpha-d = \frac{\varepsilon^2}{54},
    \label{eq:eta_rg}
\end{equation}
in agreement with Wilson's prediction \cite{Wilson}. The solution of Eq.\ \eqref{eq:u*r*} for the dynamic exponent $z$ is
\begin{equation}
    \begin{split}
        z&=2+(6\ln(4/3)-1)(2-2\alpha-d), \\
         &=2+\left(6\ln(4/3)-1\right)\eta,
    \end{split}
    \label{eq:z_rg}
\end{equation}
in agreement with Hohenberg and Halperin's result \cite{hohenberg}. Combining these results with Onsager's exact solution $\eta=1/4$ \cite{Onsager}, the analytical predictions for the kinetic roughening exponents of the models studied in this work are collected in Table \ref{tab:exponents_KR}. This particular $(z,\eta)$ pair of values agree with the generic critical exponent inequality that is derived from general properties of the noise distribution, the correlation, and the response functions \cite{PhysRevLett.107.125701,Katzav_2011}.
\begin{table}[t!]
	{\caption{Analytical estimates for the kinetic roughening exponents of the models studied in this work, as calculated through DRG as summarized in this section. Recall Eqs.\ \eqref{alpha_eta_relation}, \eqref{eq:eta_rg}, and \eqref{eq:z_rg}, and see Sec.\ \ref{sec:im} and Appendix \ref{AppA} for further details.}
        \label{tab:exponents_KR}}
	{\begin{tabular}{|c|c|c|}
		\hline
		 Exponent & $2\text{D}$ Ising model & $2\text{D}$ Integral-GL model \\
		\hline\hline 
		  $\alpha$ & $-1/8$ (exact) & $7/8$ \\
		\hline
            $z$ & $2.18$ & $2.18$ \\
		\hline
	\end{tabular}}
\end{table}

\begin{figure}[!t]
    \includegraphics[width=0.95\linewidth]{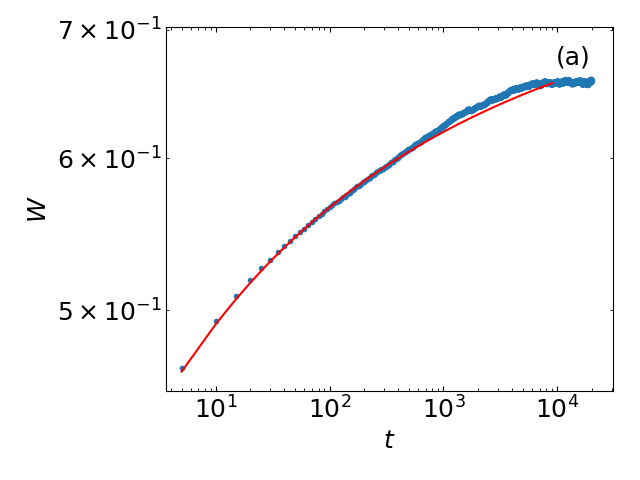}
   \includegraphics[width=0.95\linewidth]{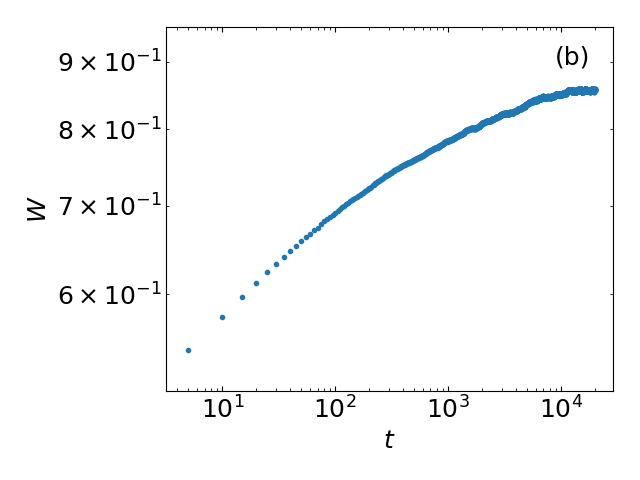}
    \caption{(a) Surface roughness as a function of time from simulations of the TDGL model after a critical quench from $T=0$. The solid line corresponds to the fit $W_{\rm fit}=0.76-0.37\; t^{-0.14}$. (b) Same as panel (a) for Glauber dynamics. }
    \label{fig:W_t0}
\end{figure}
\begin{figure}[!t]
        \includegraphics[width=0.95\linewidth]{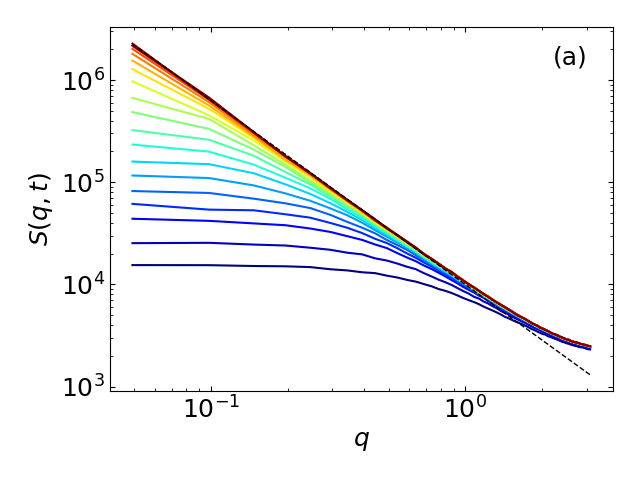}
        \includegraphics[width=0.95\linewidth]{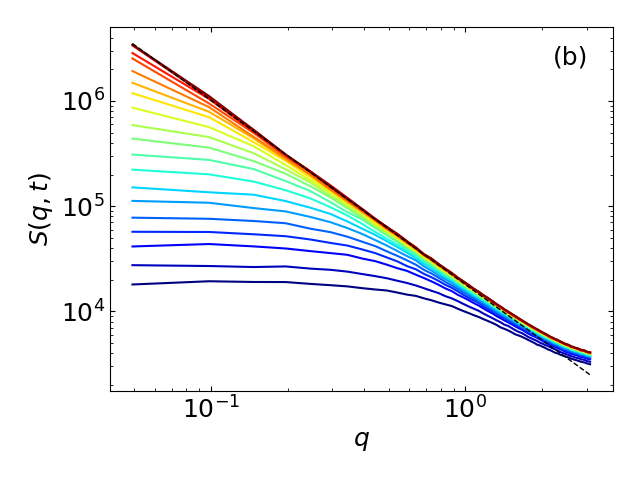}
    \caption{(a) Structure factor from simulations of the TDGL model for different times after a critical quench from $T=0$. Dashed line indicates asymptotic behavior $S(q) \sim q^{-1.80}$. (b) Same as panel (a) for Glauber dynamics. Dashed line indicates asymptotic behavior $S(q) \sim q^{-1.75}$. For both panels, time increases from blue ($t=0$) to red ($t=20000$) being log-spaced for different curves.}
    \label{fig:Sq_t0}
\end{figure}
\begin{figure}[!t]
        \includegraphics[width=0.95\linewidth]{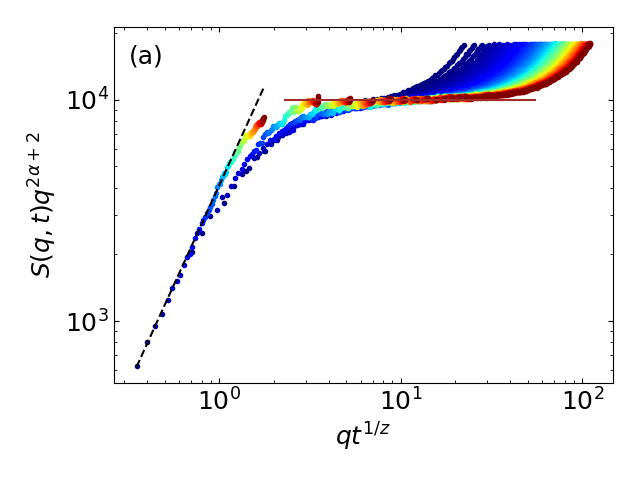}
        \includegraphics[width=0.95\linewidth]{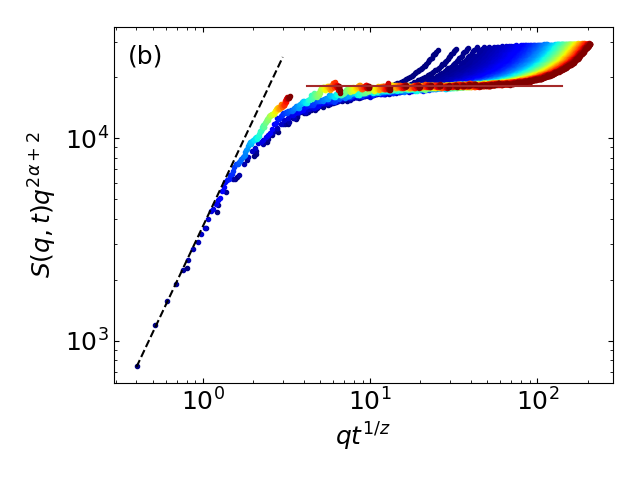}
    \caption{(a) Collapse of the data in Fig.\ \ref{fig:Sq_t0}(a) for the TDGL model following the FV dynamic scaling ansatz, Eqs.\ \eqref{ec:SFV}-\eqref{ec:SFVf}. The dashed line corresponds to $f_S \sim u^{1.80}$ and the solid line represents $f_S = \text{cnst}$. (b) Same as in panel (a) for the Glauber dynamics data in Fig.\ \ref{fig:Sq_t0}(b). The dashed line corresponds to $f_S \sim u^{1.75}$ and the solid line represents $f_S = \text{cnst}$. In both panels, time increases from blue to red (being linearly-spaced) in the growth regime $(t<10^4 )$, and the exponent values employed for collapse are $\alpha=-1/8$ and $z=2.19$.}
    \label{fig:DSA_t0}
\end{figure}

\section{Critical quench from $T=0$} \label{sec:T=0}
We start our numerical study of critical nonconserved dynamics in the Ising model as a kinetic roughening process, by assessing the time evolution of both, a discrete 2D kinetic Ising model with Glauber updates and the continuous 2D TDGL equation. We have additionally studied  other nonconserved update rules in the discrete case ---see Appendix \ref{AppB} for the Metropolis-Hastings update rule \cite{newman}---, finding very similar behavior. In this section we study a critical quench from the completely ordered state at $T=0$ up to $T=T_c$, while a different critical quench will be addressed in the next section. 

In practice, a critical quench from $T=0$ amounts to choosing all spins up as an initial condition for our Ising model with Glauber updates or, equivalently, $\phi(\mathbf{r},t=0)=1$ for the TDGL equation, while performing the time evolution at the critical temperature (whose value for each model is determined in Appendix \ref{AppC}). From this point on, all variables and numerical results are expressed in the dimensionless units defined therein. The ensuing dynamics displayed by both models is quite similar, see Fig.\ \ref{fig:W_t0} for the time evolution of $W(t)$ which, in both cases, qualitatively follows FV type behavior, with fluctuations that grow until reaching saturation to steady state (equilibrium). However, given that $\alpha < 0$, the FV scaling relations are not manifest for either of these models, since the (in principle) subdominant contributions in the asymptotic computation of the roughness are not negligible, and strong corrections to scaling occur. Negative roughness exponent values occur, e.g., for kinetically rough surfaces above their upper critical dimension \cite{Barabasi,krug}. Here, $\alpha < 0$ is unavoidable once we formulate the 2D Ising system as an interface problem [recall Eq.\ \eqref{alpha_eta_relation}, where $\eta>0$] and will partly motivate an alternative formulation which guarantees a positive roughness exponent, see Sec.\ \ref{sec:im}. In any case, an accurate measurement of $\alpha$ which is not conditioned by the sign of the roughness exponent is actually possible from the structure factor, displayed in Fig.\ \ref{fig:Sq_t0}. Again, both models show a similar behavior in which $S(q,t)$ gradually correlates from shorter to longer length scales towards a simple asymptotic power-law behavior, followed by saturation, with all of this as dictated by FV scaling behavior, Eqs.\ \eqref{ec:SFV} and \eqref{ec:SFVf}, qualitatively as e.g.\ for the EW equation of rough surfaces, see Fig.\ \ref{fig:EW} in Appendix \ref{App:EW}.

Indeed, the asymptotic structure factors yield the equilibrium power laws $S(q) \sim q^{-1.80}$ for the TDGL model and $S(q) \sim q^{-1.75}$ for Glauber dynamics, close to the theoretical expectation, $S_{\rm theor}(q) \sim q^{-(2\alpha+2)} = q^{-1.75}$. Deviations for the continuum model result from finite-size effects, since the source originates at the longest length scales. Figure \ref{fig:DSA_t0} shows the best collapse of the full data from Fig.\ \ref{fig:Sq_t0}, obtained for $\alpha\approx -1/8$ and $z\approx 2.19$, now very close to theoretical predictions and indeed, fully consistent with FV dynamic scaling. Some slight deviations may be noted at intermediate length scales, suggesting crossover behavior.

\begin{figure}[!t]
        \includegraphics[width=0.95\linewidth]{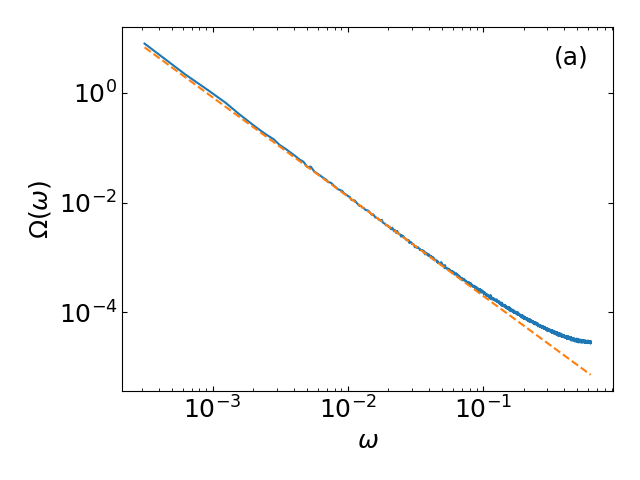}
        \includegraphics[width=0.95\linewidth]{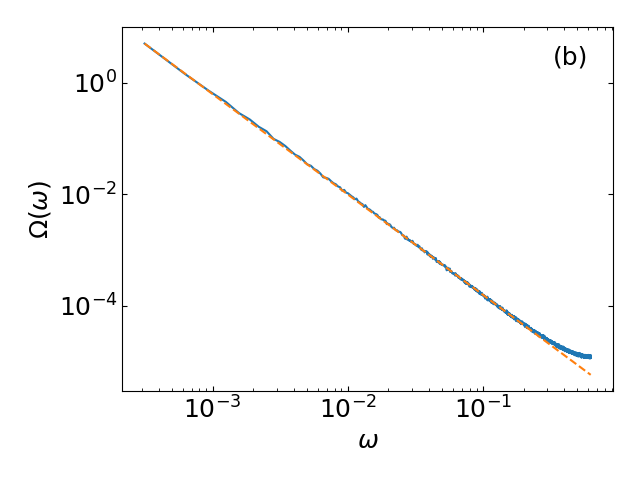}
    \caption{(a) Time power spectrum as a function of frequency from simulations of the TDGL model after a critical quench from $T=0$ (blue solid line). The dashed orange line corresponds to $\Omega(\omega) \sim \omega^{-1.805}$. (b) Same as panel (a) for Glauber dynamics (blue solid line). The dashed orange line corresponds to $\Omega(\omega) \sim \omega^{-1.80}$.}
    \label{fig:Omega_t0}
\end{figure}

As a consistency check, the dynamic exponent value is also measured through the time power spectrum method, and again turns out to be consistent with DRG predictions, see Fig.\ \ref{fig:Omega_t0}. Our results agree quantitatively with those already obtained in Ref.\ \cite{Kent} $(z=2.16-2.19)$, yielding $\mu \approx 1.805\Rightarrow z\approx 2.17$ for the TDGL model and $\mu \approx 1.80\Rightarrow z\approx 2.19$ for Glauber dynamics, confirming sub-diffusive dynamics, as $z>2$.

\begin{figure}[!t]
        \includegraphics[width=0.95\linewidth]{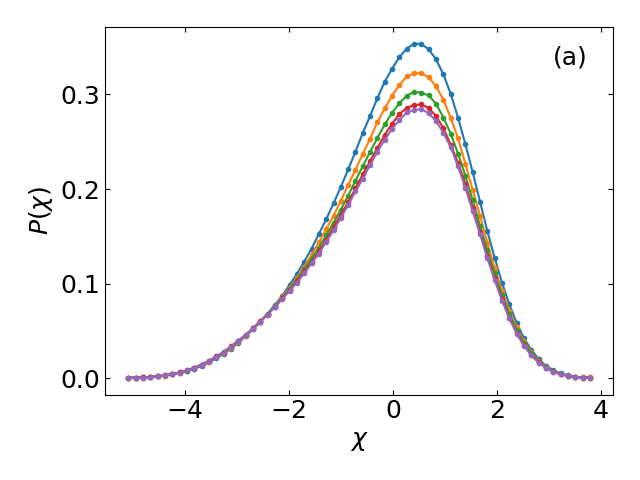}
        \includegraphics[width=0.95\linewidth]{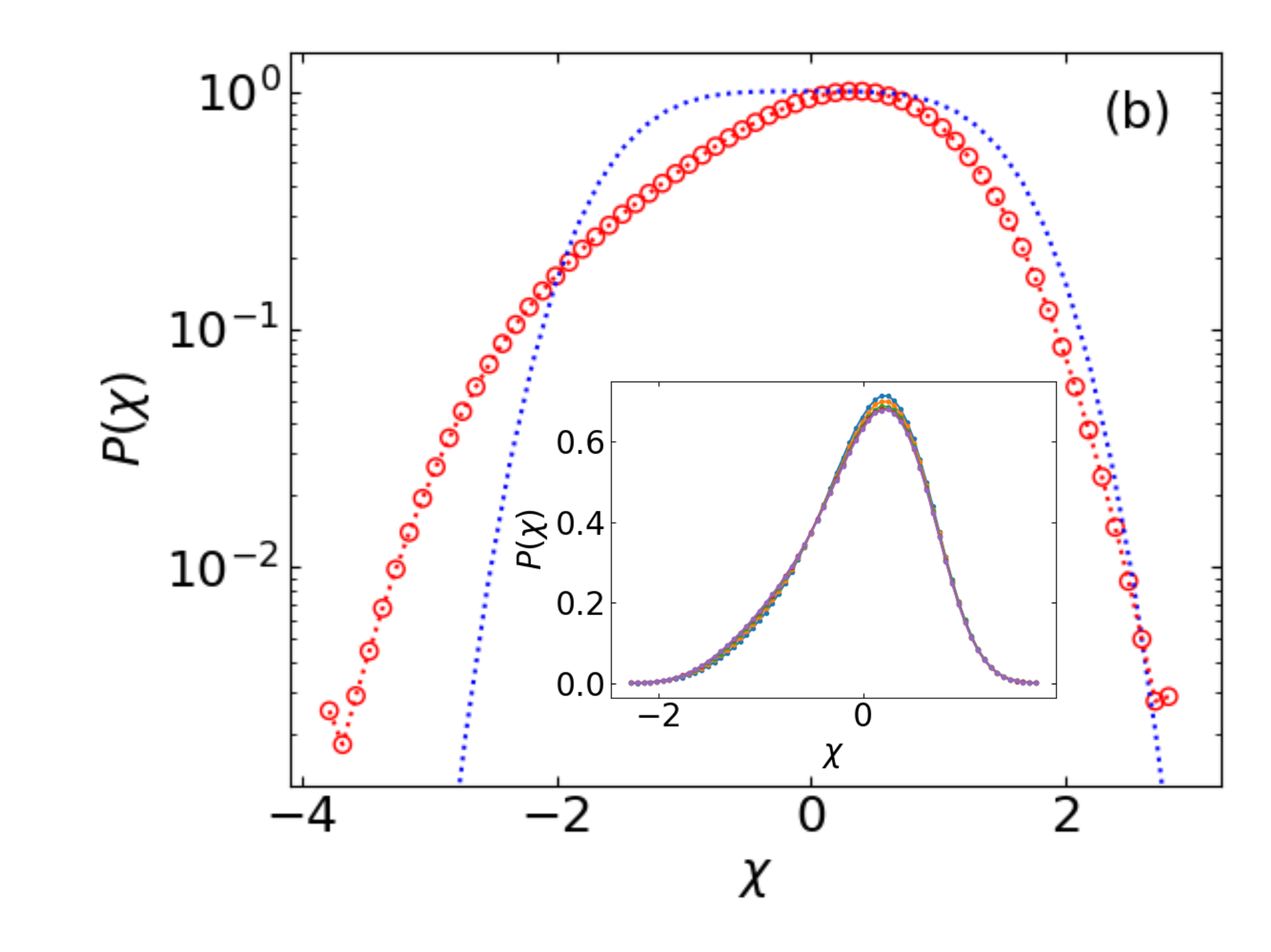}
    \caption{(a) Fluctuation PDF from simulations of the TDGL model in its growth regime $(t<10^4)$ after a critical quench from $T=0$, according to Eq.\eqref{eq:Pra-Spohn} using $\beta'=-0.14$. (b) Time-averaged PDF in log-scale (data have been normalized to zero mean and unit variance) and inset with PS collapse for $\beta=0$. Different colors correspond to different (linearly-spaced) times. The dashed blue line corresponds to a stretched exponential distribution $P(\mathcal{X})\propto e^{-\mathcal{X}^4}$, for comparison.}
    \label{fig:PDF_GL_t0}
\end{figure}

We close this section with an analysis of the fluctuation PDF. A detailed analysis is only provided for the TDGL model, since the dynamical variables in the Glauber dynamics system can only take the values $\sigma_i =\pm 1$, 
and the results just shown on $W(t)$ and $S(q,t)$ already confirm the same universality class. Since the quench from $T=0$ displays a single growth regime, its dynamics are characterized by a single PDF, shown in Fig.\ \ref{fig:PDF_GL_t0}. This PDF is asymmetric, being biased in the direction implied by the initial configuration of the local magnetization field, $\phi(\mathbf{r},t=0)=+1$. 
The PDF obtained has a non-trivial shape, characterized by three distinct behaviors, which are best identified in log-scale representation. The left tail decays similar to a Gaussian distribution. Its transition to the most frequent fluctuation amplitude is given by an exponential function, rendering linear behavior in log-scale. Finally, the right tail decays faster than the Gaussian distribution, approximately as an stretched exponential $P(\mathcal{X})\propto e^{-\mathcal{X}^4}$.
As a technical detail, again related with the fact that $\alpha, \beta <0$, note that the PS formula, Eq.\ \eqref{eq:Pra-Spohn}, is satisfied better (in the sense of defining a time-independent fluctuation variable) by approximating ${\rm std}(\phi) \sim t^{\beta}$ for $\beta=0$ (which dominates the roughness behavior) than for the theoretical $\beta=-1/(8 \cdot 2.18) \approx -0.06$ or the roughness fit $\beta' = -0.14$, see Fig.\ \ref{fig:PDF_GL_t0}.

\section{Critical quench from $T=\infty$}
\label{sec:T=infty}

Remarkably, the paradigmatically simple FV dynamic scaling assessed in the previous section does not carry over to different initial conditions at criticality. Specifically, as will be seen in this section, it is manifestly not seen for a critical quench from infinite temperature, namely, using at $T=T_c$ an initial condition in which the discrete spins for the Glauber dynamics take on the $\pm 1$ values randomly or, equivalently, starting out from a zero initial local magnetization, $\phi(\mathbf{r},t=0)=0$, in the TDGL model. 

\begin{figure}[!t]
        \includegraphics[width=0.95\linewidth]{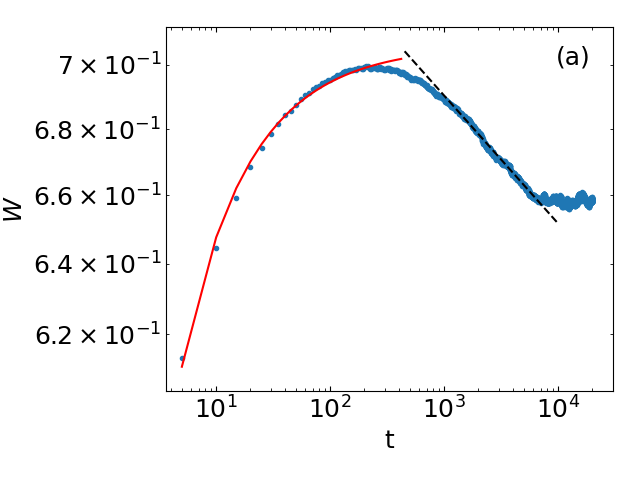}
        \includegraphics[width=0.95\linewidth]{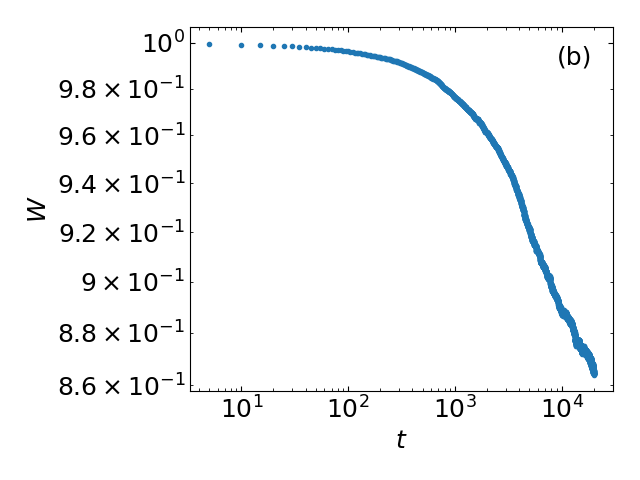}
    \caption{(a) Surface roughness as a function of time from simulations of the TDGL equation after a critical quench from $T=\infty$. The solid line corresponds to the overgrowth regime fit $W_{\rm fit} = 0.71-0.29\; t^{-0.70}$, while the dashed line corresponds to the relaxation regime behavior $W \sim t^{-0.025}$. (b) Same as panel (a) for Glauber dynamics.}
    \label{fig:W_tinf}
\end{figure}

For the TDGL equation, the roughness displays strongest fluctuations during a short-time overgrowth regime (not so conspicuous for Glauber dynamics) and then relaxes towards saturation (equilibrium) for longer times which follows the power-law $W(t) \propto t^{-0.025}$, see Fig.\ \ref{fig:W_tinf}. Again, the roughness shows corrections to scaling due to the negative roughness exponent value.

\begin{figure}[!t]
        \includegraphics[width=0.95\linewidth]{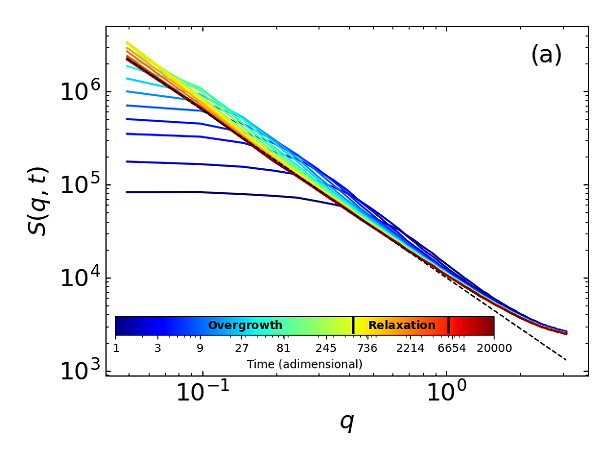}
        \includegraphics[width=0.97\linewidth]{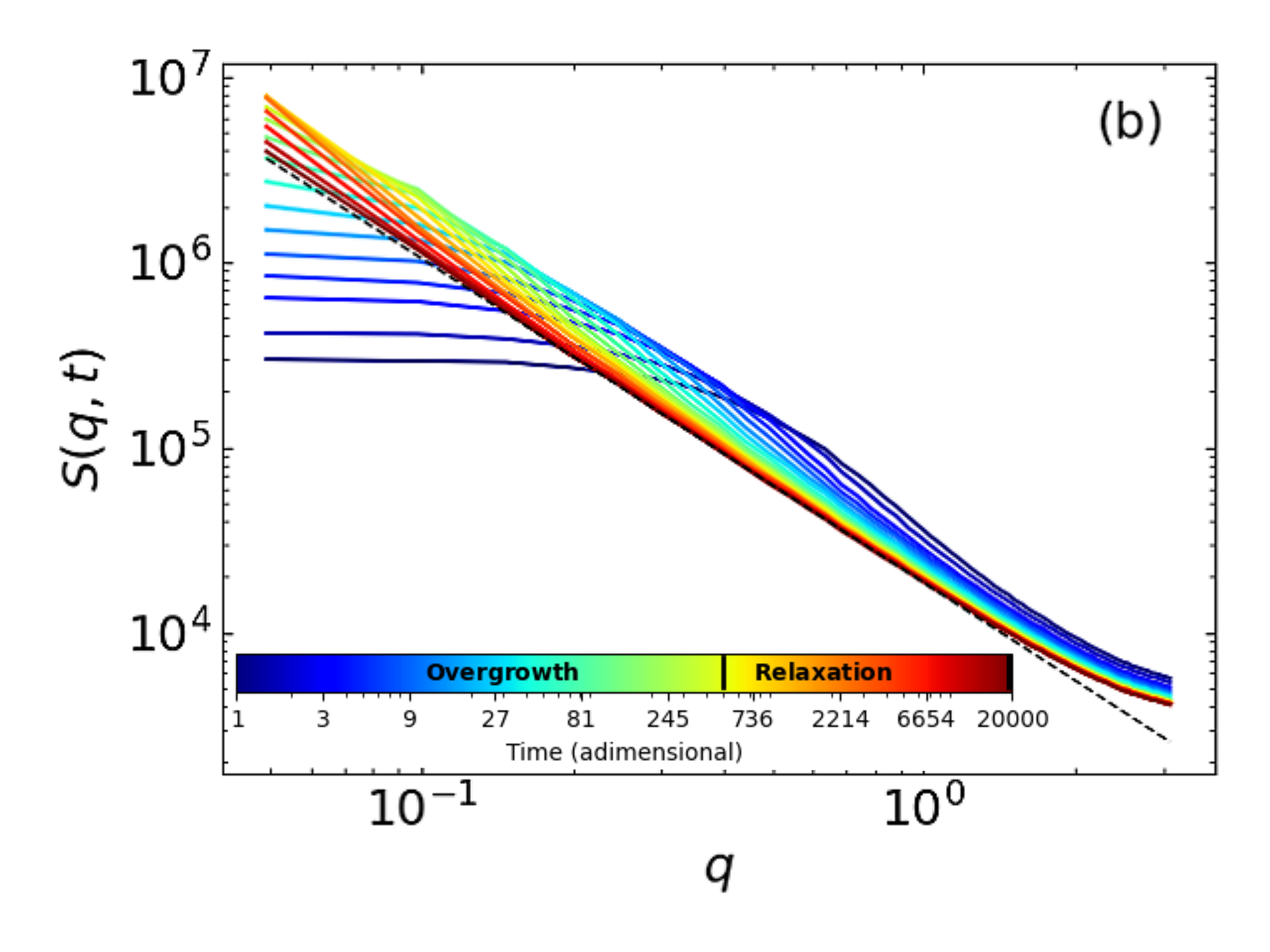}
    \caption{Time evolution of the structure factor after a critical quench from $T=\infty$ for (a) the TDGL equation (the dashed line indicates $S(q) \sim q^{-1.80}$ asymptotic behavior) and (b) Glauber dynamics (the dashed line indicates $S(q) \sim q^{-1.75}$ asymptotic behavior). The time arrow goes from blue to red (log-spaced) for both panels, as indicated in their legends.}
    \label{fig:Sq_tinf}
\end{figure}

In terms of the time evolution of correlation functions, the lattice Ising and the TDGL models also display strong similarities under the quench from the completely disordered state. Thus, both models show an asymptotic structure factor that yields the equilibrium power-law $S(q) \sim q^{-1.80}$ for the TDGL and $S(q) \sim q^{-1.75}$ for Glauber dynamics, not far from the theoretical expectations and as expected from the long-time behavior of the roughness. Notably, the two distinct non-equilibrium regimes suggested by the roughness in the continuum case are clearly observed for the two models: an initial overgrowth regime followed by a long-time relaxation to saturation, see Fig.\ \ref{fig:Sq_tinf}. Moreover, in the two models the value of $q$ at which the structure factor levels off displaces with time, within the overgrowth regime, from short to longer length scales (high to low wave vectors), until reaching a cutoff value from which the relaxation regime takes over. This process is described in further detail in Sec.\ \ref{sec:disc}.

\begin{figure}[!t]
        \includegraphics[width=0.95\linewidth]{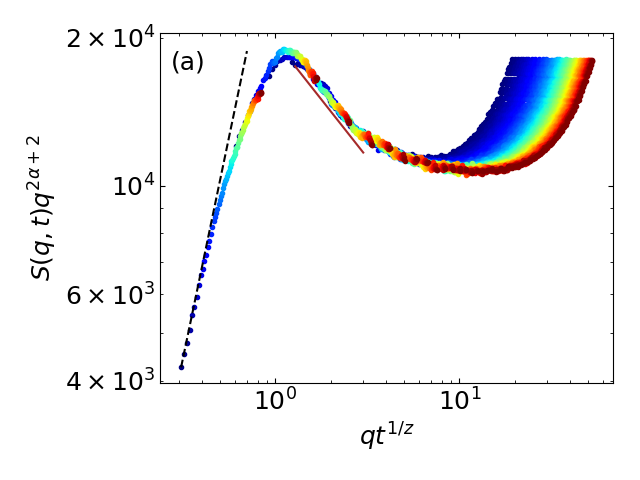}
        \includegraphics[width=0.95\linewidth]{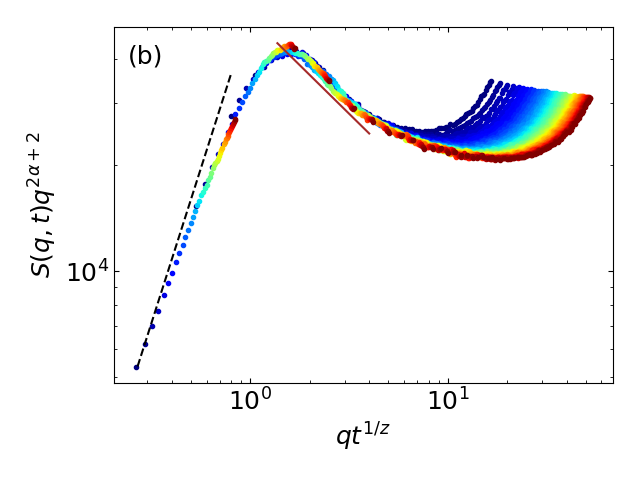}
    \caption{Collapse of the structure factor data shown in Fig.\ \ref{fig:Sq_tinf} according to the anomalous dynamic scaling ansatz, Eqs. \eqref{eq:S_anomalous_scaling}-\eqref{eq:anomalous_scaling}, using $\alpha=-1/8$ and $z=2.19$, for (a) the TDGL equation (the dashed line corresponds to power-law scaling as $f_{S'} \sim u^{1.80}$; the solid line corresponds to $f_{S'} \sim u^{-0.475}$) and (b) Glauber dynamics (the dashed line corresponds to power-law scaling as $f_{S'} \sim u^{1.75}$; the solid line corresponds to $f_{S'} \sim u^{-0.55}$). The time arrow goes from blue to red (linearly-spaced) for both panels, limited to the overgrowth regime.}
    \label{fig:DSA_tinf}
\end{figure}

The time-dependent structure factor data collapse onto a master curve under the dynamic scaling ansatz, only during the overgrowth regime, and does so for $\alpha\approx -1/8$ and $z\approx 2.19$, very close to the theoretical predictions, see Fig.\ \ref{fig:DSA_tinf}. The power-law fits show two different non-zero slopes, corresponding to intrinsic anomalous roughening as described by Eq.\ \eqref{eq:anomalous_scaling}, as can also be seen in Fig.\ \ref{fig:DSA_tinf}. The positive slope is consistent with $\alpha\approx -1/8$, while the negative slope yields a spectral exponent value $\alpha_s\approx 0.11$ for the TDGL and $\alpha_s\approx 0.15$ for the lattice Ising model. In both cases $\alpha_s <1$ and $\alpha_s \neq \alpha$, which indeed corresponds to intrinsic anomalous roughening. Hence, it is $\alpha_s$ that dictates the power-law behavior of the structure factor with $q$ during the overgrowth regime. Note that, although we are interpreting these behaviors through the physical image of a kinetically rough surface, the actual time-dependent structure factor data that we are analyzing are directly those obtained for the Glauber simulation or the TDGL model. Hence, it is these models as such which display intrinsic anomalous scaling in the growth regime for a quench from $T=\infty$.

\begin{figure}[!t]
        \includegraphics[width=0.95\linewidth]{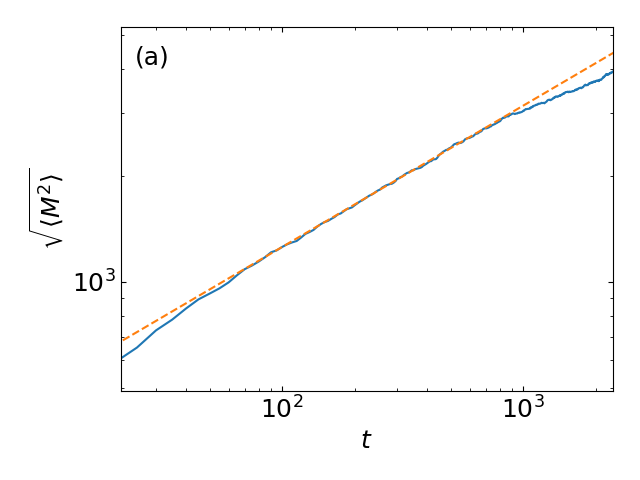}
        \includegraphics[width=0.95\linewidth]{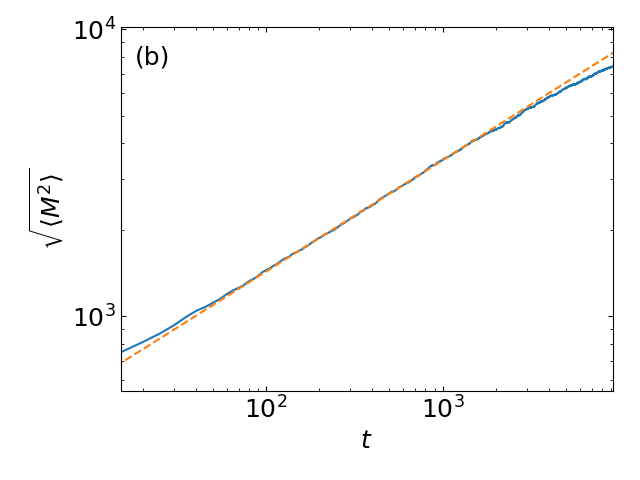}
    \caption{Root mean squared magnetization (blue) as a function of time after a critical quench from $T=\infty$ for (a) the TDGL equation (the dashed orange line corresponds to $\sqrt{\langle M^2\rangle}\sim t^{0.40}$) and (b) Glauber dynamics (the dashed orange line corresponds to $\sqrt{\langle M^2\rangle} \sim t^{0.39}$.}
    \label{fig:z_measure}
\end{figure}

As an additional check on the value of the dynamic exponent, we have applied the domain size method to both models, finding values which are again consistent with the DRG predictions and with the results previously obtained in Ref.\ \cite{domain_size}, namely, $n_c \approx 0.40\Rightarrow z\approx 2.19$ for the TDGL equation and $n_c \approx 0.39\Rightarrow z \approx 2.24$ for the lattice Ising model; see Fig.\ \ref{fig:z_measure}. 

\begin{figure}[!t]
    \centering
    \includegraphics[width=0.45\textwidth]{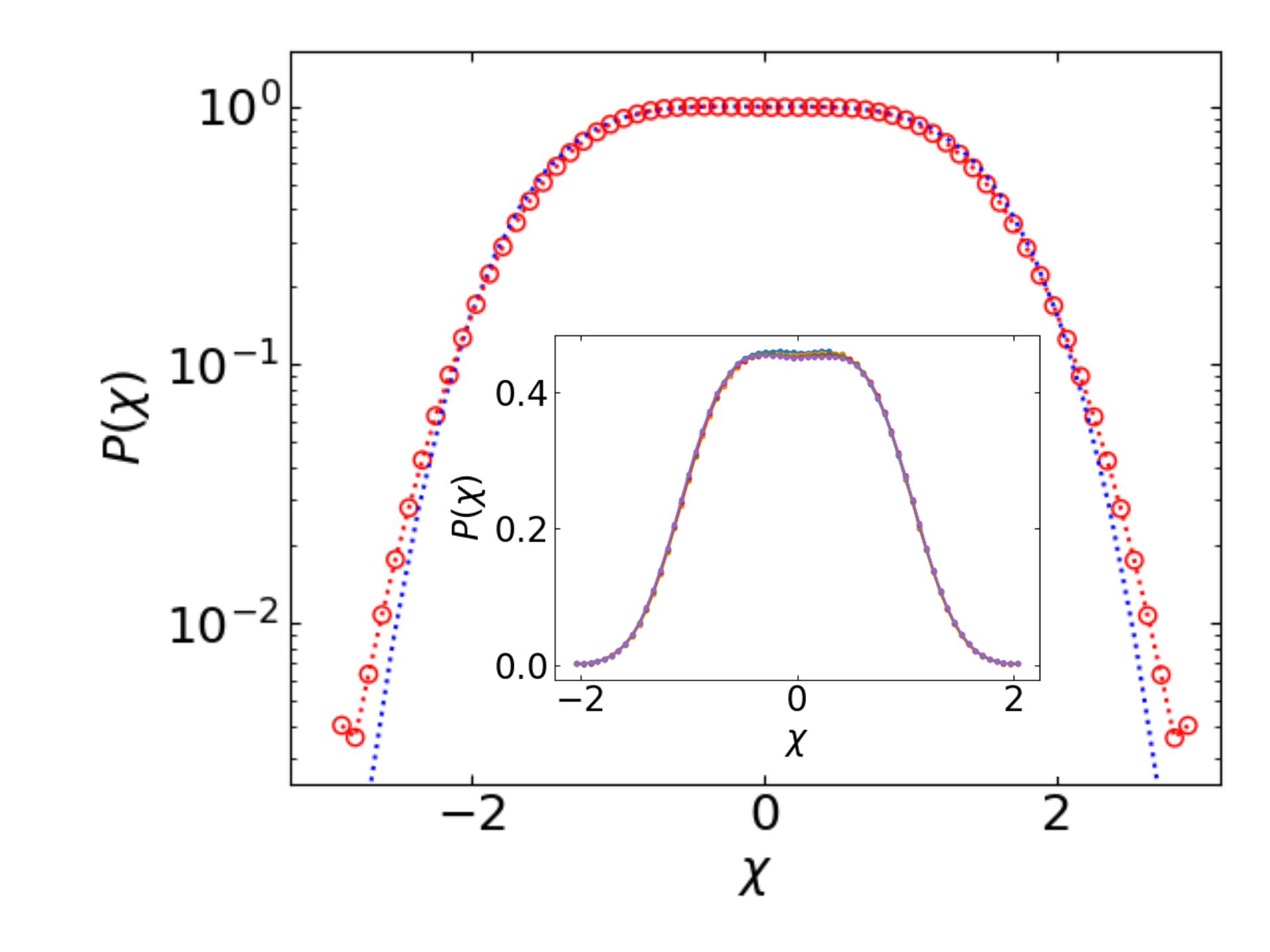}
    \caption{Time-averaged fluctuation PDF in log-scale from simulations of the TDGL model after a critical quench from $T=\infty$, within the overgrowth regime. The data have been normalized to zero mean and unit variance, and the dashed blue line corresponds to a stretched exponential distribution $P(\mathcal{X})\propto e^{-\mathcal{X}^4}$, for comparison. Inset: Pr\"ahofer-Spohn collapse of the data in the main panel for $\beta=0$. Different colors correspond to different (linearly-spaced) times.}    
    \label{fig:PDF_GL_tinf}
\end{figure}

\begin{figure}[!t]
        \includegraphics[width=0.95\linewidth]{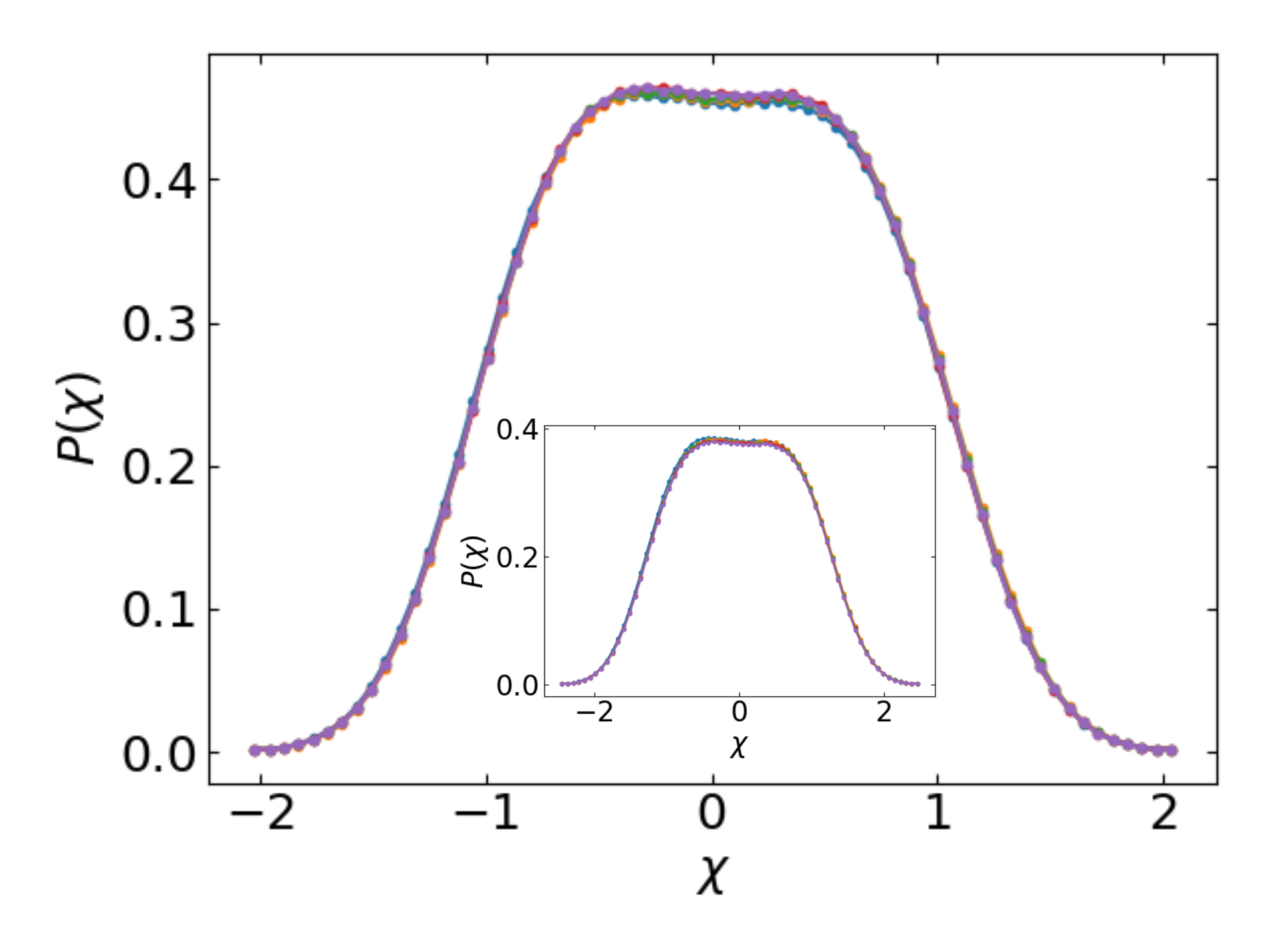}
    \caption{Pr\"ahofer-Spohn collapse of the fluctuation PDF from simulations of the TDGL model after a critical quench from $T=\infty$, within the relaxation regime. The main panel (inset) displays the collapse obtained for $\beta=0$ ($\beta'=-0.025$). Different colors correspond to different (linearly-spaced) times.}
    \label{fig:PS2_tinf}
\end{figure}

In parallel with Sec.\ \ref{sec:T=0}, we now consider the fluctuation PDF for the TDGL equation after a critical
quench from infinite temperature. As suggested by Figs.\ \ref{fig:PDF_GL_tinf} and \ref{fig:PS2_tinf}, the time-averaged non-equilibrium PDF is the same for both, the overgrowth and the relaxation regimes, exhibiting up-down symmetry and tails similar to those of an stretched exponential, $P(\mathcal{X})\propto e^{-\mathcal{X}^4}$. Moreover, the maximum is not centered at zero. There are, rather, two symmetric maxima, of approximately the same height, corresponding to positive and negative fluctuations of similar amplitude. This seems consistent with general expectations \cite{Binney,kardar,Goldenfeld}, as at the critical point the magnetization is zero on (spatial) average, imposing the (statistical) symmetry between negative and positive fluctuations. 
The PDF for the overgrowth regime (Fig.\ \ref{fig:PDF_GL_tinf}) is best collapsed through the PS formula using the effective exponent value $\beta=0$, since it is the dominant contribution to the roughness in the growth regime. For the relaxation regime, (Fig.\ \ref{fig:PS2_tinf}) both collapses are consistent, since the fit exponent $\beta'$ is very close to zero.

\section{Integral GL model}
\label{sec:im}

The interpretation, pursued in the two previous sections, of the correlation data of the critical 2D TDGL equation in terms of a rough surface with roughness exponent given by Eq.\ \eqref{alpha_eta_relation}, unavoidably renders both, $\alpha$ and $\beta$, negative for $d=2$. Nonetheless, the dynamic scaling behavior that ensues can still be consistently accounted for by suitable forms of kinetic roughening depending on the type of critical quench which is performed. However, it would still be interesting to introduce a mapping from the 2D TDGL equation to a rough surface which has positive roughness and growth exponents. This is explored in the present section.

The key idea is to perform an analogy with the known relation between the KPZ [Eq.\ \eqref{eq:KPZ_eq}] and the noisy Burgers [Eq.\ \eqref{eq:Burgers_cn}] equations, already mentioned in Sec.\ \ref{sec:em_stat}. Indeed, the latter equation for a scalar field $u(x,t)$ is the space derivative of the former [as an equation for another scalar field $h(x,t)$], provided $u(x,t)=\partial_x h(x,t)$ or, viceversa, $h(x,t)=\int^x_{x_0} u(x',t) dx'$, with the corresponding kinetic roughening exponents being related as $\alpha_{h}=\alpha_{u}+1$ and $z_h=z_u$ \cite{Rodriguez-Fernandez20}, as one might intuitively expect. Thus, we conclude that space integration of a fluctuating field can increase the roughness exponent by one unit while keeping the dynamic exponent unchanged. Hence, our next goal is to define a scalar field which scales as the space integral of the TDGL local magnetization $\phi(\mathbf{r},t)$. The ensuing time evolution equation for such a field 
will be referred to as the integral Ginzburg-Landau (integral GL) model, and our expectations for its kinetic roughening exponent values are $\alpha=-1/8+1=7/8 \simeq 0.88$ and $z=2.18$, as collected in Table \ref{tab:exponents_KR}. Note that, owing to the vector character of the gradient in $2\text{D}$, the scalar TDGL equation cannot be directly interpreted as the derivative of an underlying partial differential equation, unlike the relation between the $1\text{D}$ KPZ and noisy Burgers equations. Hence, a modified approach is necessary for our present case.

The derivative $\boldsymbol{\psi}(\textbf{r},t)$ of a scalar field $f(\textbf{r},t)$ is defined as
\begin{equation}
    \boldsymbol{\psi}(\textbf{r},t)=\nabla f(\textbf{r},t),
\end{equation}
which in Fourier space reads,
\begin{equation}
    \hat{\boldsymbol{\psi}}(\textbf{q},t)=i\textbf{q} \hat{f}(\textbf{q},t) .
\end{equation}
Then, we define the ``integral field'' [denoted $h(\textbf{r},t)$] of $\phi(\textbf{r},t)$ through its Fourier components as
\begin{equation}\label{eq:integral_model}
\begin{split}
    &\hat{h}(\textbf{q},t)=\frac{\hat{\phi}(\textbf{q},t)}{i|q|}, \quad \textbf{q} \neq \textbf{0}, \\
    &\hat{h}(-\textbf{q},t)=\hat{h}^*(\textbf{q},t)\Rightarrow h(\textbf{r},t)\in \mathbb{R},
\end{split}
\end{equation}
where the conjugate symmetry guarantees that the integral field $h(\textbf{r},t)$ is real valued.

The previous definition is close in spirit to inverting the space derivative operator in Fourier space. However, the zeroth mode $\hat{h}(\textbf{q}=\textbf{0},t)$ [the space average of the $h(\textbf{r},t)$ field] remains undefined. Hence, an additional prescription is needed to compute the space average of the integral GL field. If $\hat{D}$ denotes the $1\text{D}$ differential operator $\partial_x$ in finite differences with periodic boundary conditions, we compute
\begin{equation}
    h = \tilde{D}^{-1} \phi,
\end{equation}
where $\tilde{D}^{-1}$ denotes the Moore-Penrose pseudo-inverse \cite{meyer01} of $\hat{D}$. The zeroth mode is then approximated as the sum of $h$ at each 1D cut along the $x$-direction for constant $y$, summed over each $y$ value along the $y$-direction. 

The various mappings introduced between models are summarized in Fig.\ \ref{fig:mapping}, which also provides on its rightmost panel an explicit example of the integral GL field rough surface $h(\textbf{r},t)$, as obtained from one realization of the local magnetization $\phi(\mathbf{r},t)$ for the TDGL equation.

\begin{figure}[t!]
        \includegraphics[width=0.95\linewidth]{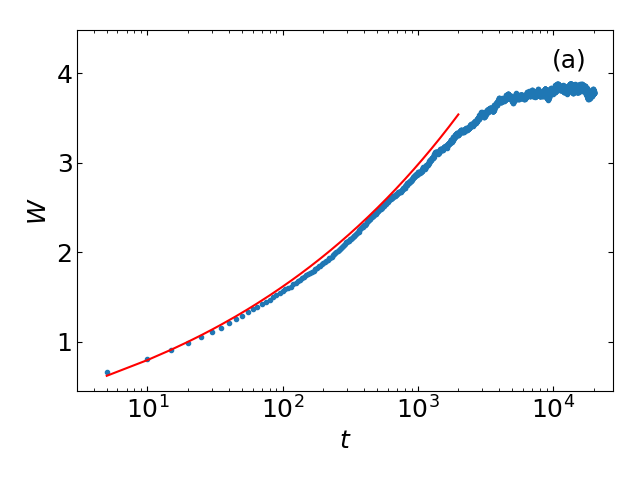}
        \includegraphics[width=0.95\linewidth]{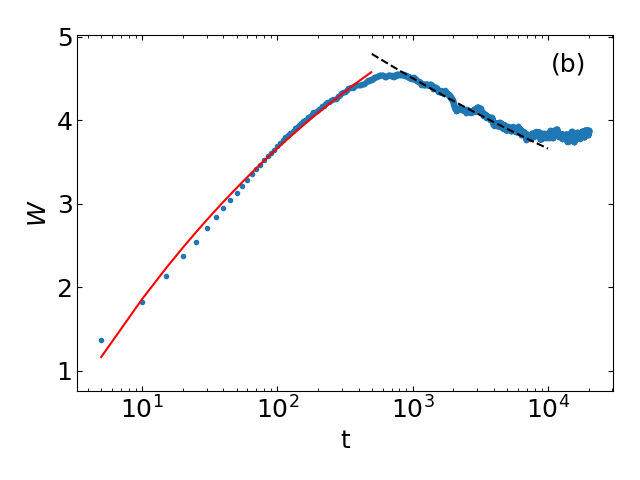}
    \caption{Roughness of the integral GL model as a function of time for a quench from (a) $T=0$ (the solid line represents $W_{\rm fit} =-0.45+0.75\; t^{0.22}$) and (b) $T=\infty$ (the solid line represents $W_{\rm fit} = 7.64-8.42\; t^{-0.16}$, while the dashed line corresponds to $W \sim t^{-0.09}$). The same time regimes as for the TDGL equation can be identified.}
    \label{fig:W_int}
\end{figure}

We next study the same quenches as in Secs.\ \ref{sec:T=0} and \ref{sec:T=infty}, now for the integral GL model. Thus, after a quench from $T=0$, the roughness of the integral field $h$ grows monotonically until saturation [see Fig.\ \ref{fig:W_int}(a)], following a power law with a positive growth exponent $\beta\approx 0.22$, which however is still far from our present expectation, namely, $7/(8\times 2.18) \simeq 0.40$, but see below. For now, note that the scaling corrections in the roughness of the original field $\phi$ are expected to influence the roughness of the integral field $h$ too. On the other hand, for the quench from $T=\infty$, the roughness reflects [see Fig.\ \ref{fig:W_int}(b)] all the time regimes identified in Sec.\ \ref{sec:T=infty}, namely, an initial increase (overgrowth), followed by relaxation to steady state as a power law, $W(t) \sim t^{-0.09}$, displaying scaling corrections as well.

\begin{figure}[!t]
        \includegraphics[width=0.95\linewidth]{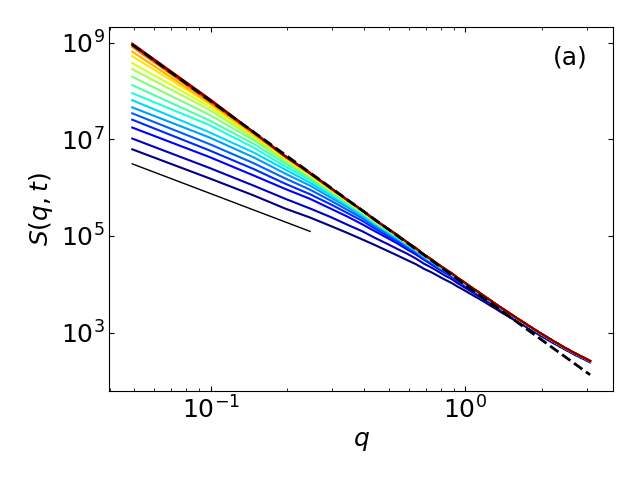}
        \includegraphics[width=0.97\linewidth]{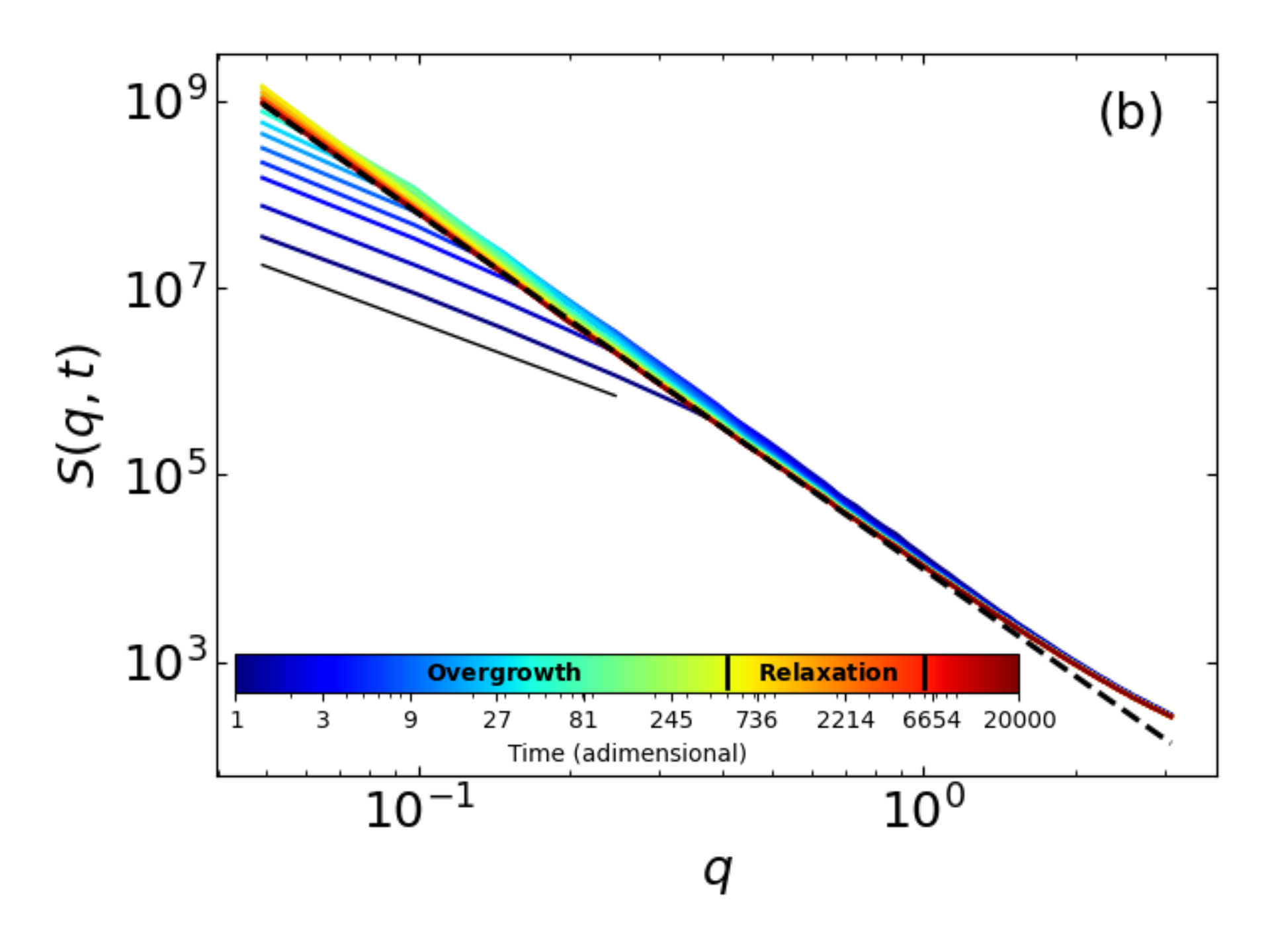}
    \caption{Structure factor of the integral GL model as a function of $q$ for different times after a quench from (a) $T=0$ and (b) $T=\infty$. The dashed lines represent an asymptotic $S(q) \sim q^{-3.80}$ behavior, while the solid black line corresponds to power-law decay as $S(q) \sim q^{-2}$. The time arrow goes from blue ($t=0$) to red ($t=20000$) being log-spaced for different curves, with the color code being common to both panels.}
    \label{fig:Sq_int}
\end{figure}

\begin{figure}[!t]
        \includegraphics[width=0.95\linewidth]{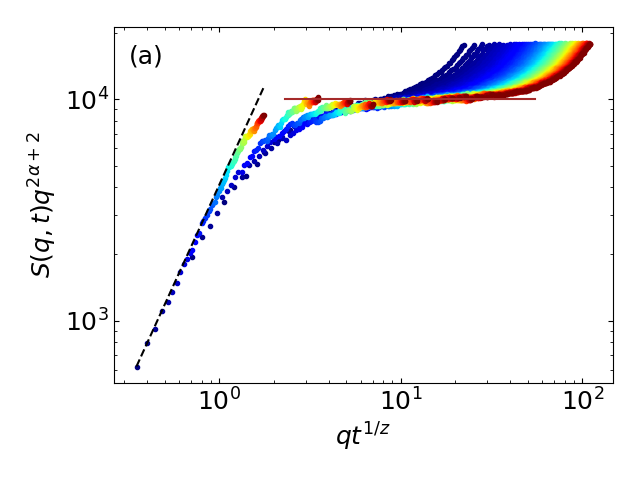}
        \includegraphics[width=0.95\linewidth]{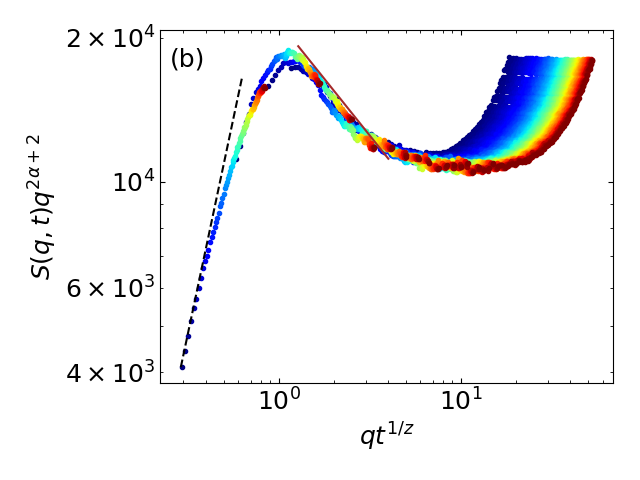}
    \caption{Collapse of the structure factor data shown in Fig.\ \ref{fig:Sq_int} for (a) the $T=0$ critical quench $(t<10^4 )$ for the integral GL model (the dashed line represents power law behavior as $f_S \sim u^{1.80}$, while the solid line corresponds to $f_S = \text{cnst.}$) and (b) the $T=\infty$ critical quench for the integral GL model, constrained to the overgrowth regime (the dashed line represents power-law as $f_{S'} \sim u^{1.80}$, while the solid line corresponds to $f_{S'} \sim u^{-0.475}$). The time arrow goes from blue to red (linearly-spaced) for both panels. The exponent values employed for the collapses are $\alpha=7/8$ and $z=2.19$, as in Table \ref{tab:exponents_KR}.}
    \label{fig:DSA_int}
\end{figure}

The time evolution of two-point correlations (in Fourier space) is shown in Fig.\ \ref{fig:Sq_int}. At saturation (equilibrium) and for both quenches, the structure factor displays the asymptotic behavior given by $S(q) \sim q^{-3.80}$, which suggests $\alpha\approx 0.9$, close to the theoretical expectation, recall Table \ref{tab:exponents_KR}. For the quench from $T=0$ at preasymptotic times [see Fig.\ \ref{fig:Sq_int}(a)], length scales below the correlation length follow the asymptotic $S(q) \sim q^{-3.80}$ behavior, while $S(q) \sim q^{-2}$ for larger length scales (smaller $q$). This is because, by construction of the integral GL model through Eq.\ \eqref{eq:integral_model}, noise is no longer uncorrelated but features, rather, $q^{-2}$ correlations. Indeed, for a fixed time within the growth regime and for length scales above the correlation length (small wave vectors) for which the structure factor of the TDGL equation is uncorrelated ($q$-independent), the structure factor of the integral model unavoidably picks up a $q^{-2}$ dependence through Eq.\ \eqref{eq:integral_model}, as seen in Fig.\ \ref{fig:Sq_int}(a). This is in full analogy to the mapping between the 1D KPZ equation (for which noise is non-conserved) and its space integral, the noisy Burgers equation, for which noise is conserved \cite{Rodriguez-Fernandez20}. Detailed confirmation of this whole interpretation of Fig.\ \ref{fig:Sq_int}(a) is provided by the data collapse of those same data, provided in Fig.\ \ref{fig:DSA_int}(a). Indeed, it is achieved using the theoretically expected exponent values $\alpha=7/8$ and $z=2.19$, and takes the form of a standard FV dynamic scaling ansatz, Eq.\ \eqref{ec:SFV}. Note that, due to the present noise correlations, the behavior of the scaling function for $qt^{1/z}\ll 1$ is characterized by an $1.80$ exponent value, as $q^{-2}q^{2\alpha+2}=q^{2\alpha}=q^{1.80}$.

For the quench from $T=\infty$, and in analogy to the TDGL case, the non-equilibrium behavior of the structure factor for the integral model consists of an overgrowth regime, followed by a long-time relaxation which does not satisfy any dynamic scaling ansatz, see Fig.\ \ref{fig:Sq_int}(b). As seen in Fig.\ \ref{fig:DSA_int}(b), the structure factor data within the overgrowth regime do collapse using $\alpha\approx 7/8$ and $z\approx 2.19$, but now displaying an anomalous scaling behavior with spectral exponent $\alpha_s\approx 1.1$ corresponding to faceted anomalous roughening, as $\alpha\neq\alpha_s >1$.

Overall, for both quenches the evolution of $S(q,t)$ for the integral GL model follows the same temporal dynamics and displays the same time regimes as for the TDGL equation, ultimately because the mapping between them, Eq.\ \eqref{eq:integral_model}, only involves the wave-vector dependence, having no explicit effect on the time dependence. Moreover, the collapse of the $S(q,t)$ data using the theoretical values of the kinetic roughening exponents $\alpha$ and $z$ for the integral model as in Table \ref{tab:exponents_KR} suggests that the deviations found in the value of $\beta$ for the roughness $W(t)$ may be attributed to stronger scaling corrections for the latter in our finite-size simulations.

\begin{figure}[!t]
        \includegraphics[width=0.95\linewidth]{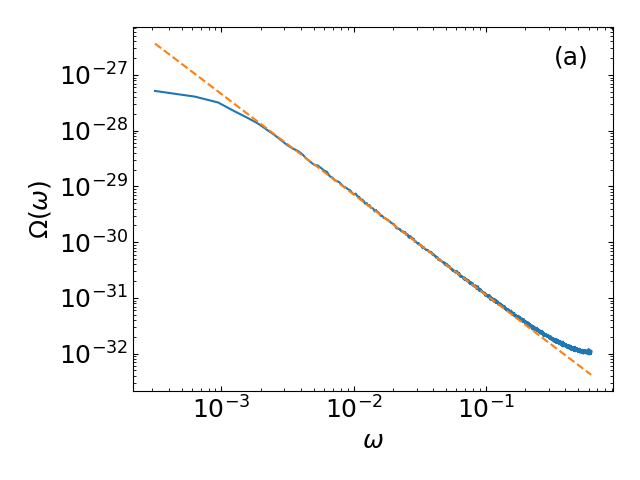}
        \includegraphics[width=0.95\linewidth]{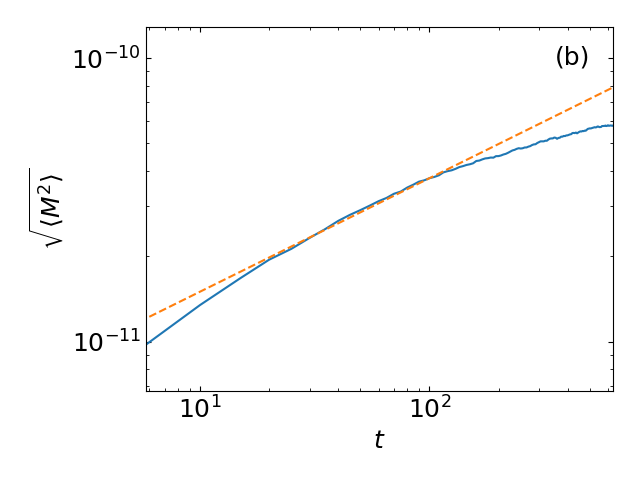}
    \caption{(a) Measurement of the dynamic exponent for the integral-GL model after a critical quench from $T=0$ through the power spectrum method (the orange dashed line corresponds to $\Omega(\omega) \sim \omega^{-1.805}$). (b) Measurement of the dynamic exponent for the integral-GL model after a critical quench from $T=\infty$ through the domain size method (the orange dashed line represents $\sqrt{\langle M^2\rangle}\sim t^{0.40}$).}
    \label{fig:z_int}
\end{figure}

\begin{figure}[h!]
\includegraphics[scale=0.47]{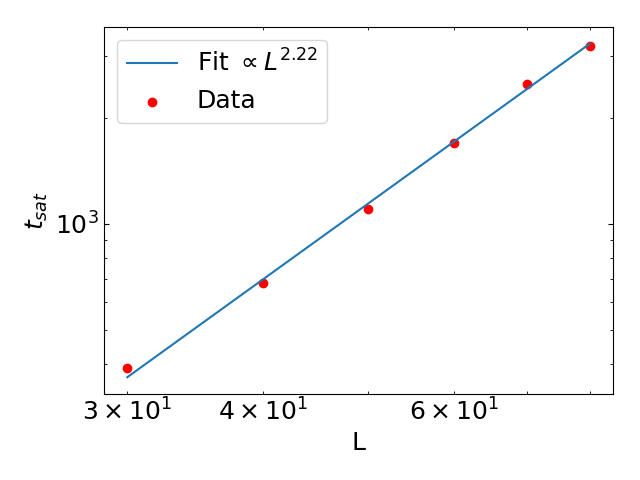}
\vspace{-0.3cm}
\caption{Saturation time versus system size for the integral GL model after a critical quench from $T=\infty$. The red points are the numerical data while the blue line corresponds to $t_{\rm sat}\sim L^{2.22}$.}
\label{fig:t_cutoff_sat}
\end{figure}

We have assessed scaling behavior for additional physical quantities in the integral  GL model. Thus, for the critical quench from $T=0$, the frequency power spectrum deviates from a power law for small frequencies, see Fig.\ \ref{fig:z_int}(a), possibly due to the approximation used to compute the spatial average of the field $h$. Nevertheless, the exponent value in the $\Omega_\omega \sim \omega^{-1.805}$ power law measured coincides with that found for the TDGL equation. This would, a priori, suggest $z\neq 2.19$ for the integral model according to Eq.\ \eqref{eq:mu_vs_eta}, since the roughness exponent of the integral model differs from the TDGL value. Nevertheless, the consistent dynamic scaling ansatz demonstrated in Fig.\ \ref{fig:DSA_int}(a) shows otherwise, and suggests revision of Eq.\ \eqref{eq:mu_vs_eta}. Analogously, for the quench from $T=\infty$ the same exponent as in the TDGL equation is also found by applying the domain size method to the integral model, namely, $n_c \approx 0.40$, see Fig.\ \ref{fig:z_int}(b). This suggests a similar inconsistency, this time through Eq.\ \eqref{eq:z_vs_nc}. We have explicitly checked the validity of the theoretical expectation of the dynamic exponent for the integral model by directly measuring the dependence of the saturation time $t_{\rm sat}\sim L^z$ with system size $L$, recall Sec.\ \ref{sec:dsa} and see e.g.\ Fig.\ \ref{fig:t_cutoff_sat} for the quench from $T=0$. The resulting value $z\approx 2.2$ is again quite consistent with the data collapse of the structure factor and with the theoretical estimate. All this reinforces the straightforward inapplicability of Eqs.\ \eqref{eq:mu_vs_eta} and \eqref{eq:z_vs_nc}, leading us to expect corrections to be required for this type of equations for $z$ in terms of $\mu$ or $n_c$. Such corrections possibly arise from the mapping, Eq.\ \eqref{eq:integral_model}, between the TDGL and the integral GL models, as those equations depend explicitly on the value of the roughness exponent $\alpha$. 


\begin{figure}[!t]
        \includegraphics[width=0.95\linewidth]{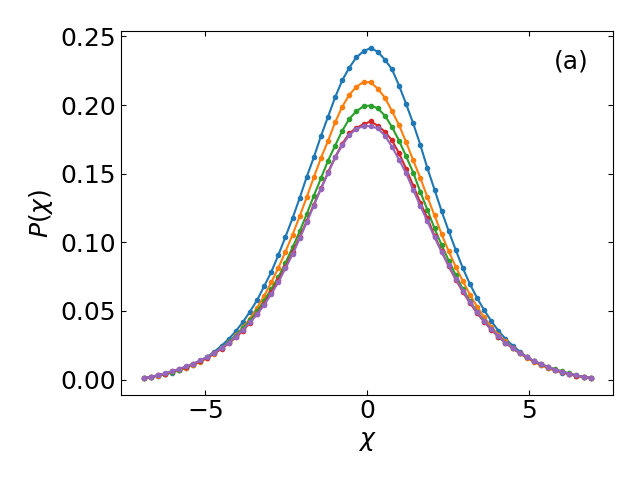}
        \includegraphics[width=0.97\linewidth]{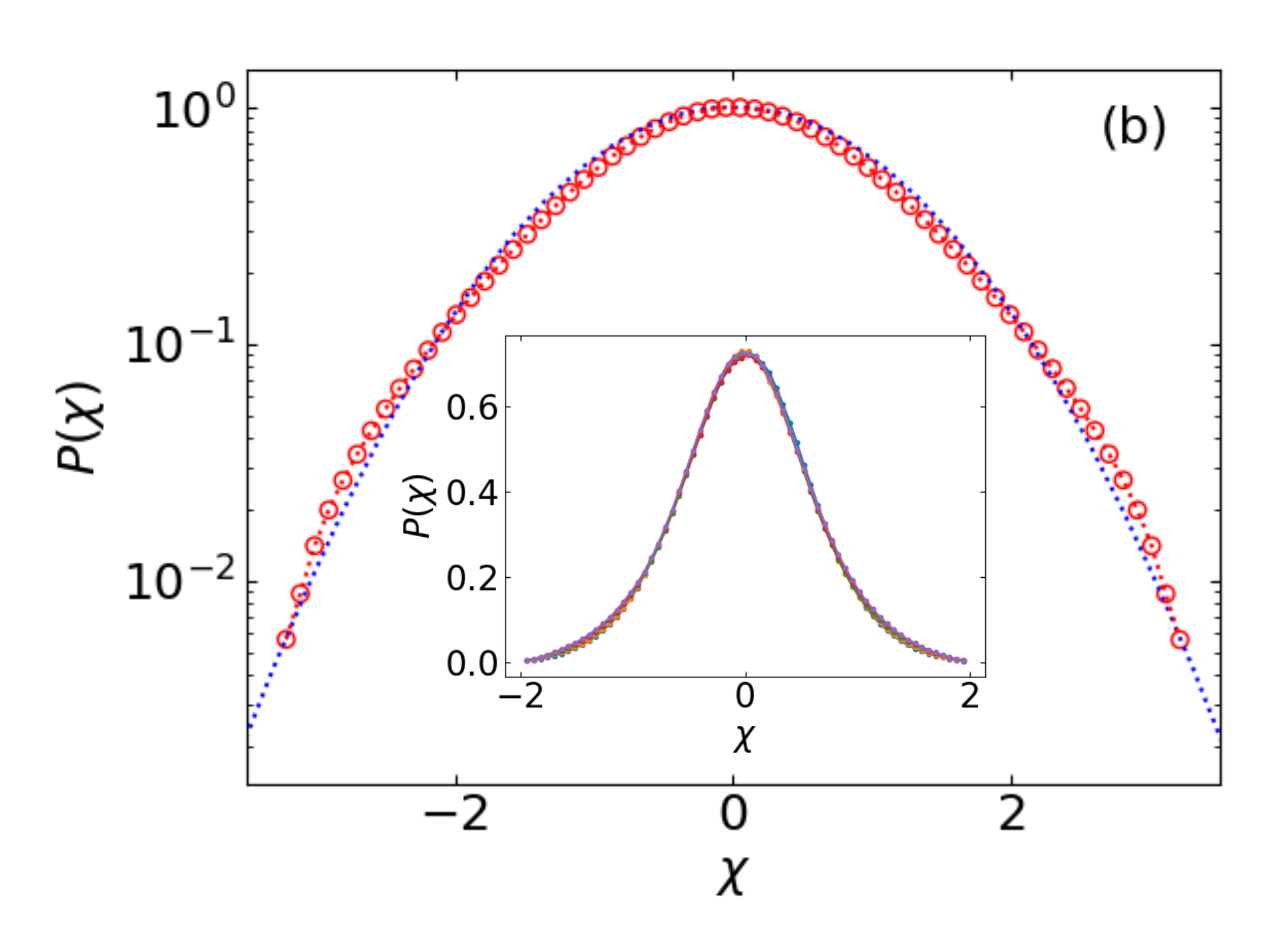}
    \caption{PDF of height fluctuations for the integral GL model after a critical quench from $T=0$, in the non equilibrium regime $(t<10^4)$. (a) PDF curves prior to normalization by the roughness.  (b) Time-averaged PDF in log-scale. The data have been normalized to zero mean and unit variance. The dashed blue line corresponds to a Gaussian distribution, for comparison. Inset: PS collapse of the data in (a) using $\beta'=0.22$. Different colors correspond to different (linearly-spaced) times.}
    \label{fig:PS_t0_int}
\end{figure}

We have also assessed the PDF of fluctuations of the integral GL model. In the non-equilibrium regime of the critical quench from $T=0$, the PDF is close to the equilibrium distribution, with the most important difference being the value of the kurtosis, see Fig.\  \ref{fig:PS_t0_int}. Indeed, the PDF is symmetric with a maximum at $\mathcal{X}=0$, away from which it decays first exponentially fast, to then develop tails that widen approximately similar to those of a Gaussian. By applying the PS formula with $\beta'=0.22$, the PDF curves for different times collapse onto a single master curve, reinforcing the accuracy of this value of the growth exponent to describe the time increase of the roughness in this case.

\begin{figure}[!h]
        \includegraphics[width=0.95\linewidth]{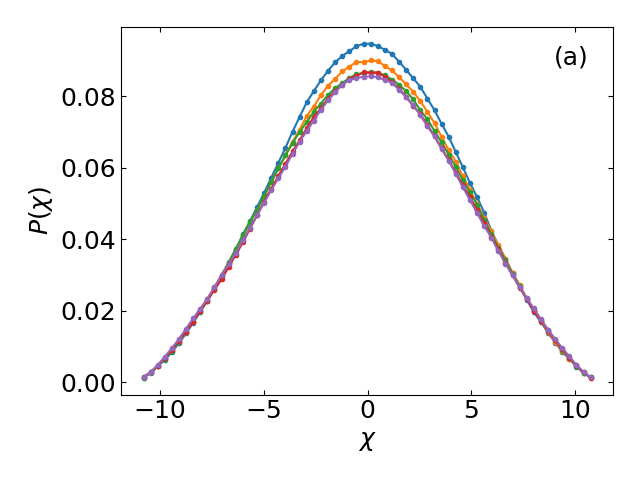}
        \includegraphics[width=0.95\linewidth]{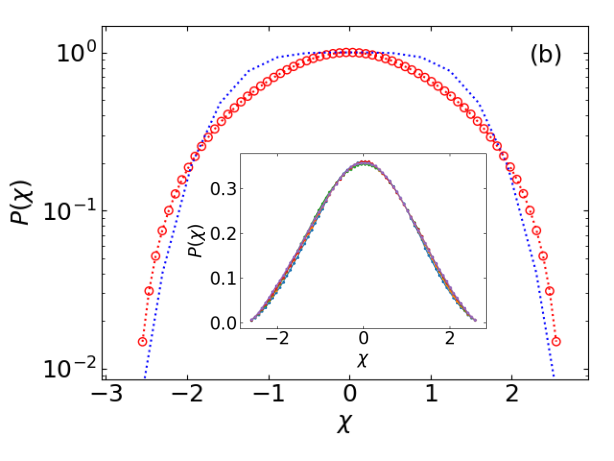}
    \caption{PDF of height fluctuations for the integral GL model after a critical quench from $T=\infty$, within the overgrowth regime. (a) PDF curves for different times without normalizing by the roughness. (b) Time-averaged PDF in log-scale. The data have been normalized to zero mean and unit variance. The dashed blue line corresponds to a stretched exponential distribution $P(\mathcal{X})\propto e^{-\mathcal{X}^4}$, for comparison. Inset: PS collapse of the data in (a) through normalization by the roughness $W(t)$. Different colors correspond to different (linearly-spaced) times. }
    \label{fig:PDF_int_inf}
\end{figure}

\begin{figure}[!h]
        \includegraphics[width=0.95\linewidth]{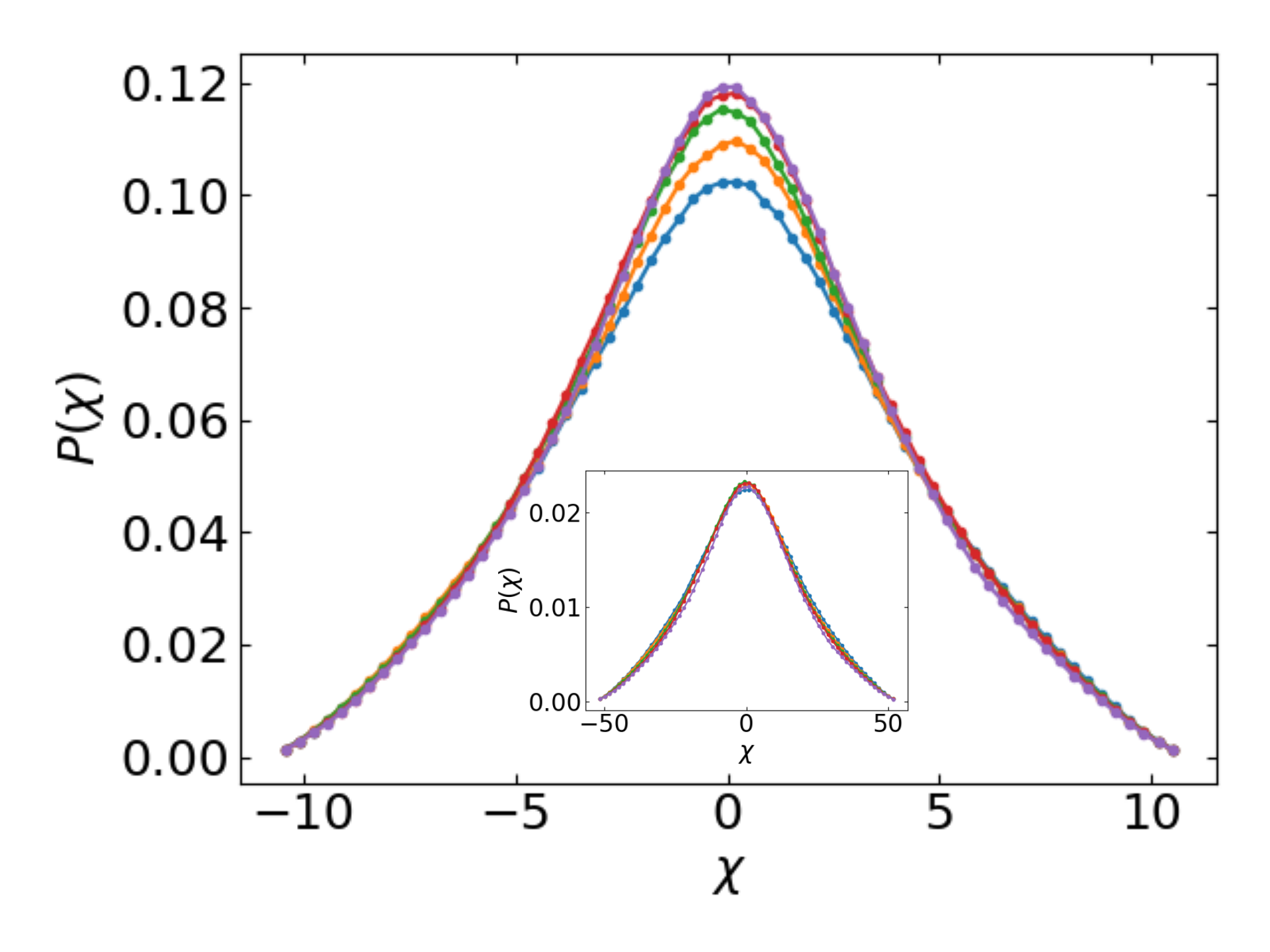}
    \caption{PDF of height fluctuations for the integral GL model after a critical quench from $T=\infty$, within the relaxation regime, without normalizing by the roughness. Inset: Pr\"ahofer-Spohn collapse of the data in the main panel, using $\beta'=-0.09$. Different colors correspond to different (linearly-spaced) times.}  
    \label{fig:PS2_tinf_int}
\end{figure}

Regarding the critical quench from $T=\infty$, within the overgrowth regime (see Fig.\ \ref{fig:PDF_int_inf}), the 
PDF curves for different times collapse consistently onto a single one by directly applying the PS formula, Eq.\ \eqref{eq:Pra-Spohn}, in which the normalization factor is the full roughness $W(t)$, rather than its dominant $t^{\beta}$ term only. 
For the relaxation regime, a consistent PS collapse can be achieved using the effective exponent value $\beta'= -0.09$, as can be seen in Fig.\ \ref{fig:PS2_tinf_int}. In analogy to the TDGL equation, the PDF in this relaxation regime differs from that in the overgrowth regime by the kurtosis only. Indeed, while both PDF functions are similar to a symmetric stretched exponential with a maximum at $\mathcal{X}=0$, the tails are relatively shorter (wider) in the overgrowth (relaxation) regime. 

\section{Summary and Discussion}\label{sec:disc}
\subsection{Ising critical quench from $T=0$}
 
The results obtained in Sec.\ \ref{sec:T=0} for the critical quench from the ordered state show a high degree of agreement between the behaviors of the lattice Ising model and of its continuum description through the TDGL equation. Consistent with the most straightforward expectations, indeed both models exhibit the same critical dynamics, characterized by numerical values for the critical exponents $\alpha$ and $z$ which agree with the theoretical predictions collected in Table \ref{tab:exponents_KR}. In particular, the power spectrum method employed to independently measure the dynamic exponent also reproduces previous results in the literature \cite{Kent}.

While the equilibrium state of the discrete and continuous models displays the predicted power law for the two-point correlation function, the non-equilibrium (or growth, in the kinetic roughening context) regime follows in both systems a standard FV dynamic scaling ansatz, qualitatively similar to that satisfied e.g.\ by the EW and the KPZ equations.
The roughness displays corrections to scaling, possibly as a consequence of $\alpha<0$: 
In this case, subdominant terms in the asymptotic expansion of $W(t)$ are not negligible as compared to the $t^{\beta}$ contribution. This was further stressed by the PDF curves collapsed through the PS formula, suggesting dominant contributions with $\beta'=0>\beta=\alpha/z$.

Regarding fluctuation statistics, the non-equilibrium PDF is biased towards the direction given by the magnetization of the initial condition, displaying an asymmetry (non-zero skewness) which approaches zero with increasing time, finally reaching a symmetric configuration at equilibrium, indeed consistent with the model symmetries. The kurtosis decays over time until saturation, yielding shorter tails than those of the Gaussian distribution, and consistent with those of an stretched exponential $P(\mathcal{X})\propto e^{-\mathcal{X}^4}$, similar to previous results in the literature \cite{ojalvo}.

\subsection{Ising critical quench from $T=\infty$}

For the critical quench from the disordered state, the lattice Ising model and the TDGL equation are also found in Sec.\ \ref{sec:T=infty} to exhibit the same critical dynamics, with critical exponents $\alpha$ and $z$ which are consistent with the theoretical predictions in Table \ref{tab:exponents_KR}. The domain size method applied to measure the dynamic exponent for this case also reproduces the results found in the literature \cite{domain_size}.

The equilibrium behavior is the same as for the critical quench from $T=0$, yielding the same structure factor and PDF since, as expected for a system verifying detailed balance, the equilibrium state should not depend on the initial condition. In contrast, the non-equilibrium dynamics of the critical quench from $T=\infty$ is structured into two time regimes, namely, an initial overgrowth of the roughness followed by a long-time relaxation to equilibrium. As seen in Fig.\ \ref{fig:Sq_tinf}, the structure factor curves for different times during the overgrowth regime do not overlap at intermediate values of $q$ as in the standard FV behavior but, rather, shift upwards systematically with increasing time. This behavior is accounted for by an intrinsically anomalous dynamic scaling ansatz with spectral exponent values $\alpha_s = 0.15$ and 0.10 for Glauber dynamics and the TDGL equation, respectively, as assessed by the data collapse discussed in Sec.\ \ref{sec:T=infty}. 

Intrinsic anomalous roughening is very frequently found in the context of morphologically unstable surfaces (see, e.g.\ Refs.\ \cite{Cuerno04,Barreales22,Marcos22} and other therein) and we believe this is also the case here. To understand the source of this behavior, recall that, at the critical point, the TDGL equation, Eq.\ \eqref{eq:TDGL}, features 
$r<0$ as obtained from the DRG computation and hence becomes, within a linear approximation such that $|\phi| \ll 1$,
\begin{equation}
    \partial_t \phi=|r|\phi+\nu\nabla^{2}\phi+ \zeta.
    \label{eq:tdgl_unst}
\end{equation}
Equation \eqref{eq:tdgl_unst} is expected to be valid for short times after the critical quench from $T=\infty$ is performed, and implies that the homogeneous solution $\phi(\mathbf{r})=0$ is actually unstable to periodic perturbations $\phi_{\rm per}(\mathbf{r},t) \propto e^{\omega(q) t} e^{{\rm i} \mathbf{q} \cdot \mathbf{r}}$, whose amplitude evolves with the linear dispersion relation
\begin{equation}\label{eq:dispersion}
    \omega(q)=\nu\left(\frac{|r|}{\nu}-q^2\right),
\end{equation}
where a characteristic, time-independent wave-vector is identified as $q_{\rm corr} = \sqrt{r/\nu}$, or equivalently, a characteristic time-independent wavelength $\xi_{\rm corr} = 2\pi/q_{\rm corr}$. With the parameter values used in our numerical simulations, $q_{\rm corr}=1$. 
In view of Eq.\ \eqref{eq:dispersion}, a Type III pattern forming instability takes place in the general classification of pattern formation \cite{cross}. The presence of noise in Eq.\ \eqref{eq:tdgl_unst} is not expected to alter this interpretation.

\begin{figure}[!t]
        \includegraphics[width=0.95\linewidth]{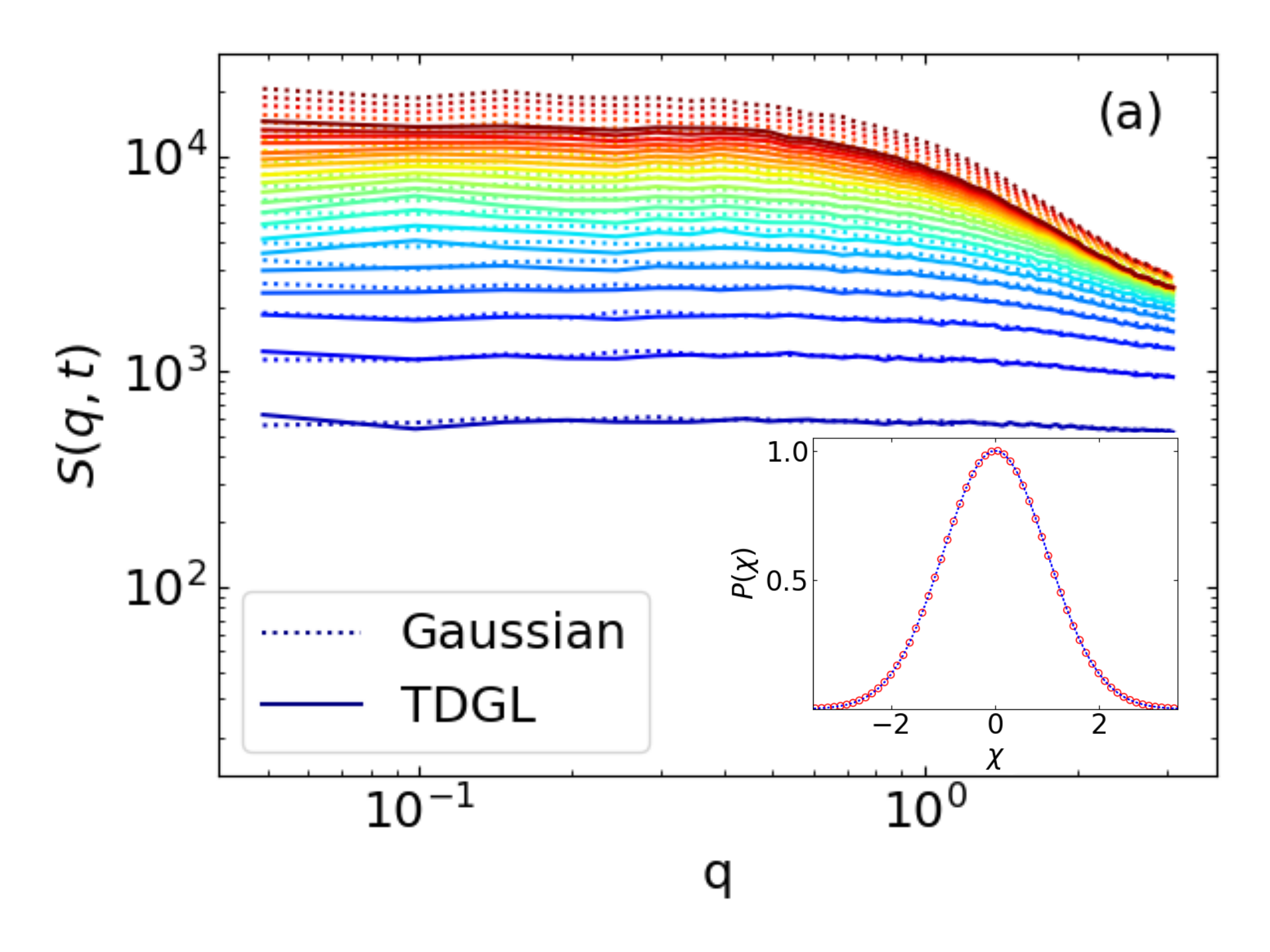}
        \includegraphics[width=0.95\linewidth]{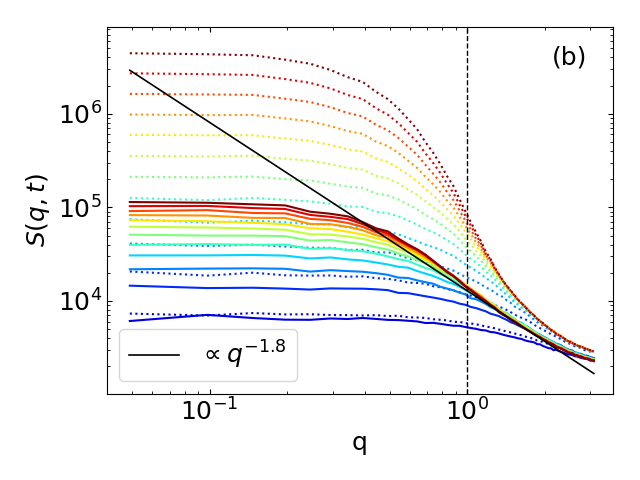}
    \caption{Comparison of the structure factors, as functions of $q$, for the full TDGL equation (solid lines) and its linear approximation, Eq.\ \eqref{eq:tdgl_unst}, for a critical quench from $T=\infty$ and times prior to and at the beginning of the overgrowth regime. The time arrow goes from blue to red. (a) For an initial short time interval, the structure factor and the fluctuation PDF (inset) of the full TDGL equation and its linear approximation coincide. (b) For times where the correlation length of the full TDGL equation is larger than $2\pi/q_{\rm corr}$ ($q_{\rm corr}=1$ is marked by the dashed vertical line), its structure factor no longer coincides with that of the linear approximation. The black solid line represents asymptotic behavior as $S(q) \sim 1/q^{1.8}$.}
    \label{fig:instability}
\end{figure}

The morphological instability introduced by the linear term $|r|\phi$ in Eq.\ \eqref{eq:tdgl_unst} is due to its positive coefficient, which physically induces exponential growth of perturbations around the $\phi(\mathbf{r},t=0)=0$ initial condition. The overgrowth regime is a consequence of such an instability. In Fig.\ \ref{fig:instability} we compare the structure factor of the linear approximation, Eq.\ \eqref{eq:tdgl_unst}, with that of the full TDGL equation, Eq.\ \eqref{eq:TDGL}. At very short times prior to the overgrowth regime, both equations indeed display the same behavior with respect to $S(q,t)$ and also in terms of the fluctuation PDF, which is Gaussian to a good approximation [see Fig.\ \ref{fig:instability}(a)].
However, the unstable amplitude growth of $\phi$ renders the linear approximation $|\phi| \ll 1$ eventually invalid, and the $\phi^3$ nonlinearity of the full TDGL equation can no longer be neglected. This implies the departure of the structure factor of the full system from its linear approximation, with the value of $S(q,t)$ at intermediate $q$ overshooting the asymptotic behavior, see the yellow to red curves in Fig.\ \ref{fig:instability}(b). This is the beginning of the overgrowth regime.

\begin{figure}[!t]
    \centering
    \includegraphics[width=0.47\textwidth]{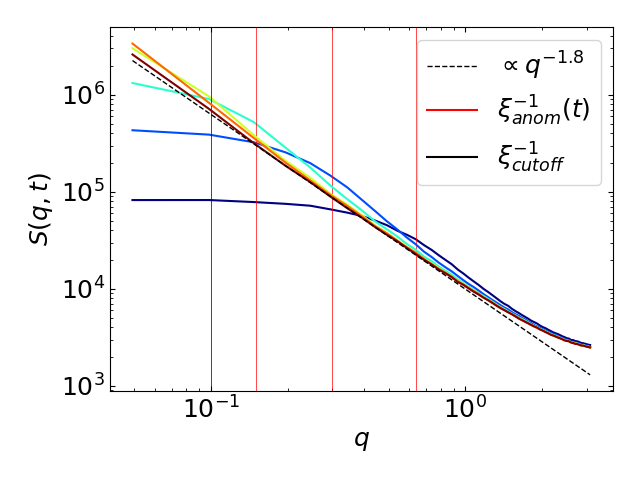}
    \caption{Structure factor of the TDGL equation after a critical quench from $T=\infty$ from numerical simulations, for several times within the overgrowth regime, increasing from blue to dark red. For the first three times shown, the vertical red lines indicate the approximate inverse value of the correlation length. Once the correlation length has increased past the (inverse) cutoff wavevector (black) the relaxation regime starts and the instability is suppressed. The dashed line indicates the asymptotic $S(q) \sim q^{-1.80}$ behavior.}
    \label{fig:q_anomalous}
\end{figure}

\begin{figure}[!t]
        \includegraphics[width=0.95\linewidth]{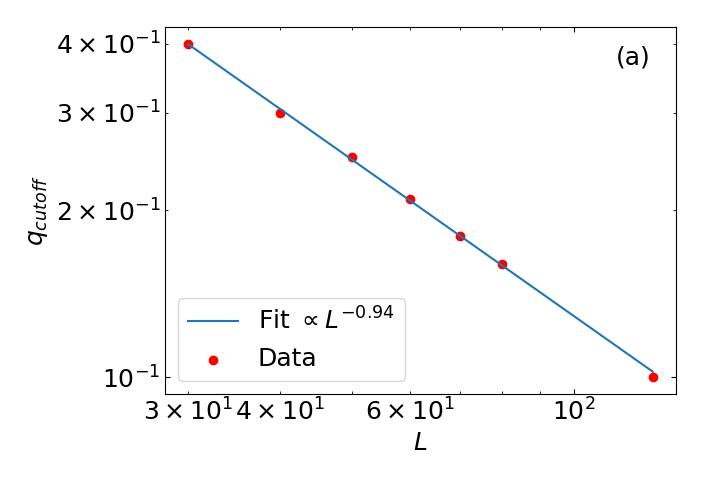}
        \includegraphics[width=0.95\linewidth]{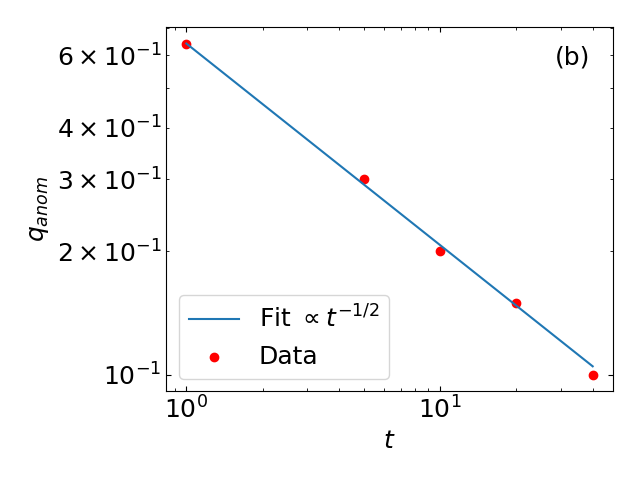}
    \caption{(a) Cutoff wave vector versus system lateral size from simulations of the TDGL equation after a critical quench from $T=\infty$, in the overgrowth regime. Red dots corresponds to numerical data and the blue line to the power-law fit $q_{\rm cutoff} \sim L^{-0.94}$. (b) Anomalous wave vector versus time. Red points correspond to numerical data, blue line to fit $q_{\rm anom} \sim t^{-1/2}$.  }
    \label{fig:q_cutoff_vs_L}
\end{figure}

The structure factor during the overgrowth regime is considered in detail in Fig.\ \ref{fig:q_anomalous}, which repeats a subset of the curves earlier presented in Fig.\ \ref{fig:Sq_tinf}(a). Figure \ref{fig:q_anomalous} identifies two relevant length scales. One is the correlation length $\xi_{\rm anom}(t)=2\pi/q_{\rm anom}(t)$, where $q_{\rm anom}(t)$ is the wave-vector value that separates $q$-independent from $q$-dependent behavior of the structure factor for a fixed $t$. This length scale grows [equivalently, $q_{\rm anom}(t)$ decreases] with increasing time, as indicated in the figure. A different, time-independent, scale is $\xi_{\rm cutoff}$. As seen in Fig.\ \ref{fig:q_anomalous}, once $\xi_{\rm anom}(t) > \xi_{\rm cutoff}$, the overgrowth regime finishes and the system enters the relaxation regime towards equilibrium, as reflected in the behavior of the roughness, recall Fig.\ \ref{fig:W_tinf}(a). To verify whether this cutoff length scale is a finite-size effect, we have analyzed the relation between the cutoff wave vector $q_{\rm cutoff} = 2\pi/\xi_{\rm cutoff}$ and the lateral system size $L$, see Fig.\ \ref{fig:q_cutoff_vs_L}(a). The numerical data are consistent with $q_{\rm cutoff}\sim L^{-0.94}$, i.e., $\xi_{\rm cutoff}\sim L^{0.94}$. Hence, the cutoff correlation length would never be reached in the thermodynamic limit, where overgrowth fluctuations would grow indefinitely. On the other hand, $q_{\rm anom}(t)=2\pi/\xi_{\rm anom}(t) \sim t^{-1/2}$ according to Fig.\ \ref{fig:q_cutoff_vs_L}(b), hence $\xi_{\rm anom}(t) \sim t^{1/2}$ grows diffusively so that this length scale can be associated with the linear behavior.

All this suggests the competition between several different length scales in the system, namely, the correlation length $\xi(t)\sim t^{1/z}$, the anomalous growth length scale $\xi_{\rm anom}(t)$, and the cutoff length scale $\xi_{\rm cutoff}$, which are not equivalent.
Thus, during the initial time evolution, the nonlinear term in the TDGL equation is negligible as compared with the linear term. Later on, the non-equilibrium regime is a result of the competition between the linear and the nonlinear terms in the TDGL equation: the former drives unstable growth of fluctuations, while the latter penalizes such growth and stabilizes it. This competition allows one to understand the non-equilibrium regimes: {\em (i)} During the overgrowth regime, the linear term still competes with the nonlinear term, increasing the amplitude of fluctuations. As the local magnetization field becomes increasingly correlated in space, the nonlinear contribution grows. The correlation length $\xi(t)\sim t^{1/z}$ controls the data collapse of the time-dependent $S(q,t)$ curves, while $\xi_{\rm anom}(t) \sim t^{1/2} < \xi_{\rm cutoff}$ within the overgrowth regime. {\em (ii)} Once $\xi_{\rm anom}(t) > \xi_{\rm cutoff}$, the nonlinear term completely dominates over the the linear instability, inducing decay to the equilibrium state in the form of a long-time relaxation. 
Overall, one can associate the overgrowth regime of the TDGL equation after a critical quench from $T=\infty$ with the transient time during which the nonlinear term stabilizes a Type III morphological instability. Let us note again the occurrence of anomalous scaling in other kinetic roughening systems that similarly undergo morphological instabilities; see, e.g.\ Refs.\ \cite{Cuerno04,Barreales22,Marcos22} and other therein.

The mechanism just described is consistent with previous literature and aligns with previous qualitative theoretical predictions of such unstable growth by e.g.\ K.\ Binder \cite{binder} and D.\ A.\ Huse \cite{huse}. It moreover also allows us to understand why the dynamics are different for the two critical quenches we have studied. In terms of the local magnetization field, we find two different cases at $t=0$:
\begin{equation}
    \begin{array}{ll}
        \text{Completely ordered state:}\; &|\phi| \approx 1\Rightarrow |\phi|^3 \approx |\phi|,\\
        \text{Completely disordered state:}\; &|\phi|\approx 0\Rightarrow |\phi|^3\ll |\phi|.
    \end{array}
\end{equation}
Thus, in the quench from $T=0$ the nonlinear term in the TDGL equation competes with and stabilizes the linear instability already from $t=0$, eventually avoiding the anomalous overgrowth while, for the quench from $T=\infty$, the linear term competes with the nonlinear term for a finite time interval until $\xi_{\rm anom}(t)$ reaches the cutoff length scale, which is fixed by parameter conditions. Qualitatively, and according to the simulations, a similar behavior seems to hold for the discrete Ising model with Glauber dynamics. Note that the present dependence of the dynamics with the value of the initial condition is consistent with the system symmetries, since e.g.\ the TDGL equation is {\em not} invariant under arbitrary shifts in the value of the local magnetization, $\phi\rightarrow \phi+K$. This contrasts with many important models of surface kinetic roughening, like e.g.\ the KPZ, EW, or CKPZ equations, Eqs.\ \eqref{eq:KPZ_eq}, \eqref{eq:EW}, or \eqref{eq:cKPZ_eq}, respectively, which do remain invariant under arbitrary global shifts in the value of the surface height, $h\rightarrow h+H$ \cite{Barabasi,kardar}.

\subsection{Integral GL model}
The integral GL model follows qualitatively the same behavior as the TDGL equation, displaying its expected values for the critical exponents $\alpha$ and $z$, as reflected in Table \ref{tab:exponents_KR}. As in the TDGL equation, the dynamic scaling ansatz is of a simple FV type for the critical quench from $T=0$, but it changes to an anomalous form in the overgrowth regime of the critical quench from $T=\infty$. Due to the increase by one unit in the value of the roughness exponent with respect to the TDGL case, now the dynamic scaling ansatz is faceted instead of intrinsic anomalous roughening. At any rate, the relation in terms of the $\alpha$, $\alpha_s$, and $z$ exponent values between the integral and the TDGL models is similar to the analogous relation between exponent values for the KPZ and the noisy Burgers equations,
for the space-time properties of a field and those of its space derivative \cite{Rodriguez-Fernandez20}.

Regarding the statistics of fluctuations, for the quench from $T=\infty$ both the TDGL and the integral GL models display symmetric PDF. However, for the quench from $T=0$, the TDGL equation yields an asymmetric PDF for the non-equilibrium regime while the integral GL model displays a symmetric PDF. Hence, in this case space integration is inducing a symmetry in the field fluctuations. Moreover, normalizing the fluctuation histogram by the roughness $W(t)$ yields universal fluctuation PDF curves, as for the TDGL equation.

\section{Conclusions and Outlook}\label{sec:concl}

The first conclusion that can be drawn from our results is that, in spite of the mathematical difficulties with the continuum limit of the TDGL equation \cite{Hairer2012,Ryser12}, finite numerical simulations of this system do still provide an accurate approximation of the nonconserved critical dynamics of the lattice Ising model \cite{ojalvo,Monroy21}, underscoring the universality of the critical behavior which is found.

By further interpreting the behavior of the model from an interfacial perspective, the critical dynamics of the Ising system displays a morphological instability which is stabilized by the nonlinear interaction in ways that do depend on the initial condition. Thus, while a $T=0$ initial condition leads to non-equilibrium dynamics satisfying a standard FV scaling ansatz that reaches saturation into equilibrium, a critical quench from $T=\infty$ presents a much more complex non-equilibrium dynamics towards the same equilibrium state. In spite of the fact that the values of $\eta$ (and, equivalently, the roughness exponent $\alpha$) and the dynamic exponent $z$ are, as expected, shared with those found in the simpler critical quench, now the dynamic scaling ansatz that ensues is intrinsically anomalous, being characterized by an additional independent roughness exponent, $\alpha_s$. This peculiar scaling behavior thus characterizes the space-time correlations of the nonconserved critical dynamics of the Ising universality class, restricted to an intermediate time (overgrowth) regime, and is fully analogous to observations in models and experiments of kinetically rough surfaces, see, e.g.\ Ref.\ \cite{Cuerno04} and other therein. In that context, intrinsic anomalous roughening is frequently associated with transient behavior related with, e.g.\ morphological instabilities and/or quenched noise \cite{Lopez2005}, although recently a continuous model (namely, the tensionless KPZ equation \cite{Cartes2022,Rodriguez22,Fontaine2023}) has been found that, while featuring simple time-dependent noise, asymptotically displays intrinsic anomalous roughening  \cite{Rodriguez22}. From the RG point of view, one would expect an additional scaling field to account for the spectral roughness exponent; within current understanding, the derivative field having an independent dynamics is expected to play some role in this aspect \cite{krug,Dassarma94,Schroeder2,Lopez1999}. 

Another bonus of the reinterpretation of the critical dynamics of the Ising model in terms of surface kinetic roughening is the assessment of universal forms of the fluctuation PDF along the non-equilibrium dynamics, which become time-independent once suitably rescaled by the time-dependent roughness, in full analogy with results recently undercovered for paradigms of kinetic roughening like, e.g., the KPZ \cite{Takeuchi18} and CKPZ \cite{Carrasco2016} equations. According to our present results, for the nonconserved critical dynamics of the Ising model these PDF functions depend on the initial condition but, in analogy with the kinetic roughening case, one could conceive of additional dependencies like additional dynamical constraints \cite{Takeuchi18} and/or the source of randomness in the system \cite{Rodriguez21}. In analogy to the equilibrium case \cite{Plascak2013,Landau_book}, the assessment of these probability distribution functions is becoming increasingly useful to obtain the critical properties of statistical mechanical systems far from equilibrium too, where, as discussed in Sec.\ \ref{sec:kr} and Appendix \ref{App_cKPZ}, they are becoming important even to identify universality classes non-ambiguously. Moreover, bearing in mind that we are considering strongly correlated (critical) systems, the study of these PDF provides an interesting context for stochastic behavior beyond the central limit theorem \cite{Rodriguez-Fernandez20,Balog2022}.

The formulation and study of the integral GL model has allowed us to confirm the above picture on a system which by construction has a positive roughness exponent. The critical exponents are found to be consistent with the expectations, while the type of dynamic scaling ansatz depends on the initial condition in full parallel to the case of the TDGL equation, although for the critical quench from $T=\infty$ the anomalous scaling within the overgrowth regime is faceted, rather than intrinsic, due to the increase in the value of the roughness exponents implied by the definition of the integral GL model. Likewise, a non-trivial symmetry emerges in the fluctuation statistics due to the space integration. This is the opposite behavior with respect to the KPZ-Burgers case \cite{Rodriguez-Fernandez20}, where space integration leads instead to non-trivial breaking of the symmetry of the fluctuation PDF.

A natural extension of the present work would be to the critical dynamics of the 3D Ising model. Indeed, a recent numerical study \cite{dynamics_d3} of $n_{th}$-order time correlation functions, $M^n (t)$, for a critical quench from $T=\infty$, reports similar behavior to our present findings under such type of initial condition. Namely, the time evolution does not follow simple scaling with the dynamic exponent $z$, while it involves an early time rise followed by late stage relaxation. We expect that the argument discussed in Sec.\ \ref{sec:disc} also applies to 3D, since the required conditions are satisfied. Indeed, the system also exhibits a critical point with long-range correlations while the linear term is unstable, i.e., $r<0$. Huse provides a criterion for the overgrowth dynamics based on a non-equilibrium critical exponent $\lambda_c$ together with data which supports the validity of this picture for 3D \cite{huse}, while Binder's argument \cite{binder} does not rely on a specific dimension but, rather, on the existence of a critical point where correlations are long-ranged.

Additional extensions of our present work include the case of conserved (model B) critical dynamics and, on a more technical side, improved  interpretations of the time power spectrum and domain size methods that circumvent limitations like those found in Sec.\ \ref{sec:im}.

\begin{acknowledgments}
We would like to thank J.\ Rodr\'{\i}guez-Laguna for suggesting the study of the critical quench from $T=0$. This work has been partially supported by Ministerio de Ciencia e Innovaci\'on (Spain), by Agencia Estatal de Investigaci\'on (AEI, Spain, 10.13039/501100011033), and by European Regional Development Fund (ERDF, A way of making Europe) through Grants No.\ PID2021-123969NB-I00 and No.\ PID2024-159024NB-C21. H.\ V.\ del P.\ acknowledges a scholarship from Ministerio de Educaci\'on y Formaci\'on Profesional, Spain, as well as hospitality at Departamento de Matem\'aticas and GISC, Universidad Carlos III de Madrid, Legan\'es, Spain, where most of this work has been performed.
\end{acknowledgments}

\appendix

\section{Determination of the effective critical point}\label{AppC}
Finite size effects induce size-dependent corrections to the critical temperature, hence we need to determine the precise critical point for the each one of the systems addressed in our numerical simulations.

\begin{figure}[t]
\includegraphics[scale=0.5]{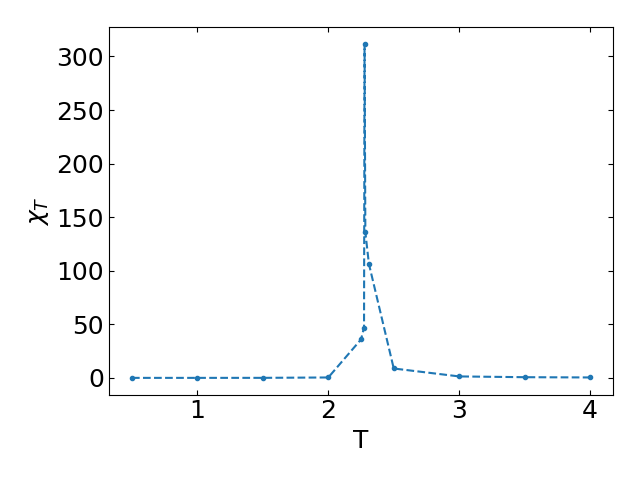}
\vspace{-0.3cm}
\caption{Magnetic susceptibility vs temperature for the Ising model with Glauber dynamics on a $N=128\times 128$ square lattice. The peak in the susceptibility indicates the location of the phase transition. The dashed line is a guide to the eye.}
\label{fig:Glauber_suscep}
\end{figure}

In the case of the discrete Ising model with Glauber dynamics, for maximum accuracy the effective $T_c$ was computed by setting the reference system size ($N=128\times 128$), and computing the dependence of the fluctuations amplitude with temperature. Then, a narrow range of temperatures enclosing the divergence which indicates criticality is scanned in $\delta T = 0.001$ steps. For each temperature, all the observables are measured, and the value which yields critical or power-law behavior in all of its observables is selected as the effective critical temperature. The structure factor has been seen to be most sensitive to temperature deviations from the effective critical value. As an example, the temperature dependence of the magnetic susceptibility obtained following this methodology is shown in Fig.\ \ref{fig:Glauber_suscep}.
Fine tuning over a narrow temperature range was performed, yielding the effective temperature $T_c =2.277$ on a square lattice of size $N=128\times 128$.

With respect to the critical temperature for the TDGL equation, note that the temperature dependence is contained in the coefficients appearing in the continuous model, Eq.\ \eqref{eq:TDGL}. Hence, it is convenient to rescale coordinates and fields to render the equation nondimensional, grouping all parameter dependencies into the noise amplitude, thus \cite{ojalvo}
\begin{equation}
    \frac{\partial \mathbf{\phi}(\mathbf{r},t)}{\partial t}=\frac{1}{2}\left(\phi(\mathbf{r},t)-\phi^{3}(\mathbf{r},t)+\nabla^{2}\phi(\mathbf{r},t)\right)+D\mathbf{\eta}(\mathbf{r},t),
    \label{eq:tdglapp}
\end{equation}
where the nondimensional variables $t,\mathbf{r},\phi$ in terms of the original ones (indetified with tildes here), are defined as 
\begin{equation}
    t=2|r|\tilde{t},\quad \mathbf{r} = \sqrt{\frac{|r|}{\nu}}\tilde{\mathbf{r}}, \quad \phi = \sqrt{\frac{u}{|r|}}\tilde{\phi}.
\end{equation}
Then, the critical point is determined by the noise amplitude $D_c$. Note the sign of the linear non-derivative term, in accordance with the DRG study described in Sec.\ \ref{sec:DRG} and Appendix \ref{AppA}, which predicts a negative sign for the coefficient of that term at the critical point, i.e., $r^{*}<0$, see Eq.\ \eqref{eq_sign_r}.

\begin{figure}[t]
\includegraphics[scale=0.5]{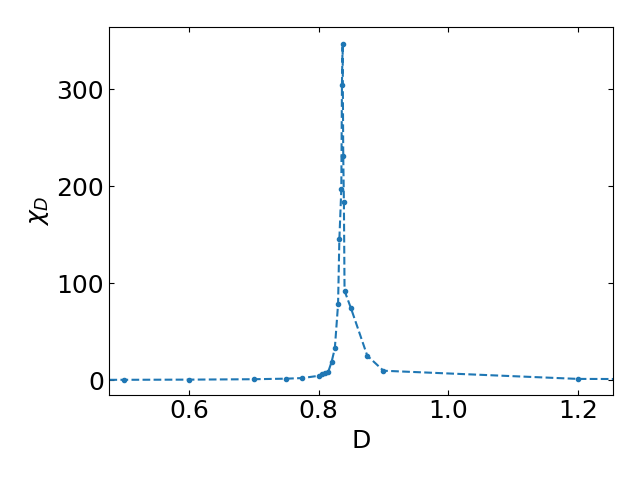}
\vspace{-0.3cm}
\caption{Susceptibility vs noise amplitude for the TDGL equation, Eq.\ \eqref{eq:tdglapp}. The peak in the susceptibility indicates the critical noise amplitude. The dashed line is a guide to the eye.}
\label{fig:TDGL_suscep}
\end{figure}

Following the same methodology as for the Glauber dynamics, the dependence of the magnetic susceptibility on the noise amplitude is given in Fig. \ref{fig:TDGL_suscep}.
Accordingly, the critical noise amplitude is given by $D_c=0.835$ for the nondimensional TDGL equation. 

The critical quenches are thus simulated by setting $T_c =2.277$ and $D_c =0.835$ for the lattice and continuum Ising model, respectively. Furthermore, simulations of TDGL were performed setting the time and spatial steps $\delta t = 0.1$ and $\delta x =\delta y = 1$. For Glauber dynamics, $\mathcal{J}=J/k_B =1$ was set for simulations. Observables were averaged over 1000 noise realizations, and the numerical scheme was benchmarked against the linear model presented in Appendix \ref{App:EW}. The uncertainty in the fitted critical exponents is indicated by the last reported digit (e.g., $z = 2.17 \pm 0.01$).

\begin{figure}[!t]
        \includegraphics[width=0.95\linewidth]{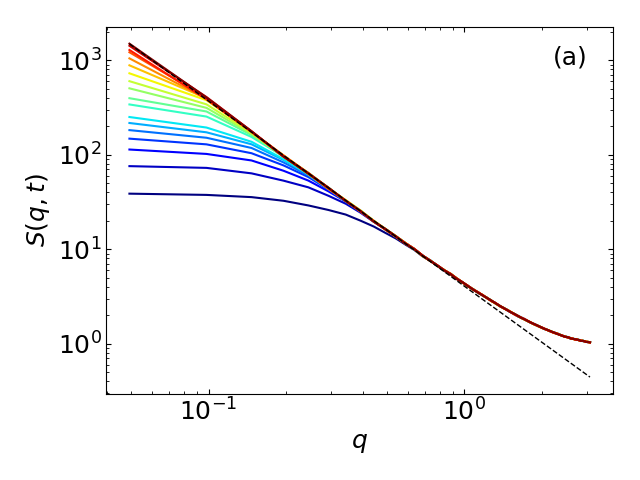}
        \includegraphics[width=0.95\linewidth]{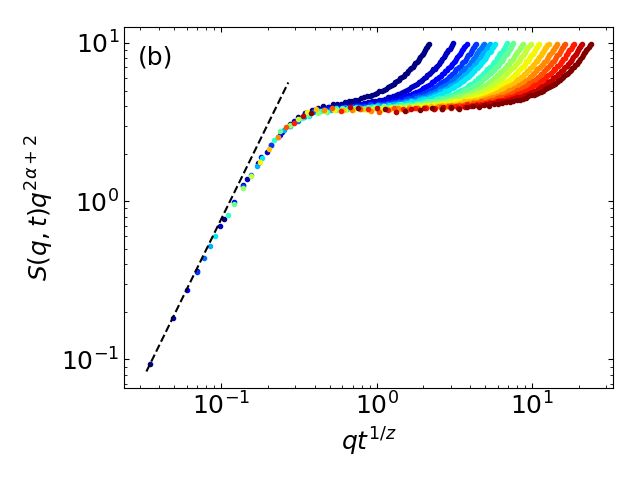}
        \includegraphics[width=0.95\linewidth]{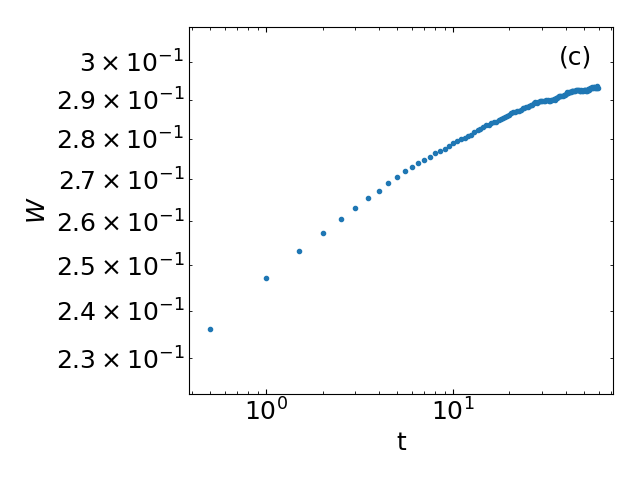}
    \caption{Time evolution of the 2D EW equation from numerical simulations of Eq.\ \eqref{eq:EW} for a $\phi(\mathbf{r},t=0)=0$ flat initial condition. (a) Structure factor as a function of $q$ for different times, with time increasing from blue to red (log-spaced). The dashed line corresponds to the asymptotic $S(q) \sim q^{-2}$ behavior, Eq.\ \eqref{eq:SqEW} for $qt^{1/2}\ll 1$. (b) Collapse of the data in panel (a) following the FV scaling ansatz, Eqs.\ \eqref{ec:SFV}-\eqref{ec:SFVf}, for $\alpha=0$ and $z=2$. The dashed line corresponds to $f_S \sim u^{2}$. (c) Surface roughness as a function of time. Note the slow growth rate, as $\beta_{\rm EW}=0$ (log) for $d=2$ \cite{Barabasi}.}
    \label{fig:EW}
\end{figure}

\section{Edwards-Wilkinson equation}\label{App:EW}

The Gaussian approximation of the TDGL equation, Eq.\ \eqref{eq:TDGL}, is \cite{kardar}
\begin{equation}\label{eq:TDGLG}
    \partial_t \phi(\mathbf{x},t)=-r\phi(\mathbf{x},t)+\nu\nabla^{2}\phi(\mathbf{x},t)+\zeta(\mathbf{x},t) ,
\end{equation}
where $\zeta(\mathbf{x},t)$ is delta-correlated Gaussian noise, as in Eq.\ \eqref{eq:noise}. This system is critical for $r=0$ \cite{kardar}, where it simply becomes
\begin{equation}\label{eq:EW}
    \partial_t \phi(\mathbf{x},t)=\nu\nabla^{2}\phi(\mathbf{x},t)+\zeta(\mathbf{x},t) .
\end{equation}
In the surface kinetic roughening context, Eq.\ \eqref{eq:EW} is known as the Edwards-Wilkinson (EW) equation, and it defines an important universality class of its own \cite{Barabasi,krug}. Being linear, its solution can be readily found analytically by Fourier transforms \cite{kardar}. Thus, e.g.\ the structure factor reads, for a null initial condition $\phi(\mathbf{r},t=0)=0$,
\begin{equation}
    S_{\rm EW}(\mathbf{q},t) = \frac{(2\pi)^2 \Gamma}{\nu } \frac{1-e^{-2\nu t q^2}}{q^2},
    \label{eq:SqEW}
\end{equation}
wherefrom, comparing with Eq.\ \eqref{ec:SFV}, the kinetic roughening exponents are $\alpha_{\rm EW}=(2-d)/2$ and $z_{\rm EW}=2$. As a graphical illustration to be compared with e.g.\ the results discussed in Sec.\ \ref{sec:T=0} for a critical quench of the 2D Ising model from the ordered phase, we collect in Fig.\ \ref{fig:EW} numerical results for the time evolution of the 2D EW model using $\phi(\mathbf{r},t=0)=0$. 
Qualitative agreement can be observed with the full TDGL dynamics for the critical quench from $T=0$. In both cases, a standard FV dynamic scaling ansatz is satisfied, although naturally with different values for the kinetic roughening exponents.

\section{Scaling exponents vs height statistics: an example}\label{App_cKPZ}

The conserved KPZ (CKPZ) equation of kinetically rough interfaces reads \cite{Barabasi,krug}
\begin{equation}\label{eq:cKPZ_eq}
    \partial_t \phi(\mathbf{x},t)=-\nu_4 \nabla^{4}\phi(\mathbf{x},t)+\frac{\lambda_4}{2}\nabla^2 \left(\nabla\phi(\mathbf{x},t)\right)^2+\mathbf{\zeta}(\mathbf{x},t),
\end{equation}
where $\zeta(\mathbf{x},t)$ is delta-correlated Gaussian noise, as in Eq.\ \eqref{eq:noise}, and $\nu_4>0$ and $\lambda_4$ are parameters quantifying linear and nonlinear surface diffusion, respectively. The CKPZ universality class applies to a wide range of systems \cite{Barabasi,krug} for which interfacial dynamics are conserved (namely, the total amount of material at the interface is time-independent) and noise is not, as e.g.\ for ultrathin film deposition by Molecular Beam Epitaxy \cite{Pimpinelli98}. To simplify this discussion, we henceforth consider Eq.\ \eqref{eq:cKPZ_eq} for one-dimensional interfaces. Our argument can readily be generalized to higher-dimensional interfaces for which $d>1$. 

The scaling exponents of the CKPZ equation are analytically known from a two-loop DRG calculation to be \cite{Janssen97}
\begin{equation}
    \alpha_{\scriptscriptstyle{\rm CKPZ,1D}}=1-\delta, \; z_{\scriptscriptstyle{\rm CKPZ,1D}}=3-2\delta,
\label{eq:exp_ckpz_1d}
\end{equation}
where $\delta_{\scriptscriptstyle{\rm CKPZ,1D}} \simeq 0.03$ is a small correction to the one-loop result (note, alternative analytical approaches like the self-consistent expansion of Ref.\ \cite{PhysRevE.65.032103} suggest that the one-loop values might actually hold, in which case $\delta=0$; nevertheless, our argument does not require $\delta\neq 0$). The following scaling relation, termed hyperscaling, is believed to hold exactly \cite{Janssen97}, 
\begin{equation}
    2\alpha_{\scriptscriptstyle{\rm CKPZ}}+d = z_{\scriptscriptstyle{\rm CKPZ}},
\label{eq:hyperscaling}
\end{equation}
with $d=1$ for the exponents in Eq.\ \eqref{eq:exp_ckpz_1d}, due to the known fact that, for conserved interface equations like CKPZ, non-conserved noise does not renormalize \cite{Barabasi,Janssen97}. The (approximate) analytical exponent values of Eq.\ \eqref{eq:exp_ckpz_1d} compare moreover very well with numerical estimates \cite{Carrasco2016}. Now, consider the following {\em linear equation} for a different height field $h(x,t)$,
\begin{equation}
\partial_t \hat{h}(q,t) = -\tilde{\nu}_4 |q|^{z_{\scriptscriptstyle{\rm CKPZ,1D}}} \hat{h}(q,t) + \hat{\zeta}(q,t), 
\label{eq:fakeckpz}
\end{equation}
where hat denotes space Fourier transform, $q$ is 1D wave-vector, and $\tilde{\nu}_4>0$ may or not equal $\nu_4$ because the scaling behaviors of Eqs.\ \eqref{eq:cKPZ_eq} and \eqref{eq:fakeckpz} do not depend on the precise numerical values of these coefficients. Notably, one can solve Eq.\ \eqref{eq:fakeckpz} analytically \cite{krug,Vivo2012}. In particular, the structure factor reads
\begin{equation}
    S_{\rm (C4)}(\mathbf{q},t) \propto \frac{1-e^{-2\tilde{\nu}_4 t |q|^{z_{\scriptscriptstyle{\rm CKPZ,1D}}}}}{|q|^{z_{\scriptscriptstyle{\rm CKPZ,1D}}}},
    \label{eq:SqC4} 
\end{equation} 
wherefrom, comparing again with Eq.\ \eqref{ec:SFV}, $z_{\rm (C4)}=z_{\scriptscriptstyle{\rm CKPZ,1D}}$ and $\alpha_{\rm (C4)}=(z_{\scriptscriptstyle{\rm CKPZ,1D}}-1)/2$, so that hyperscaling holds again \cite{Vivo2012}, namely,
\begin{equation}
    2\alpha_{\rm (C4)}+d = z_{\rm (C4)},
\label{eq:hyperscaling_b}
\end{equation}
where $d=1$. Hence, $\alpha_{\rm (C4)}=\alpha_{\scriptscriptstyle{\rm CKPZ,1D}}$, so that the linear Eq.\ \eqref{eq:fakeckpz} shares the exact same set of exponent values as the nonlinear 1D CKPZ equation, Eq.\ \eqref{eq:cKPZ_eq}! For this reason, one can consider the former to be a Gaussian approximation of the latter \cite{Vivo2012,Rodriguez-Fernandez19,Rodriguez-Fernandez20}. Further analogous cases exist in which e.g.\ the dynamic exponent of a nonlinear model is exactly the same as in the corresponding linear theory \cite{PhysRevE.107.025002,Steinbock_2023}, as confirmed by simulations, yet the nonlinear model differs in other respects.

Does the previous discussion imply that the Gaussian approximation, Eq.\ \eqref{eq:fakeckpz}, is in the same universality class as the 1D CKPZ equation? Of course not. Here is where the statistics of height fluctuations becomes crucial. For the linear equation, the PDF of height fluctuations is exactly Gaussian \cite{Barabasi,krug}, in particular with zero skewness, while for the 1D CKPZ equation it is numerically known not to be Gaussian, in particular with a non-zero skewness \cite{Carrasco2016}.

\section{Dynamic Renormalization Group}\label{AppA}

Although the DRG was successfully applied to the study of the stochastic TDGL equation \cite{hohenberg}, in the main text the RG flow equations for this model are presented in the context of surface kinetic roughening, see Sec.\ \ref{sec:DRG}. The original RG flow equations are fully reproduced in this Appendix, from which the flow equations in the interfacial context, Eqs.\ \eqref{eq:RG_flow}, are finally obtained. The roughness and dynamic exponents $\alpha$ and $z$ are then obtained from the DRG flow, leading to the results collected in Table \ref{tab:exponents_KR}. As the full details of this derivation spread over a number of different references in the literature, here we collect them together for the reader's convenience. This calculation is also reviewed in general sources like, e.g., Refs.\ \cite{kardar,Ma2018,Mazenko,ojalvo,Tauber} and many other. However, here we intend to make a detailed, while compact, presentation that fills in some gaps in presentation and/or results. Readers with previous familiarity with the DRG in the context of kinetic roughening \cite{Medina89,Barabasi} may readily appreciate the similarities and the differences in the method for the present model, in particular with respect to the renormalization of the noise.

The original convention \cite{hohenberg} writes down the mobility $\Gamma$ explicitly in the TDGL equation which, by applying the Fluctuation-Dissipation theorem, is equivalent to the noise amplitude (setting $k_B T=1$), the bare model reading 
\begin{align}
& \partial_{t} \phi=-\Gamma \left\{ r \phi +u \phi^3-\nu \nabla^2 \phi \right\}+\zeta, \label{eq:tdgl_hohen} \\
& \langle \zeta(\textbf{r}, t) \zeta(\textbf{r}^{\prime}, t^{\prime})\rangle =2 \Gamma \delta\left(\textbf{r}-\textbf{r}^{\prime}\right) \delta\left(t-t^{\prime}\right). \label{eq:noise_hohen}
\end{align}
The DRG operates in wave vector and time frequency space. Recall that, through the coarse-graining of the lattice Ising model, an upper cutoff $\Lambda=1/a$ is introduced, where $a$ is the lattice spacing. Hence, after multiplication by $G_0(\textbf{q},\omega)$, the Fourier transform of Eq.\ \eqref{eq:tdgl_hohen} becomes
\begin{equation}
    \begin{split}
    & \hat{\phi}(\textbf{q},\omega)=  G_0(\textbf{q},\omega)\hat{\zeta}(\textbf{q},\omega) -uG_0(\textbf{q},\omega) \\ & \times \iint_{\substack{\textbf{q}_1,\omega_1\\  \textbf{q}_2,\omega_2}} \hat{\phi}(\textbf{q}_1,\omega_1)\hat{\phi}(\textbf{q}_2,\omega_2)\hat{\phi}(\textbf{q}-\textbf{q}_1-\textbf{q}_2,\omega-\omega_1-\omega_2),
    \end{split}
    \label{eq:drg_basic}
\end{equation}
where hats here denote space-time Fourier transform, the integral shorthand notation stands for
\begin{equation}
    \iint_{\substack{\textbf{q}_1,\omega_1\\  \textbf{q}_2,\omega_2}} = \iint_{-\infty}^{\infty}\frac{d\omega_1}{2\pi}\frac{d\omega_2}{2\pi} \iint_{0}^{\Lambda}\frac{d\textbf{q}_1}{(2\pi)^d}\frac{d\textbf{q}_2}{(2\pi)^d},
\end{equation}
and $G_0$ stands for the linear or free bare propagator,
\begin{equation}
    G_0 (\textbf{q},\omega)=\frac{1}{r+\nu q^2-i\omega/\Gamma}.
\end{equation}
For convenience, the equation is further shortened to
\begin{equation}\label{eq:fourier_shorthand}
\hat{\phi}(\textbf{q},\omega)=\hat{\phi}_0(\textbf{q},\omega)-u G_0 (\textbf{q},\omega)\mathcal{N}[\hat{\phi}(\textbf{q},\omega)],
\end{equation}
where $\mathcal{N}[\hat{\phi}(\textbf{q},\omega)]$ stands for the nonlinear term, and 
\begin{equation}
   \hat{\phi}_0(\textbf{q},\omega)=G_0 (\textbf{q},\omega)\hat{\zeta}(\textbf{q},\omega) 
   \label{eq:phi0}
\end{equation}
is the zeroth-order approximation (in powers of $u$) of the field.

\subsection{Coarse-graining}
The next step involves a coarse-graining by separating Fourier modes into slow ($\hat{\phi}^{<}(q)=\hat{\phi}(q)$ for $0<q<\Lambda/b$) and fast ($\hat{\phi}^{>}(q)=\hat{\phi}(q)$ for $\Lambda/b<q<\Lambda$) modes for the $\hat{\phi}$ and $\hat{\zeta}$ fields, where $b>0$ is an arbitrary constant. 
At this point, it is very convenient to introduce the diagrammatic notation, to simplify further steps:
\begin{itemize}
    \item A thick line stands for the field $\hat{\phi}$.
    \item A thin line stands for the zeroth order approximation $\hat{\phi}_0$. According to Eq.\ \eqref{eq:phi0}, this contribution is proportional to a noise term.
    \item A thin line with an arrow represents a free propagator $G_0 (\textbf{q},\omega)$.
    \item A vertex stands for the double integral and the parameter $(-u)$. Wave vector $\textbf{q}$ must be conserved at every vertex.
    \item Fast (slow) modes are represented by slashed (unslashed) lines.
\end{itemize}

Hence, the equations for the slow and fast modes can be represented diagrammatically as in Fig.\ \ref{fig:split_eq_diag}.
\begin{figure}[h!]
\scalebox{1.2}{\begin{tikzpicture}
\node [name = n1, scale=0.5] at (0,0) {};
\draw [ultra thick] ($(n1)$) -- ($(n1)+(0.5,0)$);
\node [scale=0.5] at ($(n1)+(0.7,0)$) {=};
\draw [thick] ($(n1)+(0.9,0)$) -- ($(n1)+(1.4,0)$);;
\node [scale=0.5] at ($(n1)+(1.6,0)$) {+};

\node [name = n2, scale=0.5] at (1.8,0) {};
\draw [thick, arrows=-stealth] ($(n2)$) -- ($(n2)+(0.5,0)$);
\draw [ultra thick] ($(n2)+(0.5,0)$) -- ($(n2)+(1,0)$);
\draw [ultra thick] ($(n2)+(0.5,0)$) -- ($(n2)+({0.5+0.5/sqrt(2)},{0.5/sqrt(2)})$);
\draw [ultra thick] ($(n2)+(0.5,0)$) -- ($(n2)+({0.5+0.5/sqrt(2)},{-0.5/sqrt(2)})$);;
\node [scale=0.5] at ($(n2)+(1.2,0)$) {+};

\node [scale=0.8] at (3.2,0) {3};
\node [name = n3, scale=0.5] at (3.3,0) {};
\draw [thick, arrows=-stealth] ($(n3)$) -- ($(n3)+(0.5,0)$);
\draw [ultra thick] ($(n3)+(0.5,0)$) -- ($(n3)+(1,0)$);
\node [scale=0.6] at ($(n3)+(0.8,0)$) {/};;
\draw [ultra thick] ($(n3)+(0.5,0)$) -- ($(n3)+({0.5+0.5/sqrt(2)},{0.5/sqrt(2)})$);
\draw [ultra thick] ($(n3)+(0.5,0)$) -- ($(n3)+({0.5+0.5/sqrt(2)},{-0.5/sqrt(2)})$);;
\node [scale=0.5] at ($(n3)+(1.2,0)$) {+};

\node [scale=0.8] at (4.7,0) {3};
\node [name = n4, scale=0.5] at (4.8,0) {};
\draw [thick, arrows=-stealth] ($(n4)$) -- ($(n4)+(0.5,0)$);
\draw [ultra thick] ($(n4)+(0.5,0)$) -- ($(n4)+(1,0)$);;
\draw [ultra thick] ($(n4)+(0.5,0)$) -- ($(n4)+({0.5+0.5/sqrt(2)},{0.5/sqrt(2)})$);
\node [scale=0.6] at ($(n4)+({0.5+0.3/sqrt(2)},{0.3/sqrt(2)})$) {$\backslash$};
\draw [ultra thick] ($(n4)+(0.5,0)$) -- ($(n4)+({0.5+0.5/sqrt(2)},{-0.5/sqrt(2)})$);
\node [scale=0.6] at ($(n4)+({0.5+0.3/sqrt(2)},{-0.3/sqrt(2)})$) {/};;
\node [scale=0.5] at ($(n4)+(1.2,0)$) {+};

\node [name = n5, scale=0.5] at (6.2,0) {};
\draw [thick, arrows=-stealth] ($(n5)$) -- ($(n5)+(0.5,0)$);
\draw [ultra thick] ($(n5)+(0.5,0)$) -- ($(n5)+(1,0)$);
\node [scale=0.6] at ($(n5)+(0.8,0)$) {/};;
\draw [ultra thick] ($(n5)+(0.5,0)$) -- ($(n5)+({0.5+0.5/sqrt(2)},{0.5/sqrt(2)})$);
\node [scale=0.6] at ($(n5)+({0.5+0.3/sqrt(2)},{0.3/sqrt(2)})$) {$\backslash$};
\draw [ultra thick] ($(n5)+(0.5,0)$) -- ($(n5)+({0.5+0.5/sqrt(2)},{-0.5/sqrt(2)})$);
\node [scale=0.6] at ($(n5)+({0.5+0.3/sqrt(2)},{-0.3/sqrt(2)})$) {/};;

\node [name = n1, scale=0.5] at (0,-1) {};
\draw [ultra thick] ($(n1)$) -- ($(n1)+(0.5,0)$);
\node [scale=0.6] at ($(n1)+(0.25,0)$) {/};;
\node [scale=0.5] at ($(n1)+(0.7,0)$) {=};
\draw [thick] ($(n1)+(0.9,0)$) -- ($(n1)+(1.4,0)$);;
\node [scale=0.6] at ($(n1)+(1.15,0)$) {/};;
\node [scale=0.5] at ($(n1)+(1.6,0)$) {+};

\node [name = n2, scale=0.5] at (1.8,-1) {};
\draw [thick, arrows=-stealth] ($(n2)$) -- ($(n2)+(0.5,0)$);
\node [scale=0.6] at ($(n2)+(0.2,0)$) {/};;
\draw [ultra thick] ($(n2)+(0.5,0)$) -- ($(n2)+(1,0)$);
\draw [ultra thick] ($(n2)+(0.5,0)$) -- ($(n2)+({0.5+0.5/sqrt(2)},{0.5/sqrt(2)})$);
\draw [ultra thick] ($(n2)+(0.5,0)$) -- ($(n2)+({0.5+0.5/sqrt(2)},{-0.5/sqrt(2)})$);;
\node [scale=0.5] at ($(n2)+(1.2,0)$) {+};

\node [scale=0.8] at (3.2,-1) {3};
\node [name = n3, scale=0.5] at (3.3,-1) {};
\draw [thick, arrows=-stealth] ($(n3)$) -- ($(n3)+(0.5,0)$);
\node [scale=0.6] at ($(n3)+(0.2,0)$) {/};;
\draw [ultra thick] ($(n3)+(0.5,0)$) -- ($(n3)+(1,0)$);
\node [scale=0.6] at ($(n3)+(0.8,0)$) {/};;
\draw [ultra thick] ($(n3)+(0.5,0)$) -- ($(n3)+({0.5+0.5/sqrt(2)},{0.5/sqrt(2)})$);
\draw [ultra thick] ($(n3)+(0.5,0)$) -- ($(n3)+({0.5+0.5/sqrt(2)},{-0.5/sqrt(2)})$);;
\node [scale=0.5] at ($(n3)+(1.2,0)$) {+};

\node [scale=0.8] at (4.7,-1) {3};
\node [name = n4, scale=0.5] at (4.8,-1) {};
\draw [thick, arrows=-stealth] ($(n4)$) -- ($(n4)+(0.5,0)$);
\node [scale=0.6] at ($(n4)+(0.2,0)$) {/};;
\draw [ultra thick] ($(n4)+(0.5,0)$) -- ($(n4)+(1,0)$);;
\draw [ultra thick] ($(n4)+(0.5,0)$) -- ($(n4)+({0.5+0.5/sqrt(2)},{0.5/sqrt(2)})$);
\node [scale=0.6] at ($(n4)+({0.5+0.3/sqrt(2)},{0.3/sqrt(2)})$) {$\backslash$};
\draw [ultra thick] ($(n4)+(0.5,0)$) -- ($(n4)+({0.5+0.5/sqrt(2)},{-0.5/sqrt(2)})$);
\node [scale=0.6] at ($(n4)+({0.5+0.3/sqrt(2)},{-0.3/sqrt(2)})$) {/};;
\node [scale=0.5] at ($(n4)+(1.2,0)$) {+};

\node [name = n5, scale=0.5] at (6.2,-1) {};
\draw [thick, arrows=-stealth] ($(n5)$) -- ($(n5)+(0.5,0)$);
\node [scale=0.6] at ($(n5)+(0.2,0)$) {/};;
\draw [ultra thick] ($(n5)+(0.5,0)$) -- ($(n5)+(1,0)$);
\node [scale=0.6] at ($(n5)+(0.8,0)$) {/};;
\draw [ultra thick] ($(n5)+(0.5,0)$) -- ($(n5)+({0.5+0.5/sqrt(2)},{0.5/sqrt(2)})$);
\node [scale=0.6] at ($(n5)+({0.5+0.3/sqrt(2)},{0.3/sqrt(2)})$) {$\backslash$};
\draw [ultra thick] ($(n5)+(0.5,0)$) -- ($(n5)+({0.5+0.5/sqrt(2)},{-0.5/sqrt(2)})$);
\node [scale=0.6] at ($(n5)+({0.5+0.3/sqrt(2)},{-0.3/sqrt(2)})$) {/};;

\end{tikzpicture}}
\vspace{-0.3cm}
\caption{Top: Diagrammatic representation of Eq.\ \eqref{eq:drg_basic} for the slow modes. Bottom: Same as top, but for the fast modes.}
\label{fig:split_eq_diag}
\end{figure}
Next, fast modes are expanded in a perturbation series in powers of the nonlinear coupling parameter $u$ as 
\begin{equation}
    \hat{\phi}^{>} = \hat{\phi}_0^{>}+u\hat{\phi}_1^{>}+\mathcal{O}(u^2),
\end{equation}
which can be readily implemented by iteration of its diagrammatic expression shown in Fig.\ \ref{fig:split_eq_diag}. At first order in $u$, the result takes the form depicted in Fig.\ \ref{fig:fast_modes_diag}.
\begin{figure}[h!]
\scalebox{1.2}{\begin{tikzpicture}
\node [name = n1, scale=0.5] at (0,-1) {};
\draw [ultra thick] ($(n1)$) -- ($(n1)+(0.5,0)$);
\node [scale=0.6] at ($(n1)+(0.25,0)$) {/};;
\node [scale=0.5] at ($(n1)+(0.7,0)$) {$\approx$};
\draw [thick] ($(n1)+(0.9,0)$) -- ($(n1)+(1.4,0)$);;
\node [scale=0.6] at ($(n1)+(1.15,0)$) {/};;
\node [scale=0.5] at ($(n1)+(1.6,0)$) {+};

\node [name = n2, scale=0.5] at (1.8,-1) {};
\draw [thick, arrows=-stealth] ($(n2)$) -- ($(n2)+(0.5,0)$);
\node [scale=0.6] at ($(n2)+(0.2,0)$) {/};;
\draw [ultra thick] ($(n2)+(0.5,0)$) -- ($(n2)+(1,0)$);
\draw [ultra thick] ($(n2)+(0.5,0)$) -- ($(n2)+({0.5+0.5/sqrt(2)},{0.5/sqrt(2)})$);
\draw [ultra thick] ($(n2)+(0.5,0)$) -- ($(n2)+({0.5+0.5/sqrt(2)},{-0.5/sqrt(2)})$);;
\node [scale=0.5] at ($(n2)+(1.2,0)$) {+};

\node [scale=0.8] at (3.2,-1) {3};
\node [name = n3, scale=0.5] at (3.3,-1) {};
\draw [thick, arrows=-stealth] ($(n3)$) -- ($(n3)+(0.5,0)$);
\node [scale=0.6] at ($(n3)+(0.2,0)$) {/};;
\draw [thick] ($(n3)+(0.5,0)$) -- ($(n3)+(1,0)$);
\node [scale=0.6] at ($(n3)+(0.8,0)$) {/};;
\draw [ultra thick] ($(n3)+(0.5,0)$) -- ($(n3)+({0.5+0.5/sqrt(2)},{0.5/sqrt(2)})$);
\draw [ultra thick] ($(n3)+(0.5,0)$) -- ($(n3)+({0.5+0.5/sqrt(2)},{-0.5/sqrt(2)})$);;
\node [scale=0.5] at ($(n3)+(1.2,0)$) {+};

\node [scale=0.8] at (4.7,-1) {3};
\node [name = n4, scale=0.5] at (4.8,-1) {};
\draw [thick, arrows=-stealth] ($(n4)$) -- ($(n4)+(0.5,0)$);
\node [scale=0.6] at ($(n4)+(0.2,0)$) {/};;
\draw [ultra thick] ($(n4)+(0.5,0)$) -- ($(n4)+(1,0)$);;
\draw [thick] ($(n4)+(0.5,0)$) -- ($(n4)+({0.5+0.5/sqrt(2)},{0.5/sqrt(2)})$);
\node [scale=0.6] at ($(n4)+({0.5+0.3/sqrt(2)},{0.3/sqrt(2)})$) {$\backslash$};
\draw [thick] ($(n4)+(0.5,0)$) -- ($(n4)+({0.5+0.5/sqrt(2)},{-0.5/sqrt(2)})$);
\node [scale=0.6] at ($(n4)+({0.5+0.3/sqrt(2)},{-0.3/sqrt(2)})$) {/};;
\node [scale=0.5] at ($(n4)+(1.2,0)$) {+};

\node [name = n5, scale=0.5] at (6.2,-1) {};
\draw [thick, arrows=-stealth] ($(n5)$) -- ($(n5)+(0.5,0)$);
\node [scale=0.6] at ($(n5)+(0.2,0)$) {/};;
\draw [thick] ($(n5)+(0.5,0)$) -- ($(n5)+(1,0)$);
\node [scale=0.6] at ($(n5)+(0.8,0)$) {/};;
\draw [thick] ($(n5)+(0.5,0)$) -- ($(n5)+({0.5+0.5/sqrt(2)},{0.5/sqrt(2)})$);
\node [scale=0.6] at ($(n5)+({0.5+0.3/sqrt(2)},{0.3/sqrt(2)})$) {$\backslash$};
\draw [thick] ($(n5)+(0.5,0)$) -- ($(n5)+({0.5+0.5/sqrt(2)},{-0.5/sqrt(2)})$);
\node [scale=0.6] at ($(n5)+({0.5+0.3/sqrt(2)},{-0.3/sqrt(2)})$) {/};;

\end{tikzpicture}}
\vspace{-0.3cm}
\caption{Diagrammatic expansion of the equation for the fast modes, to first order in the perturbative expansion.}
\label{fig:fast_modes_diag}
\end{figure}

The equation for the fast modes represented in Fig.\ \ref{fig:fast_modes_diag} can then be substituted into the equation for the slow modes, since now the only unknown is $\hat{\phi}^{<}$. This procedure yields a large number of diagrams, with those contributing non-negligibly to the slow modes equation being sketched in Fig.\ \ref{fig:slow_modes_diag}. Additional diagrams are neglected because they are of higher orders in $u$ and for other reasons that will be justified later. For the full diagrammatic expansion at one-loop order, see Ref.\ \cite{Mazenko}. 
\begin{figure}[h!]
\scalebox{1.5}{\begin{tikzpicture}
\node [name = n1, scale=0.5] at (0,0) {};
\draw [ultra thick] ($(n1)$) -- ($(n1)+(0.5,0)$);
\node [scale=0.5] at ($(n1)+(0.7,0)$) {=};
\draw [thick] ($(n1)+(0.9,0)$) -- ($(n1)+(1.4,0)$);;
\node [scale=0.5] at ($(n1)+(1.6,0)$) {+};

\node [name = n2, scale=0.5] at (1.8,0) {};
\draw [thick, arrows=-stealth] ($(n2)$) -- ($(n2)+(0.5,0)$);
\draw [ultra thick] ($(n2)+(0.5,0)$) -- ($(n2)+(1,0)$);
\draw [ultra thick] ($(n2)+(0.5,0)$) -- ($(n2)+({0.5+0.5/sqrt(2)},{0.5/sqrt(2)})$);
\draw [ultra thick] ($(n2)+(0.5,0)$) -- ($(n2)+({0.5+0.5/sqrt(2)},{-0.5/sqrt(2)})$);;
\node [scale=0.5] at ($(n2)+(1.2,0)$) {+};

\node [scale=0.8] at (3.2,0) {3};
\node [name = n3, scale=0.5] at (3.3,0) {};
\draw [thick, arrows=-stealth] ($(n3)$) -- ($(n3)+(0.5,0)$);
\draw [ultra thick] ($(n3)+(0.5,0)$) -- ($(n3)+(1,0)$);;
\draw [thick] ($(n3)+(0.5,0)$) -- ($(n3)+({0.5+0.5/sqrt(2)},{0.5/sqrt(2)})$);
\node [scale=0.6] at ($(n3)+({0.5+0.3/sqrt(2)},{0.3/sqrt(2)})$) {$\backslash$};
\draw [thick] ($(n3)+(0.5,0)$) -- ($(n3)+({0.5+0.5/sqrt(2)},{-0.5/sqrt(2)})$);
\node [scale=0.6] at ($(n3)+({0.5+0.3/sqrt(2)},{-0.3/sqrt(2)})$) {/};;
\node [scale=0.5] at ($(n3)+(1.2,0)$) {+};

\node [scale=0.8] at (0,-1) {18};
\node [name = n4, scale=0.5] at (0.2,-1) {};
\draw [thick, arrows=-stealth] ($(n4)$) -- ($(n4)+(0.5,0)$);
\draw [ultra thick] ($(n4)+(0.5,0)$) -- ($(n4)+(1,0)$);;
\draw [thick] ($(n4)+(0.5,0)$) -- ($(n4)+({0.5+0.5/sqrt(2)},{0.5/sqrt(2)})$);
\node [scale=0.6] at ($(n4)+({0.5+0.3/sqrt(2)},{0.3/sqrt(2)})$) {$\backslash$};
\draw [thick, arrows=-stealth] ($(n4)+(0.5,0)$) -- ($(n4)+({0.5+0.7/sqrt(2)},{-0.7/sqrt(2)})$);
\node [scale=0.6] at ($(n4)+({0.5+0.3/sqrt(2)},{-0.3/sqrt(2)})$) {/};;
\node [name = n4b, scale=0.5] at ($(n4)+({0.5+0.7/sqrt(2)},{-0.7/sqrt(2)})$) {};
\draw [ultra thick] ($(n4b)$) -- ($(n4b)+(0.5,0)$);;
\draw [ultra thick] ($(n4b)$) -- ($(n4b)+(0,{-0.5})$);
\draw [thick] ($(n4b)$) -- ($(n4b)+({0.5/sqrt(2)},{-0.5/sqrt(2)})$);
\node [scale=0.6] at ($(n4b)+({0.3/sqrt(2)},{-0.3/sqrt(2)})$) {/};;

\node [scale=0.5] at ($(n4)+(1.7,0)$) {+};

\node [scale=0.8] at (2.1,-1) {9};
\node [name = n5, scale=0.5] at (2.2,-1) {};
\draw [thick, arrows=-stealth] ($(n5)$) -- ($(n5)+(0.5,0)$);
\draw [thick] ($(n5)+(0.5,0)$) -- ($(n5)+({0.5+0.5/sqrt(2)},{0.5/sqrt(2)})$);
\node [scale=0.6] at ($(n5)+({0.5+0.3/sqrt(2)},{0.3/sqrt(2)})$) {$\backslash$};
\draw [thick] ($(n5)+(0.5,0)$) -- ($(n5)+({0.5+0.5/sqrt(2)},{-0.5/sqrt(2)})$);
\node [scale=0.6] at ($(n5)+({0.5+0.3/sqrt(2)},{-0.3/sqrt(2)})$) {/};;

\node [name = n5b, scale=0.5] at ($(n5)+(0.5,0)$) {};;
\draw [thick, arrows=-stealth] ($(n5b)$) -- ($(n5b)+(0.8,0)$);
\node [scale=0.6] at ($(n5b)+(0.4,0)$) {/};;
\node [name = n5c, scale=0.5] at ($(n5b)+(0.8,0)$) {};
\draw [ultra thick] ($(n5c)$) -- ($(n5c)+(0.5,0)$);;
\draw [thick] ($(n5c)$) -- ($(n5c)+({0.5/sqrt(2)},{0.5/sqrt(2)})$);
\node [scale=0.6] at ($(n5c)+({0.3/sqrt(2)},{0.3/sqrt(2)})$) {$\backslash$};
\draw [thick] ($(n5c)$) -- ($(n5c)+({0.5/sqrt(2)},{-0.5/sqrt(2)})$);
\node [scale=0.6] at ($(n5c)+({0.3/sqrt(2)},{-0.3/sqrt(2)})$) {/};;

\node [scale=0.5] at ($(n5c)+(0.9,0)$) {+ ...};

\end{tikzpicture}}
\vspace{-0.3cm}
\caption{Relevant contributions to the diagrammatic expansion of the equation for the slow modes.}
\label{fig:slow_modes_diag}
\end{figure}

\begin{figure}[h!]
\scalebox{1.5}{ 
  \begin{tikzpicture}
\node [name = n1, scale=0.5] at (0,0) {};
\draw [ultra thick] ($(n1)$) -- ($(n1)+(0.5,0)$);
\node [scale=0.5] at ($(n1)+(0.7,0)$) {=};
\draw [thick] ($(n1)+(0.9,0)$) -- ($(n1)+(1.4,0)$);;
\node [scale=0.5] at ($(n1)+(1.6,0)$) {+};

\node [name = n2, scale=0.5] at (1.8,0) {};
\draw [thick, arrows=-stealth] ($(n2)$) -- ($(n2)+(0.5,0)$);
\draw [ultra thick] ($(n2)+(0.5,0)$) -- ($(n2)+(1,0)$);
\draw [ultra thick] ($(n2)+(0.5,0)$) -- ($(n2)+({0.5+0.5/sqrt(2)},{0.5/sqrt(2)})$);
\draw [ultra thick] ($(n2)+(0.5,0)$) -- ($(n2)+({0.5+0.5/sqrt(2)},{-0.5/sqrt(2)})$);;
\node [scale=0.5] at ($(n2)+(1.2,0)$) {+};

\node [scale=0.8] at (3.2,0) {3};
\node [name = n3, scale=0.5] at (3.3,0) {};
\draw [thick, arrows=-stealth] ($(n3)$) -- ($(n3)+(0.5,0)$);
\node [name = n3b, scale=0.1] at ($(n3)+(0.5,0)$) {};
\node [scale=0.8] at ($(n3b)-(0,0.25)$) {$\Sigma_1$};
\node [fill = white, shape=circle, draw=black, scale = 0.3, name = ver3] at ($(n3b)+(0,0.6)$) {};
\draw (ver3.north west)--(ver3.south east);
\draw (ver3.north east)--(ver3.south west);

\draw [thick, arrows=-stealth] (n3b) to [out = 145, in = 220] (ver3);
\node [scale=0.6] at ($(n3b)+({-0.22/sqrt(2)},{0.3/sqrt(2)})$) {\_};;
\draw [thick, arrows=-stealth] (n3b) to [out = 35, in = 320] (ver3);
\node [scale=0.6] at ($(n3b)+({0.22/sqrt(2)},{0.3/sqrt(2)})$) {\_};;

\draw [ultra thick] ($(n3)+(0.5,0)$) -- ($(n3)+(1,0)$);;
\node [scale=0.5] at ($(n3)+(1.2,0)$) {+};

\node [scale=0.8] at (-0.3,-1.4) {18};
\node [name = n4, scale=0.5] at (-0.1,-1.4) {};
\draw [thick, arrows=-stealth] ($(n4)$) -- ($(n4)+(0.5,0)$);
\node [name = n4b, scale=0.1] at ($(n4)+(0.5,0)$) {};;
\draw [ultra thick] ($(n4b)$) -- ($(n4b)+({0.5/sqrt(2)},{0.5/sqrt(2)})$);

\draw [thick, arrows=-stealth] ($(n4b)$) -- ($(n4b)+(1.2,0)$);
\node [scale=0.6] at ($(n4b)+(0.6,0)$) {/};;
\node [fill = white, shape=circle, draw=black, scale = 0.3, name = ver4] at ($(n4b) + (0.6,-0.6)$) {};
\node [scale=0.8] at ($(ver4)-(0,0.3)$) {$\Sigma_2$};
\draw (ver4.north west)--(ver4.south east);
\draw (ver4.north east)--(ver4.south west);

\draw [thick, arrows=-stealth] (n4b) to [out = 270, in = 180] (ver4);
\node [scale=0.6] at ($(n4b)+({0.25/sqrt(2)},{-0.65/sqrt(2)})$) {/};;
\node [name = n4c, scale=0.1] at ($(n4b)+(1.2,0)$) {};
\draw [thick, arrows=-stealth] (n4c) to [out = 270, in = 0] (ver4);
\node [scale=0.6] at ($(n4c)+({-0.25/sqrt(2)},{-0.65/sqrt(2)})$) {$\backslash$};;

\draw [ultra thick] ($(n4c)$) -- ($(n4c)+({0.5/sqrt(2)},{0.5/sqrt(2)})$);
\draw [ultra thick] ($(n4c)$) -- ($(n4c)+({0.5/sqrt(2)},{-0.5/sqrt(2)})$);
\node [scale=0.5] at ($(n4c)+(0.6,0)$) {+};

\node [scale=0.8] at ($(n4c)+(0.8,0)$) {18};
\node [name = n5, scale=0.5] at ($(n4c)+(1,0)$) {};
\draw [thick, arrows=-stealth] ($(n5)$) -- ($(n5)+(0.5,0)$);
\node [name = n5b, scale=0.5] at ($(n5)+(0.5,0)$) {};;
\draw [thick, arrows=-stealth] ($(n5b)$) -- ($(n5b)+(1.2,0)$);
\node [scale=0.6] at ($(n5b)+(0.6,0)$) {/};;
\node [name = n5c, scale=0.5] at ($(n5b)+(1.2,0)$) {};
\node [fill = white, shape=circle, draw=black, scale = 0.3, name = ver5a] at ($(n5b) + (0.6,-0.6)$) {};
\node [scale=0.8] at ($(ver5a)-(0,0.3)$) {$\Sigma_3$};
\draw (ver5a.north west)--(ver5a.south east);
\draw (ver5a.north east)--(ver5a.south west);
\node [fill = white, shape=circle, draw=black, scale = 0.3, name = ver5b] at ($(n5b) + (0.6,0.6)$) {};
\draw (ver5b.north west)--(ver5b.south east);
\draw (ver5b.north east)--(ver5b.south west);

\draw [thick, arrows=-stealth] (n5b) to [out = 270, in = 180] (ver5a);
\node [scale=0.6] at ($(n5b)+({0.25/sqrt(2)},{-0.65/sqrt(2)})$) {/};;
\draw [thick, arrows=-stealth] (n5c) to [out = 270, in = 0] (ver5a);
\node [scale=0.6] at ($(n5c)+({-0.25/sqrt(2)},{-0.65/sqrt(2)})$) {$\backslash$};;

\draw [thick, arrows=-stealth] (n5b) to [out = 90, in = 180] (ver5b);
\node [scale=0.6] at ($(n5b)+({0.25/sqrt(2)},{0.65/sqrt(2)})$) {$\backslash$};;
\draw [thick, arrows=-stealth] (n5c) to [out = 90, in = 0] (ver5b);
\node [scale=0.6] at ($(n5c)+({-0.25/sqrt(2)},{0.65/sqrt(2)})$) {/};;

\draw [ultra thick] ($(n5c)$) -- ($(n5c)+(0.5,0)$);;

\end{tikzpicture}}
\vspace{-0.3cm}
\caption{Coarse-grained equation for slow modes, Eq.\ \eqref{eq:drg_basic}, up to order $u^2$.}
\label{fig:coarse_grain_diag}
\end{figure}
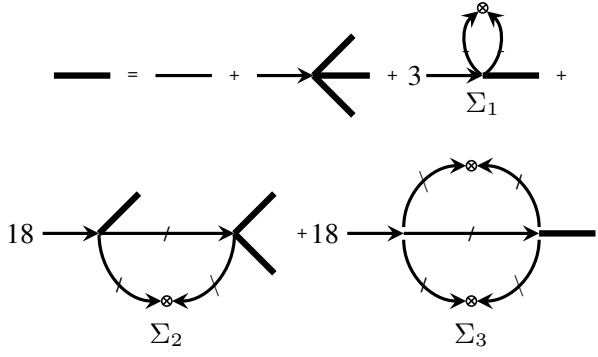

The last step to obtain the coarse-grained equation is to average out fast modes over noise realizations and integrate short length-scales, denoted as $\int^{>}$. 
The key tool for this is \textit{Wick's Theorem}, which states \cite{kardar,Tauber} that, for a Gaussian-distributed variable $\hat{\phi}$,
\begin{equation*}
    \langle\hat{\phi}_1 ... \hat{\phi}_n \rangle =
    \begin{cases}
        0,\quad 
                n\ \text{odd}, \\
        \text{sum of all pairwise contractions},\quad 
                 n\ \text{even}.
    \end{cases}
\end{equation*}
Accordingly, 
diagrams which are odd in powers of $\hat{\phi}_0=G_0\hat{\zeta}$ are neglected, since the noise is Gaussian. Furthermore, by denoting pairwise contractions over the noise as a crossed vertex ($\boldsymbol{\otimes}$), the final coarse-grained equation takes the form represented in Fig.\ \ref{fig:coarse_grain_diag}.
Analytically, this equation reads
\begin{equation}\label{eq:coarse_grained_app}
    \hat{\phi}=\hat{\phi}_0-u G_0\mathcal{N}[\hat{\phi}]+\Sigma_1 +\Sigma_2 +\Sigma_3 .
\end{equation}
To compute the diagram contributions $\{\Sigma_i\}_{i=1}^3$,  assumptions are made:
\begin{enumerate}
    \item Computations are for $d\simeq4$, to first perturbative order in a $\varepsilon$-expansion around the upper critical dimension ($d_c=4$) of the Ising universality class, with $\varepsilon=4-d$; hence, we assume $r \sim u \sim \varepsilon$. 
    \item Wave vectors verify $q^{<} \ll q^{>}$.
    \item For fast modes $\Lambda/b<q<\Lambda$, where $b=e^l$, $\Lambda$ is arbitrary, and $l\rightarrow 0$, thus \cite{kardar}
    \begin{equation}\label{eq:integral_approx}
    \begin{split}
    \int_{\Lambda/b}^{\Lambda}I(q)\frac{d\textbf{q}}{(2\pi)^d}&\approx I(\Lambda)(1-e^{-l})\Lambda \int_q \frac{d\Omega}{(2\pi)^d} \\ &\approx I(\Lambda=1)lK_d,
    \end{split}
    \end{equation}
    where $\Lambda=1$ without loss of generality, and $K_d$ is a geometric factor, 
    \begin{equation}
        K_d = \int_q \frac{d\Omega}{(2\pi)^d}=\frac{2^{1-d}}{\pi^{d/2}}\Gamma(d/2).
    \end{equation}
\end{enumerate}

To shorten the calculations, it is also worth computing the pairwise contraction when averaging over fast modes, corresponding to loops. 
By applying the noise correlation properties,
\begin{equation}
\begin{split}
    &\iint_{\substack{\textbf{q}_1,\omega_1\\  \textbf{q}_2,\omega_2}} \langle\hat{\phi}_0^{>}(\textbf{q}_1,\omega_1)\hat{\phi}_0^{>}(\textbf{q}-\textbf{q}_1-\textbf{q}_2,\omega-\omega_1-\omega_2) \rangle = \\
    &\iint_{\substack{\textbf{q}_1,\omega_1\\  \textbf{q}_2,\omega_2}} G_0^{>}G_0^{>\prime}\langle\hat{\zeta}^{>}(\textbf{q}_1,\omega_1)\hat{\zeta}^{>}(\textbf{q}-\textbf{q}_1-\textbf{q}_2,\omega-\omega_1-\omega_2) \rangle =\\
    &2\Gamma \int_{\textbf{q}_1,\omega_1} G_0^{>}(\textbf{q}_1,\omega_1)G_0^{>}(-\textbf{q}_1,-\omega_1).
\end{split}
\end{equation}
Using this result and with the various approximations in view, the first diagram, $\Sigma_1$, yields
\begin{equation}
\begin{split}
    \Sigma_1 &= 3(-u) G_0^{<}\hat{\phi}^{<}2\Gamma \int_{\textbf{q}_1,\omega_1}^{>} G_0^{>}(\textbf{q}_1,\omega_1)G_0^{>}(-\textbf{q}_1,-\omega_1) \\ &=-3u G_0^{<}\hat{\phi}^{<} \int_{\textbf{q}_1,\omega_1}^{>}\frac{2\Gamma}{(r+\nu q_1^2)^2+\omega_1^2/\Gamma^2}.
\end{split}
\end{equation}
The frequency integral 
has simple pole $\omega^{*}_{\pm}=\pm i(r+\nu q_1^2)\Gamma$. Using also Eq.\ \eqref{eq:integral_approx}, one finally obtains
\begin{equation}
    \Sigma_1 = -3u G_0^{<}\hat{\phi}^{<} \int_{\textbf{q}_1}^{>}\frac{1}{r+\nu q_1^2}\approx -\frac{3u}{\nu}\frac{K_4 l}{1+r/\nu}G_0^{<}\hat{\phi}^{<},
    \label{eq:S1}
\end{equation}
which renormalizes the linear non-derivative term in the TDGL equation.

Likewise the second diagram, $\Sigma_2$, which renormalizes the nonlinearity in the TDGL equation, corresponds to
\begin{equation}
    \begin{split}
        \Sigma_2 = &18(-u)^2 G_0^{<}\mathcal{N}[\hat{\phi}^{<}]2\Gamma \int_{\textbf{q}_3,\omega_3}^{>} G_0^{>}(\textbf{q}_3,\omega_3)\times \\ &G_0^{>}(-\textbf{q}_3,-\omega_3)G_0^{>}(\textbf{q}-\textbf{q}_1-\textbf{q}_3,\omega-\omega_1-\omega_3)\\
        =& 18u^2 G_0^{<}\mathcal{N}[\hat{\phi}^{<}]\times \\ &\int_{\textbf{q}_3,\omega_3}^{>}  \frac{2\Gamma}{(r+\nu q_3^2)^2+\omega_3^2/\Gamma^2}\times \\ &\frac{1}{r+\nu (\textbf{q}-\textbf{q}_1-\textbf{q}_3)^2-i(\omega-\omega_1-\omega_3)/\Gamma}.
    \end{split}
\end{equation}
Proceeding as for $\Sigma_1$ and using that the fast mode $\textbf{q}_3\gg \textbf{q}-\textbf{q}_1$, $q_3^2\gg q_3$, 
\begin{equation}
\begin{split}
    \Sigma_2 &= 18u^2 G_0^{<}\mathcal{N}[\hat{\phi}^{<}]\int_{\textbf{q}_3}^{>} \frac{1+\mathcal{O}(\textbf{q}-\textbf{q}_1)}{2(r+\nu q_3^2)^2+\mathcal{O}(q_3)}\\ &\approx 9\frac{u^2}{\nu^2}\frac{K_4l}{(1+r/\nu)^2}G_0^{<}\mathcal{N}[\hat{\phi}^{<}]\approx 9\frac{u^2}{\nu^2}K_4l\;G_0^{<}\mathcal{N}[\hat{\phi}^{<}]. 
\end{split}
\label{eq:S2}
\end{equation}
Finally, the third diagram, $\Sigma_3$, reads
\begin{equation}
    \begin{split}
        \Sigma_3 =& 18u^2 G_0^{<}\hat{\phi}^{<} \int_{\textbf{q}_1,\omega_1}^{>} 2\Gamma G_0^{>}(\textbf{q}_1,\omega_1)G_0^{>}(-\textbf{q}_1,-\omega_1)\times \\ &\int_{\textbf{q}_2,\omega_2}^{>}2\Gamma G_0^{>}(\textbf{q}_2,\omega_2)G_0^{>}(-\textbf{q}_2,-\omega_2)\times \\
        &G_0^{>}(\textbf{q}-\textbf{q}_1-\textbf{q}_2,\omega-\omega_1-\omega_2) \\
        =&18u^2 G_0^{<}\hat{\phi}^{<}\int_{\textbf{q}_1,\omega_1}^{>}\int_{\textbf{q}_2,\omega_2}^{>} \frac{2\Gamma}{(r+\nu q_1^2)^2+\omega_1^2/\Gamma^2} \times \\ &\frac{2\Gamma}{(r+\nu q_2^2)^2+\omega_2^2/\Gamma^2}\times\\
        &\frac{1}{r+\nu (\textbf{q}-\textbf{q}_1-\textbf{q}_2)^2-i(\omega-\omega_1-\omega_2)/\Gamma}.
    \end{split}
\end{equation}
The solution to this integral can be found in Ref.\ \cite{HohenHalpMa},
\begin{equation}
    \Sigma_3=u_0^2 \; l\left\{\frac{i \omega}{\Gamma} \frac{6(n+2)}{8\pi^4} \ln (4 / 3) - q^2 \frac{(n+2)}{8\pi^4}\right\} G_0^{<} \hat{\phi}^{<},
\end{equation}
where $n=1$ is the rank of the order parameter and $u_0 = u/(4\nu^2)$. Thus,
\begin{equation}
    \Sigma_3=\frac{u^2}{\nu^4} \; K_4^2 l\left\{\frac{i \omega}{\Gamma} 9 \ln (4 / 3) -\nu q^2 \frac{3}{2}\right\} G_0^{<} \hat{\phi}^{<}, \label{eq:S3}
\end{equation}
which can be verified comparing with Ref.\ \cite{Mazenko} (for the term proportional to $\omega$), and Ref.\ \cite{Ma2018} (for the term proportional to $q^2$), hence this diagram renormalizes the mobility and the linear diffusion term in the TDGL equation.

Substituting Eqs.\ \eqref{eq:S1}, \eqref{eq:S2}, and \eqref{eq:S3} into Eq.\ \eqref{eq:coarse_grained_app} after dividing both sides by $G_0$, leads to
\begin{equation}\label{eq:fourier_app}
\begin{split}
    &\left(r+\nu q^2-\frac{i\omega}{\Gamma}\right)\hat{\phi}^{<}=\hat{\zeta}^{<}-u\mathcal{N}[\hat{\phi}^{<}]-\frac{3 u}{\nu} \; \frac{K_4 l}{1+r/\nu} \hat{\phi}^{<}+\\
    &9 \frac{u^2}{\nu^2} K_4 l \; \mathcal{N}\left[\hat{\phi}^{<}\right]+\frac{u^2}{\nu^4} \; K_4^2 l\left\{\frac{i \omega}{\Gamma} 9 \ln (4 / 3) -\nu q^2 \frac{3}{2}\right\}  \hat{\phi}^{<},
\end{split}
\end{equation}
which can be finally rewritten as a coarse-grained equation as
\begin{equation}
    \left(\tilde{r}+\tilde{\nu} q^2-\frac{i\omega}{\tilde{\Gamma}}\right)\hat{\phi}^{<}(\textbf{q},\omega)=\hat{\zeta}^{<}(\textbf{q},\omega)-\tilde{u}\mathcal{N}[\hat{\phi}^{<}(\textbf{q},\omega)],
\end{equation}
provided the coarse-grained parameters are defined as
\begin{equation}
    \begin{split}
    & \tilde{r}=r+3 \frac{u}{\nu} \; \frac{K_4 l}{1+r/\nu},\\
    & \tilde{u}=u-9 \frac{u^2}{\nu^2} K_4 l,\\
    & \tilde{\Gamma}^{-1}=\Gamma^{-1}\left\{1+9 \ln (4 / 3) \; \frac{u^2}{\nu^4} K_4^2 l\right\} ,\\
    & \tilde{\nu}=\nu\left\{1+\frac{3}{2} \; \frac{u^2}{\nu^4} \; K_4^2 l\right\} .
\end{split}
\end{equation}
Notice that, in contrast with e.g.\ the KPZ equation for which the DRG renormalization of the noise requires an additional perturbative expansion of the two-point function \cite{Medina89,Barabasi}, this is not needed in our present case \cite{Mazenko}. 

\subsection{Rescaling}
To recover the same wave vector cut-off as in the bare system, we need to rescale wavevector as $\textbf{q}^{\prime}=b \textbf{q}$, which implies rescaling for other variables and fields as in Eq.\ \eqref{eq:rescaling} \cite{Mazenko}
\begin{equation}
        \textbf{r}^{\prime}=\textbf{r}/b, \quad t^{\prime}=b^{-z} t, \quad \phi^{\prime}=b^{-\alpha}\phi\left(b \textbf{r}^{\prime}, b^z t^{\prime}\right),
\end{equation}
Then, 
\begin{equation}
    \begin{split}
        &\partial_{t^{\prime}} \phi^{\prime}=-\tilde{\Gamma} b^z\left(\tilde{r} \phi^{\prime}+\tilde{u} b^{2 \alpha} \phi'^3-\tilde{\nu} b^{-2} \nabla^2 \phi^{\prime}\right)+\zeta^{\prime},
    \end{split}
\end{equation}
where $\zeta'\left(\textbf{r}^{\prime}, t^{\prime}\right)=b^{z-\alpha}\zeta(\textbf{r},t)$, 
with noise amplitude
\begin{equation}
\begin{split}
    \langle \zeta'\left(\textbf{r}_1^{\prime}, t^{\prime}_1\right) \zeta'\left(\textbf{r}_2^{\prime}, t^{\prime}_2\right)\rangle &= 
    2\Gamma'\delta\left(\textbf{r}_1^{\prime}-\textbf{r}_2^{\prime}\right)\delta\left( t^{\prime}_1-t^{\prime}_2\right) \\ &= 2\tilde{\Gamma}b^{z-d-2\alpha}\delta\left(\textbf{r}_1^{\prime}-\textbf{r}_2^{\prime}\right)\delta\left( t^{\prime}_1-t^{\prime}_2\right), \nonumber 
\end{split}
\end{equation}
resulting into the rescaled equation
\begin{equation}
    \partial_{t^{\prime}} \phi^{\prime}=-\Gamma' \left(r' \phi^{\prime}+u' \phi'^3-\nu'\nabla^2 \phi^{\prime}\right)+\zeta^{\prime},
\end{equation}
with the exact same shape as the original (bare) TDGL equation. 
Gathering the coarse-grained and rescaled coupling constants \cite{Mazenko}, the recursion relations for the renormalized couplings finally become
\begin{equation}
    \left\{\begin{array}{l}
 r^{\prime}=b^{2\alpha+d} \tilde{r}, \\
 u^{\prime}=b^{4\alpha +d} \tilde{u}, \\
 \nu^{\prime}=b^{2\alpha +d-2} \tilde{\nu}, \\
 \Gamma^{\prime}=b^{z-2\alpha-d} \tilde{\Gamma}.
\end{array}\right. 
\end{equation}

\subsection{Differential flow equations}
To obtain differential equations for the RG parameter flow, we take $b=e^{l}$ in the $l \rightarrow 0$ limit; thus, e.g.
\begin{equation}
\begin{split}
r^{\prime}&=e^{l(2\alpha+d)}\left\{r+\frac{3 u}{\nu} \frac{K_4 l}{1+r / \nu}\right\}\\ &\approx r+l\left\{(2\alpha+d) r+\frac{3 u}{\nu} \frac{K_4}{1+r/ \nu}\right\} +\mathcal{O}(l^2),\\
\frac{d r}{d l}&=\lim _{l \rightarrow 0} \frac{r^{\prime}-r}{l}=(2\alpha+d) r+\frac{3 u}{\nu} \; \frac{K_4}{1+r / \nu}.
\end{split}
\end{equation}
In the same fashion,
\begin{equation}
\begin{split}
& \frac{d u}{d l}=\lim _{l \rightarrow 0} \frac{u^{\prime}-u}{l}=(4\alpha +d) u-9 \frac{u^2}{\nu^2} K_4, \\
& \frac{d \nu}{d l}=\lim _{l \rightarrow 0} \frac{\nu^{\prime}-\nu}{l}=\nu\left\{(2\alpha +d-2)+\frac{3}{2} \frac{u^2}{\nu^4} K_4^2\right\}.
    \end{split}
\end{equation}
Note $(1+x)^{-1} \approx 1-x+O\left(x^2\right)$, hence
\begin{equation}
\begin{aligned}
& \tilde{\Gamma} \approx \Gamma\left[1-9 \ln (4 / 3) \frac{u^2}{\nu^4} K_4^2 l\right] , \\
& \frac{d \Gamma}{d l}=\lim _{l \rightarrow 0} \frac{\Gamma^{\prime}-\Gamma}{l}=\Gamma\left\{(z-2\alpha-d)-9 \ln (4 / 3) \frac{u^2}{\nu^4} K_4^2\right\}.
\end{aligned}
\end{equation}


\subsection{Mapping to the kinetic roughening convention}

Our ultimate goal is to obtain the DRG flow for the coupling constants corresponding to Eq.\ \eqref{eq:TDGL} of the main text. Comparing it with Eq.\ \eqref{eq:tdgl_hohen}, whose couplings are now denoted with a 0 subscript, we can make the following identification,
\begin{equation}
    \left\{\begin{array}{l}
        r = \Gamma_0 r_0,\\
        u = \Gamma_0 u_0,\\
        \nu = \Gamma_0 \nu_0,\\
        \Gamma = \Gamma_0 .
    \end{array} \right.
\end{equation}
The differential flow is computed by application of the chain rule,
$dr/dl=\Gamma_0 dr_0/dl+r_0 d\Gamma_0/dl$, and similarly for $du/dl$, $d\nu/dl$, finally yielding the DRG flow equations for the TDGL equation, Eq.\ \eqref{eq:TDGL}, provided by Eq.\ \eqref{eq:RG_flow} in Sec.\ \ref{sec:DRG}.

\section{Metropolis-Hastings algorithm}\label{AppB}

\begin{figure}[h!]
        \includegraphics[width=0.95\linewidth]{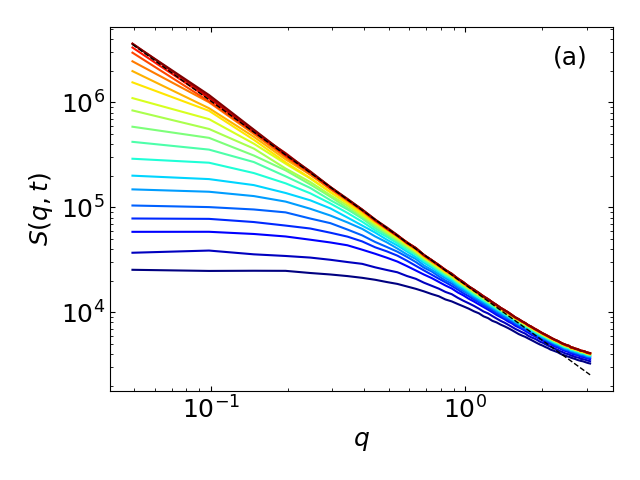}
        \includegraphics[width=0.95\linewidth]{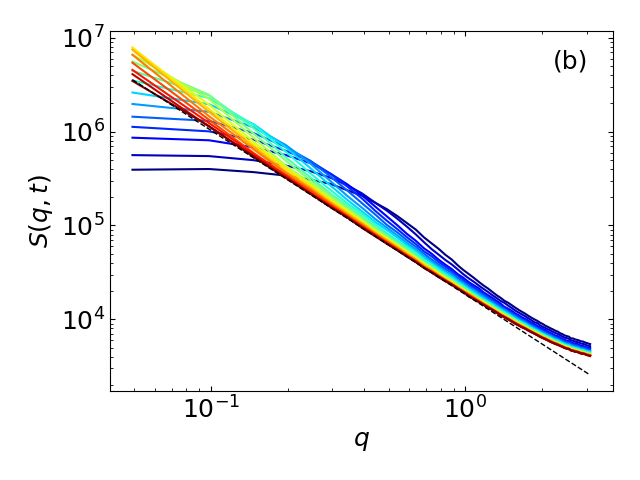}
    \caption{Dynamics of the structure factor from simulations of the Ising model with Metropolis-Hastings dynamics for a critical quench from (a) $T=0$ and (b) $T=\infty$. For both panels, the dashed line corresponds to asymptotic behavior as $S(q) \sim q^{-1.75}$ and the time arrow goes from blue to red (log-spaced).}
    \label{fig:MH_SF}
\end{figure}

\begin{figure}[h!]
        \includegraphics[width=0.95\linewidth]{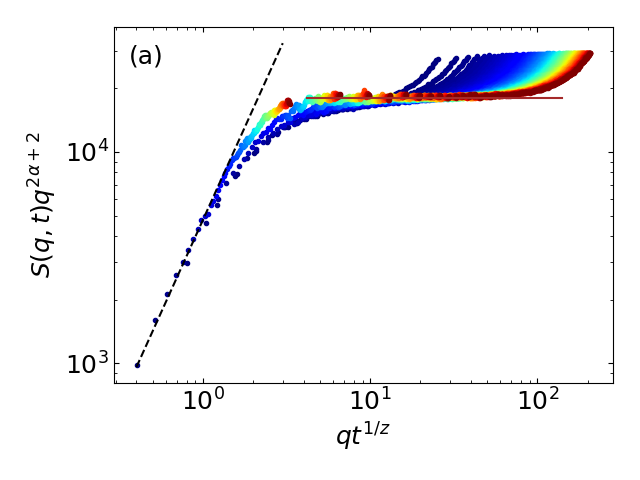}
        \includegraphics[width=0.95\linewidth]{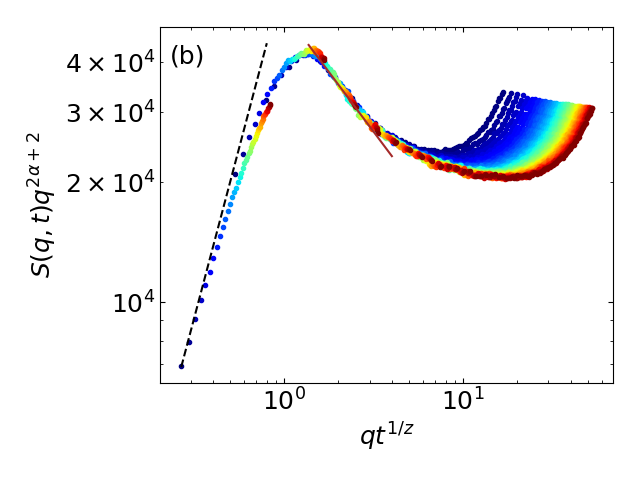}
    \caption{Collapse of the structure factor data shown in Fig.\ \ref{fig:MH_SF} for the critical quench from (a) $T=0$, using $\alpha=-1/8$ and $z=2.19$ [the dashed (solid) line corresponds to $f_S \sim u^{1.75}$ ($f_S \sim \text{csnt}$)] and from (b) $T=\infty$, using $\alpha=-0.15$ and $z=2.30$ [the dashed (solid) line corresponds to $f_{S'} \sim u^{1.70}$ ($f_{S'} \sim u^{-0.6}$)]. The time arrow goes from blue to red (linearly-spaced) for both panels.}
    \label{fig:MH_DSA}
\end{figure}

In addition to  Glauber's rule, the same methodology for the evolution of the discrete Ising model has been applied using the Metropolis-Hastings algorithm \cite{newman,Toral}. The dynamics of the structure factor for both critical quenches is shown in Fig.\ \ref{fig:MH_SF}, and the corresponding data collapses appear in Fig.\ \ref{fig:MH_DSA}.

For the critical quench from $T=0$, the qualitative behavior is analogous to that of Glauber dynamics, namely, a standard FV dynamic scaling ansatz characterized by $\alpha\approx -0.125$ and $z\approx 2.19$, close to the theoretical expectations in Table \ref{tab:exponents_KR}.

For the critical quench from $T=\infty$, both overgrowth and relaxation regimes can be identified, the former being characterized by intrinsic anomalous scaling with $\alpha\approx -0.15$, $\alpha_s\approx 0.15$, and $z\approx 2.30$. Thus, the critical dynamics of the Metropolis-Hastings algorithm seems to be more affected by corrections to scaling than the Glauber update rule, see Sec.\ \ref{sec:T=infty}.

\clearpage

\bibliography{referencias}{}

\end{document}